\documentclass[a4paper,11pt]{article}
\usepackage{jheppub} 
 \usepackage{amssymb,amsthm}
 \usepackage{amsmath,amssymb,amsfonts,bm,amscd}
 \usepackage{graphicx}
 
 \usepackage{caption, subcaption}
 \usepackage{multirow}
 \usepackage{tensor}
 
 \usepackage{tikz}
 
 \usepackage{amsthm}
 
 \theoremstyle{definition}
 \newtheorem{example}{\bf Example}[section]
 
\def\rank{\mathrm{rank}}

\def\fkf{\mathfrak{f}}
\def\fkg{\mathfrak{g}}
\def\fkh{\mathfrak{h}}
\def\fkj{\mathfrak{j}}
\def\fkl{\mathfrak{l}}
\def\fkn{\mathfrak{n}}
\def\fkp{\mathfrak{p}}

\def\CI{\mathcal{I}}
\def\CM{\mathcal{M}}
\def\CN{\mathcal{N}}
\def\CO{\mathcal{O}}
\def\CT{\mathcal{T}}

\def\bbC{\mathbb{C}}

\def\bbR{\mathbb{R}}
\def\bbZ{\mathbb{Z}}

\def\CA{{\cal A}}

\def\CI{{\cal I}}

\def\CM{{\cal M}}
\def\CN{{\cal N}}
\def\CO{{\cal O}}

\def\rank{\mathop{\rm rank}}

\def\beq#1\eeq{\begin{align}#1\end{align}}

\usepackage{enumitem}

\def\fg{\mathfrak{g}}

\def\fsl{\mathfrak{sl}}

\def\bbC{\mathbb{C}}
\def\bbS{\mathbb{S}}
\def\bbT{\mathbb{T}}
\def\bbZ{\mathbb{Z}}

\def\bfG{\mathbf{G}}
\def\bfI{\mathbf{I}}

\def\bfP{\mathbf{P}}

\def\ch{\mathrm{ch}}

\def\half{\frac{1}{2}}

\def\gcd{\mathrm{gcd}}

\def\beq#1\eeq{\begin{align}#1\end{align}}

\title{Mirror symmetry for circle compactified 4d $\mathcal{N}=2$ SCFTs}

\author[a,b]{Peng Shan}
\author[b]{Dan Xie}
\author[a]{Wenbin Yan}

\affiliation[a]{Yau Mathematics Science center, Tsinghua University, Beijing, 100084, China}
\affiliation[b]{Department of Mathematics, Tsinghua University, Beijing, 100084, China}

\abstract{We propose a mirror symmetry for 4d $\mathcal{N}=2$ superconformal field theories (SCFTs) compactified on a circle with finite size. The mirror symmetry involves vertex operator algebra (VOA) describing the Schur sector (containing Higgs branch) of 4d theory, and 
the Coulomb branch of the effective 3d theory. The basic feature of the mirror symmetry is that many representational properties of VOA are matched with geometric properties of the Coulomb branch moduli space. Our proposal is verified for a large class of 
Argyres-Douglas (AD) theories engineered from M5 branes, whose VOAs are W-algebras, and Coulomb branches are  the Hitchin moduli spaces.  VOA data such as simple modules, Zhu's algebra, and modular properties are matched with
 geometric properties like $\bbC^\ast$-fixed varieties in Hitchin fibers, cohomologies, and some DAHA representations. We also mention relationships to 3d symplectic duality.}

\begin{document} 
\maketitle
\flushbottom

\section{Introduction}

Mirror symmetry plays an important role in modern theoretical physics and mathematics as it connects a large number of disciplines including string theory, geometry, algebra, representation theory and etc. 
The two dimensional mirror symmetry \cite{Greene:1990ud} involves a pair of Calabi-Yau (CY) manifold $X, \check{X}$ which can be used to define a pair of two dimensional $(2,2)$ superconformal field theories (SCFTs) $\CT(X)$ and $\CT(\check{X})$. The statement is then that $\CT(X)$ and $\CT(\check{X})$ are dual in the infrared (IR)
\begin{equation}
\CT(X)\simeq \CT(\check{X}).
\end{equation} 
The basic feature of the mirror symmetry is that: the same physical quantities (such as prepotential) can be  computed  from different geometrical data of $X$ or $X^\vee$ \cite{Candelas:1990rm}, which leads to many interesting correspondences in mathematics. More importantly, things which are difficult to compute on one side might become easier by looking at its mirror. 

Three dimensional $\mathcal{N}=4$ SCFTs also have similar mirror symmetric properties \cite{Intriligator:1996ex}, which often involves two hyper-K\"ahler manifolds $X$ and $Y$ acting as   moduli spaces of vacua of the 3d theory. 
The basic feature of  3d mirror symmetry discussed in \cite{Intriligator:1996ex} is that $X$ (resp. $Y$) can be realized either as the Higgs (resp. Coulomb) branch of one theory $\CT_1$ or the Coulomb (resp. Higgs) branch of another theory $\CT_2$.  Again, the manifold which is difficult to describe on one side may have a simpler description in its mirror. It was further realized in \cite{braden2010gale,braden2012quantizations, braden2014quantizations,Bullimore:2016nji} that there are duality involving  geometric properties of $X$ and $Y$. For example, one can get an algebra ${\cal A}_X$ through the quantization of $X$ (and its resolution), and the representation 
theory of ${\cal A}_X$ is closed related to the geometric property of $Y$
\begin{equation}
\CA_X\longleftrightarrow Y.
\end{equation} 
This kind of duality is called symplectic duality \cite{braden2012quantizations, braden2014quantizations}.

Now consider  a four dimensional $\mathcal{N}=2$ SCFT compactified on a circle $S^1$ with finite radius. One may wonder whether there is a similar mirror symmetry. The resulting 3d effective theory has 
a Coulomb branch $\CM_C$ which is a hyper-K\"{a}hler manifold admitting torus fibration \cite{Seiberg:1996nz}, and a Higgs branch $\CM_H$ which is  the same hyper-K\"{a}hler cone as that of the original 4d theory. In this case, 
$\CM_H$ and $\CM_C$ are rather different and one does not expect to find a dual theory which exchanges the role of $\CM_C$ and $\CM_H$. 

However, motivated by the symplectic duality interpretation of the 3d mirror symmetry, the analog of the mirror symmetry of circle compactified 4d theories might be formulated as an algebra/geometry duality. 
Indeed, there are strong evidence \cite{Fredrickson:2017jcf,Fredrickson:2017yka, Dedushenko:2018bpp} that the algebra should be the  vertex operator algebra (VOA) associated with the 4d theory \cite{Beem:2013sza} which indeed consists of Higgs branch operators as a subset  \cite{Song:2017oew, Beem:2017ooy, arakawa2018chiral}, and 
the geometric side should be the Coulomb branch.  Given an arbitrary 4d $\mathcal{N}=2$ SCFT ${\cal T}$, we propose the following mirror symmetry between the corresponding $\text{VOA}(\CT)$ and the Coulomb branch $\CM_C(\CT)$ of $\CT$ compactified on the circle
\begin{equation}
\label{eq:4dmirrorCore}
\text{VOA}(\CT) \longleftrightarrow \CM_{C}(\CT),
\end{equation}
with the dictionary summarized in table \ref{table:mirrorDict}.  
Given a 4d $\mathcal{N}=2$ SCFT, it is in general difficult to know neither its associated VOA nor the Coulomb branch $\CM_{C}$. 
However, in a series of previous works, both the corresponding VOA \cite{Xie:2016evu, Song:2017oew, Wang:2018gvb,Xie:2019yds} and the Coulomb branch of a large class of 4d $\mathcal{N}=2$ SCFTs \cite{Xie:2012hs,Wang:2015mra,Wang:2018gvb} are known \footnote{Corresponding VOAs of many different series in this class of generalized AD theories were already studied in \cite{Buican:2015ina, Buican:2015hsa, Cordova:2015nma, Buican:2015tda,  Song:2015wta, Cecotti:2015lab, Nishinaka:2016hbw,
Buican:2016arp, Cordova:2016uwk, Song:2016yfd, Creutzig:2017qyf,  Cordova:2017ohl, Cordova:2017mhb, Buican:2017uka, Song:2017oew, Buican:2017fiq, Buican:2017rya, Choi:2017nur, Creutzig:2018lbc, Nishinaka:2018zwq, Beem:2019snk}.}, so one can thoroughly study and check the mirror symmetry for this class of theories. 

\begin{table}[h]
\begin{center}
\begin{tabular}{|c|c|}
\hline
$ \text{VOA}(\CT)$ & $\CM_C(\CT)$ \\ \hline
Simple modules & $\bbC^\ast$-fixed varieties \\ \hline
Conformal weights & Critical values of moment maps \\ \hline
Zhu's $C_2$ algebra & Cohomology ring \\ \hline
\begin{tabular}{c}Modular properties of \\space of characters\end{tabular} & \begin{tabular}{c}Modular properties of \\ cohomology of $\bbC^\ast$-fixed varieties\end{tabular} \\ \hline
\end{tabular}
\end{center}
\caption{\label{table:mirrorDict}Dictionary between $\text{VOA}(\CT)$ and $\CM_C(\CT)$ for a 4d $\CN=2$ SCFT $\CT$ compactified on a circle.}
\end{table}

This class of theories is engineered by compactification of a 6d $(2,0)$ theory of type $\mathfrak{j}=\text{ADE}$ on a sphere with a regular and an irregular singularity. For our interest, the irregular singularity 
is labelled by a rational number $\nu=\frac{u}{m}$ \footnote{$m$ takes value from a finite set given by the Lie algebra, and $u\geq 1$.} (see table \ref{table:ellnumber} for allowed values), and the regular singularity is labelled by a nilpotent orbit of $\mathfrak{j}$. 
It was found in \cite{Xie:2016evu, Song:2017oew, Xie:2019yds} that the associated VOA is the W-algebra $W_{-h^\vee+\frac{1}{\nu}}(\mathfrak{j}, f)$, and the associated $\CM_{C}$ is the Hitchin moduli space $\CM_{Hit}(\mathfrak{j}, \nu, (f^\vee, c))$ with $(f^\vee, c)$ being the dual  of $f$ and $c$ being a conjugacy class of the component group \footnote{We will omit $c$ when the component group is trivial or $c=1$.} \cite{achar2003order}. Therefore 
the mirror symmetry is the correspondence between the following two objects
\begin{equation}
\label{eq:mirrorADE}
W_{-h^\vee+\frac{1}{\nu}}(\mathfrak{j}, f)  \longleftrightarrow \CM_{Hit}(\mathfrak{j}, \nu, (f^\vee, c)).
\end{equation}
One can also get non-simply laced W-algebra by doing outer automorphism twist around the singularity \cite{Wang:2018gvb}, and the pair of objects are
\begin{equation}
\label{eq:mirrorNonSimply}
W_{-h^\vee+\frac{1}{n\nu}}(\mathfrak{g}, f)  \longleftrightarrow \CM_{Hit}((\mathfrak{j},o), \nu, (f^\vee,c)).
\end{equation}
Here $o$ is the outer automorphism of ADE Lie algebra $\mathfrak{j}$ whose invariant Lie algebra is $\mathfrak{g}^\vee$ (the Langlands dual of $\mathfrak{g}$), $n$ is the lacety of $\fkg$, summarized in table \ref{table:outm}. 
The simply laced case \eqref{eq:mirrorADE} can also be fit into \eqref{eq:mirrorNonSimply} by noticing that $\fkj=\fkg=\fkg^\vee$ when $\fkj$ is simply laced and choosing $o=\{1\}$.  The appearance of a Lie algebra and its Langlands dual on each side of the duality is a feature similar to many dualities of physical theories found before (For example, in 4d $\CN=4$ SYM theories).

In the following we briefly explain how the representation aspects of VOA is related to geometric property of Coulomb branch in our particular class of examples. Part of the statements can be formulated rigorously and will be proved in a parallel math paper \cite{Xie:2023pre}.
\begin{enumerate}
\item \textbf{Simple modules in the category $\CO$ of VOA and $\mathbb{C}^\ast$ fixed varieties of $\CM_C$}: There is a bijiection between the simple modules of $W_{-h^\vee+\frac{1}{n\nu}}(\mathfrak{g}, f)$ \footnote{To be more precise, they are simple modules in the category $\CO$ for affine vertex algebras and simple Ramond twisted modules for general W-algebras.} and the $\bbC^\ast$-fixed varieties of $\CM_{Hit}((\mathfrak{j},o), \nu, (f^\vee,c))$. It was first observed in \cite{Fredrickson:2017jcf} for cases when 4d theories are $(A_{N-1},A_{M-1})$ Argyres-Douglas (AD) theories with $N$ and $M$ coprime \footnote{These correspond to $\fkg=A_{N-1}$, $\nu=\frac{M+N}{N}$ and $f=principal$ in our notation.}, then generalized to cases when the 4d theories are $(A_1, A_N)$ and $(A_1, D_N)$ AD theories for $N\in\bbZ_{>0}$ in \cite{Fredrickson:2017yka}. To generalize this correspondence to arbitrary $\fkg$, $\nu$ and $f$, a crucial observation is that the fixed varieties of Hitchin moduli spaces $\CM_{Hit}((\mathfrak{j},o), \nu, f^\vee)$
are reduced to that of the affine Springer fibre of elliptic type, and there is a nice algebraic description of the latter. Using this description, we find a natural bijection between fixed varieties and simple modules of the corresponding affine Lie algebra when the level is boundary admissible \footnote{This happens when $\nu=\frac{u}{ h_\theta}$, where $h_\theta$ is the Coxeter number for untwisted theory, and twisted Coxeter number for twisted case.}. This will be explained in \cite{Xie:2023pre}.  For general W-algebras, it is conjectured in  \cite{kac2008rationality} that simple modules can be obtained from simple modules of the affine Lie algebra from BRST reduction. We explain also in loc.~cit.~ that this reduction is the same as a reduction of fixed varieties on the Hitchin side. Moreover, our results also provide predictions for classifications of simple modules of non-admissible W-algebras.

\item  \textbf{Conformal weight and momentum map}: One can compute the momentum map for a fixed point using the Morse theory on $\CM_{C}$, and match this with the conformal weight of the corresponding VOA \cite{Fredrickson:2017jcf,Fredrickson:2017yka}. 
In this work, we propose  a general formula relating  conformal weights of simple modules of $W_{-h^\vee+\frac{1}{n\nu}}(\mathfrak{g}, f)$ to the values of moment maps of $\bbC^\ast$-fixed points of $\CM_{Hit}((\mathfrak{j},o), \nu, (f^\vee,c))$.
\item  \textbf{Modular transformation and DAHA}: The space of characters of simple modules of some VOA's admit modular property with respect to certain $SL(2,\bbZ)$ action. This was shown for admissible AKM \cite{kac2008rationality} and W-algebras \cite{kac1988modular}. On the other hand, the cohomology of fixed varieties of $\CM_{Hit}((\mathfrak{j},o), \nu, f^\vee)$ gives a finite dimensional representation of double affine Hecke algebra (DAHA)\cite{varagnolo2009finite, oblomkov2016geometric}, and in some cases, it admits a projective action of $SL(2,\bbZ)$ which is compatible with corresponding automorphisms of DAHA \cite{cherednik2005double}. For  admissible W-algebras, we show in \cite{Xie:2023pre} that the $SL(2,\bbZ)$ representations on both sides coincide \footnote{The relation between modular matrices of minimal W-algebras of $A$ type and spherical DAHA of $A$ type was already shown in \cite{Gukov:2022gei}.}. Our result also gives interesting insights on the modular property of non-admissible W-algebras.

\item \textbf{Modular property and Coulomb branch index}: The Coulomb branch index of a 4d theory on lens space $L(k,1)$ times  $S^1$ can be computed using the Morse theory data on the fixed varieties of Coulomb branch. It was found in \cite{Fredrickson:2017yka,Kozcaz:2018usv} 
that the Coulomb branch index is related to the modular properties of the corresponding VOA. We will show that the same relation works for the admissible cases, which gives strong hint that such relation should work in general.



\item   \textbf{Zhu's $C_2$ algebra and cohomology ring}: There is a so-called Zhu's $C_2$-algebra associated with the $\text{VOA}(\CT)$ \cite{zhu1996modular}. It provides important information on the representation theory. On the Hitchin side, one naturally have 
a cohomological ring. In the context of principal admissible W-algebra, we find that Zhu's $C_2$ algebra 
is the same as the cohomology ring of $\CM_C(\CT)$. In general, we would expect that the cohomological ring should be related to some algebra on the VOA side which characterizes  simple modules.

\item  \textbf{Relation with 3d symplectic duality}: One can take the radius of the compactification circle to zero to get a 3d $\mathcal{N}=4$ SCFT from the 4d theory $\CT$.
The Higgs branch $\CM_H^{3d}(\CT)$ of the 3d theory is the same as $\CM_H(\CT)$, while the Coulomb branch $\CM_C^{3d}(\CT)$ is related to $\CM_C(\CT)$ in a less obvious way \cite{boalch2008irregular,Benini:2010uu}. The Higgs and Coulomb branch of 3d theory can also be described by its 3d mirror \cite{Xie:2021ewm}. $\CM_H^{3d}(\CT)$ and $\CM_C^{3d}(\CT)$  naturally forms a symplectic pair, and  many known symplectic pairs can be obtained in this way. Moreover, a finite W-algebra can be found as the twisted Zhu's algebra of $\text{VOA}(\CT)$ \cite{de2006finite}, which is exactly the same algebra studied in the context of 3d symplectic duality. So from 4d perspective, the appearance of an algebra in the symplectic duality is natural.

\end{enumerate}
We would like to add that there is one more interesting relation for the mirror pair: the character of VOA modules can be computed using the wall crossing data on $\CM_C(\CT)$ \cite{Cecotti:2010fi,Cordova:2015nma,Cecotti:2015lab}. We would not discuss 
this duality in this paper, but hope to study it in the future.

\begin{figure}
\begin{center}

\tikzset{every picture/.style={line width=0.75pt}} 

\begin{tikzpicture}[x=0.55pt,y=0.55pt,yscale=-1,xscale=1]

\draw   (121,133) .. controls (121,121.95) and (136.67,113) .. (156,113) .. controls (175.33,113) and (191,121.95) .. (191,133) .. controls (191,144.05) and (175.33,153) .. (156,153) .. controls (136.67,153) and (121,144.05) .. (121,133) -- cycle ;
\draw   (153,119) -- (162,119) -- (162,128) -- (153,128) -- cycle ;
\draw    (152,138.22) -- (162,148.22) ;
\draw    (151,148) -- (161,138.22) ;

\draw   (417,196) .. controls (417,184.95) and (432.67,176) .. (452,176) .. controls (471.33,176) and (487,184.95) .. (487,196) .. controls (487,207.05) and (471.33,216) .. (452,216) .. controls (432.67,216) and (417,207.05) .. (417,196) -- cycle ;
\draw   (449,182) -- (458,182) -- (458,191) -- (449,191) -- cycle ;
\draw    (448,201.22) -- (458,211.22) ;
\draw    (447,211) -- (457,201.22) ;

\draw    (233,266) -- (374,266.22) ;
\draw    (233,259) -- (374,259.22) ;

\draw (127,52.4) node [anchor=north west][inner sep=0.75pt]    {$6d\ ( 2,0) \ g$};
\draw (144,194.4) node [anchor=north west][inner sep=0.75pt]    {$S^{1}$};
\draw (143,254.4) node [anchor=north west][inner sep=0.75pt]    {$R^{1,2}$};
\draw (417,51.4) node [anchor=north west][inner sep=0.75pt]    {$6d\ ( 2,0) \ g$};
\draw (442,114.4) node [anchor=north west][inner sep=0.75pt]    {$S^{1}$};
\draw (433,253.4) node [anchor=north west][inner sep=0.75pt]    {$R^{1,2}$};

\end{tikzpicture}

\end{center}
\caption{Left: One first compactify 6d $(2,0)$ theory on a Riemann surface to get a 4d theory, and then on a circle to get an effective 3d theory; Right: One first compactify 6d $(2,0)$ theory 
on a circle to get a 5d theory and then on a Riemann surface to get an effective 3d theory. The Coulomb branch of the theory on the left is given by the Higgs branch of 
the theory on the right. }
\label{intro1}
\end{figure}
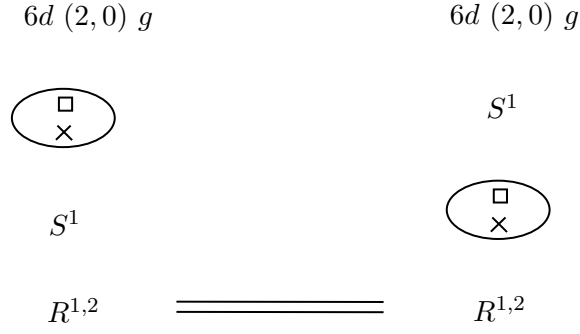

\textbf{Physical interpretation of mirror symmetry}:  Let us now justify the name of mirror symmetry, namely the Coulomb branch of circle compactified 4d theory $\CT_1$ is given by the Higgs branch of another theory $\CT_2$. The crucial difference with respect to the 3d mirror is that $\CT_2$ has to be a \textbf{five} dimensional theory.
 Following the discussion in \cite{Benini:2010uu}, one first compactifies the 6d theory on a Riemann surface $\Sigma$ and then on a circle $S^1$ to get a 4d $\CN=2$ theory on a circle. On the other hand, by changing the order of compactification, one first gets a 5d maximal SYM in the low energy from the 6d theory. The Coulomb  branch of original theory is then the 
Higgs  branch of the 5d theory compactified on $\Sigma$. This  leads to the description of the Coulomb branch of the 4d theory on a circle as the Hitchin moduli space by explicitly writing down the Higgs branch equation of motion of the 5d theory (figure \ref{intro1}).

The paper is organized as the following: in section \ref{sec:ADtheory}, we review the classification of 4d $\mathcal{N}=2$ SCFTs from 6d $(2,0$) theory and  the 
structure of their Coulomb and Higgs (Schur) branches. Section \ref{sec:repAdmWalgebra} reviews the representation theory of admissible W-algebras. Section \ref{sec:CBfixedPoints} discusses the zero fiber of Hitchin moduli space, its relation to affine Springer fibre, and the computation of  fixed varieties. Using the knowledge of previous sections, we finally check the dictionary of the mirror symmetry in table \ref{table:mirrorDict} which is the main focus of section \ref{sec:4dmirror}. We will mainly provide examples and predictions here.
Finally various generalizations are discussed in section \ref{sec:concl}.

\newpage
\section{4d $\CN=2$ SCFTs from 6d SCFTs on a sphere}
\label{sec:ADtheory}

4d $\mathcal{N}=2$ theories has two kinds of moduli spaces of vacua: the Coulomb branch and the  Higgs branch. 
The low energy effective theory of the Coulomb branch is solved by the Seiberg-Witten (SW) solution \cite{Seiberg:1994rs,Seiberg:1994aj}. Roughly speaking, the SW solution is given by 
a family of algebraic varieties fibered over a base manifold $B$ which is the Coulomb branch of the 4d theory on flat space. If we further compactify 4d theory on a circle with finite radius $R$, the effective 
 theory also has a Coulomb branch $\CM_C$ which is a hyper-K\"{a}hler manifold \cite{Seiberg:1996nz}. $\CM_C$ is 
given by an abelian variety fibered over the base $B$ in one of its complex structures.

In general, it is difficult to find the SW solution for an arbitrary 4d $\mathcal{N}=2$ theory.  
However, for models constructed using the 6d $(2,0)$ theory, one can find SW solutiongs using Hitchin moduli spaces \cite{Gaiotto:2009we,Gaiotto:2009hg}. 
Given a 6d $(2,0)$ theory of type $\fkj$, a Riemann surface $\Sigma_{g,n}$ of genus $g$ with $n$ punctures, one obtains a $4d$ $\CN=2$ SCFT by compactification of the 6d theory on $\Sigma_{g,n}$, then the Coulomb branch of this 4d theory on $S^1$ is the same as the moduli space of the Hitchin system on $\Sigma_{g,n}$. In the following section, we will review data required to specify the 4d theory when the Riemann surface is a sphere with one irregular and one regular punctures \cite{Xie:2012hs,Wang:2015mra,Wang:2018gvb}.

\subsection{Basic constructions}
\label{sec:ADandWk}

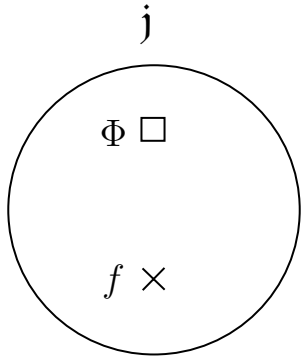
\begin{figure}
	
	\tikzset{every picture/.style={line width=0.75pt}} 
	\centering
	\begin{tikzpicture}[x=0.75pt,y=0.75pt,yscale=-1,xscale=1]
	
	\draw   (125.52,135.74) .. controls (125.52,95.57) and (158.08,63) .. (198.26,63) .. controls (238.43,63) and (271,95.57) .. (271,135.74) .. controls (271,175.92) and (238.43,208.48) .. (198.26,208.48) .. controls (158.08,208.48) and (125.52,175.92) .. (125.52,135.74) -- cycle ;
	
	\node at (195,40) {{\LARGE$\mathfrak{j}$}};
	\draw   (192,89.48) -- (203.52,89.48) -- (203.52,101) -- (192,101) -- cycle ;
	\node at (178,172) {\Large $f$};
	\draw   (192.53,165.18) -- (203.58,175.9)(203.49,164.94) -- (192.62,176.13) ;
	\node at (178,97) {\Large $\Phi$};
	\end{tikzpicture}
	\caption{A 4d AD theory is constructed by putting a 6d $(2,0)$ theory of type $\mathfrak{j}$ on a sphere with one irregular singularity and one regular singularity. The irregular singularity is labeled by $\Phi$, see \eqref{ire}, and 
	the regular singularity is labeled by $f$.}
	\label{6dconstruct}
\end{figure}

One can engineer a large class of 4d $\mathcal{N}=2$ SCFTs by putting a 6d $(2,0)$ theory of type $\mathfrak{j}=ADE$ on a sphere with an irregular singularity and a regular singularity \cite{Gaiotto:2009we,Gaiotto:2009hg,Xie:2012hs,Wang:2015mra,Wang:2018gvb} (figure \ref{6dconstruct}). 
The Coulomb branch of this 4d $\mathcal{N}=2$ theory is captured by a Hitchin system with the following boundary conditions near the irregular singularities
\begin{equation}
\Phi(z)=\left({T_k\over z^{2+{k\over b}}}+\sum_{-b\leq l<k}\frac{T_l}{z^{2+\frac{l}{b}}}+\cdots\right)dz.
\label{ire}
\end{equation}
Here one first choose a $\bbZ/b \bbZ$ grading (a positive principal grading) of Lie algebra $\fkj$ \cite{reeder2012gradings}
\begin{equation}
\fkj=\oplus_{i\in \bbZ/b\bbZ}\fkj_{i/b},
\end{equation}
then each $T_l$ is a regular semi-simple element in $\fkj_{i/b}$. Possible choices of the integer $b$ for each $\fkj$ are listed in table \ref{table:sing:b}, and the integer  $k$   is greater than $-b$. Subsequent terms of the Higgs field are chosen 
such that they are compatible with the leading order term (essentially determined by the grading). We call them irregular punctures of $\fkj^{b}[k]$ type. This choice of irregular singularities ensures that the resulting 4d $\CN=2$ theory has a $U(1)_R$ symmetry and therefore superconformal. 
Theories constructed using only these irregular singularities can also be engineered by putting type IIB string theory on a three dimensional singularity \cite{Xie:2015rpa} as summarized in table \ref{table:sing:b}.  One can add another regular singularity which is labelled by an element $f$ in a nilpotent orbit of $\mathfrak{j}$ \footnote{We use Nahm labels such that the trivial orbit corresponding to regular puncture with maximal flavor symmetry. A detailed discussion about these defects can be found in  \cite{Chacaltana:2012zy}.}. 
All in all the 4d theory in our consideration is specified by four labels  $\boxed{<\mathfrak{j}, b, k, f>}$, wtih $\mathfrak{j}$ labelling the type of 6d $(2,0)$ SCFT, $b,k$ specifying the irregular singularity, and $f$ fixing the regular singularity.

\begin{table}[!htb]\small
\begin{center}

	\begin{tabular}{ |c|c|c|c|c|c| }
		\hline
		$ \mathfrak{j}$& $b$  & Singularity  & Spectral curve at SCFT point & $\Delta[z]$ & $\mu$ \\[1pt] \hline
		$A_{N-1}$&$N$ &$x_1^2+x_2^2+x_3^N+z^k=0$ &$x^{N}+z^{k}=0$ & ${N\over N+k}$ & $(N-1)(k-1)$ \\ \hline
		$~$  & $N-1$ & $x_1^2+x_2^2+x_3^N+x_3 z^k=0$ &$x^{N}+xz^{k}=0$ & ${N-1\over N+k-1}$ & $N(k-1)+1$ \\ \hline
		
		$D_N$  &$2N-2$ & $x_1^2+x_2^{N-1}+x_2x_3^2+z^k=0$ &$x^{2N}+x^{2}z^{k}=0$ & ${2N-2\over 2N+k-2}$ & $N(k-1)$ \\     \hline
		$~$  & $N$ &$x_1^2+x_2^{N-1}+x_2x_3^2+z^k x_3=0$&$x^{2N}+z^{2k}=0$ & ${N\over N+k}$ & $2k(N-1)-N$  \\     \hline
		
		$E_6$&12  & $x_1^2+x_2^3+x_3^4+z^k=0$   &$x^{12}+z^{k}=0$ & ${12\over 12+k}$  & $6k-6$ \\     \hline
		$~$ &9 & $x_1^2+x_2^3+x_3^4+z^k x_3=0$   &$x^{12}+x^{3}z^{k}=0$ & ${9\over 9+k}$  & $8k-6$ \\     \hline
		$~$  &8 & $x_1^2+x_2^3+x_3^4+z^k x_2=0$    &$x^{12}+x^{4}z^{k}=0$ & ${8\over 8+k}$  & $9k-6$ \\     \hline
		
		$E_7$& 18  & $x_1^2+x_2^3+x_2x_3^3+z^k=0$   &$x^{18}+z^{k}=0$ & ${18\over 18+k}$   & $7k-7$\\     \hline
		$~$&14   & $x_1^2+x_2^3+x_2x_3^3+z^kx_3=0$    &$x^{18}+x^{4}z^{k}=0$ & ${14\over 14+k}$  & $9k-7$ \\     \hline

		$E_8$ &30   & $x_1^2+x_2^3+x_3^5+z^k=0$  &$x^{30}+z^{k}=0$ & ${30\over 30+k}$  & $8k-8$ \\     \hline
		$~$  &24  & $x_1^2+x_2^3+x_3^5+z^k x_3=0$ &$x^{30}+x^{6}z^{k}=0$ & ${24\over 24+k}$  & $10k-8$  \\     \hline
		$~$  & 20  & $x_1^2+x_2^3+x_3^5+z^k x_2=0$  &$x^{30}+x^{10}z^{k}=0$ & ${20\over 20+k}$  & $12k-8$ \\     \hline
	\end{tabular}

\end{center}
\caption{Three-fold isolated quasi-homogenous singularities of cDV type corresponding to the $\fkj^{b}[k]$  irregular punctures of the regular-semisimple type in \cite{Wang:2015mra}. These 3d singularity is very useful in extracting the Coulomb branch spectrum \cite{Xie:2015rpa}. }
\label{table:sing:b}
\end{table}

\begin{table}[htb]
	\begin{center}
		\begin{tabular}{ |c|c| c|c|c|c| }
			\hline
			$  j $ ~&$A_{2N}$ &$A_{2N-1}$ & $D_{N+1}$  &$E_6$&$D_4$ \\ \hline
			Outer-automorphism $o$  &$\bbZ_2$ &$\bbZ_2$& $\bbZ_2$  & $\bbZ_2$&$\bbZ_3$\\     \hline
			Invariant subalgebra  $\fg^\vee$ &$B_N$&$C_N$& $B_{N}$  & $F_4$&$G_2$\\     \hline
			Flavor symmetry $\fg$ &$C_N^{(1)}$&$B_N$& $C_{N}^{(2)}$  & $F_4$&$G_2$\\     \hline
			Lacety $n$ & $4$ & $2$ &2&$2$ & $3$ \\ \hline
		    $h_\theta$ & $4N+2$ & $4N-2$ &2N+2&$18$ & $12$ \\ \hline
		\end{tabular}
	\end{center}
	\caption{Outer-automorphisms of simple Lie algebras $ \fkj$, its invariant subalgebra $ \fkg^\vee$ and flavor symmetry $ \fkg$ from the Langlands dual of $\fkg^\vee$.}
	\label{table:outm}
\end{table}

\begin{table}[!htb]
\small
	\begin{center}
		\begin{tabular}{|c|c|c|c|c|}
			\hline
			$\mathfrak{j}/o$ & $b_t$ & SW geometry at SCFT point & Spectral curve at SCFT point & $\Delta[z]$ \\[1pt] \hline
			$A_{2N}/\mathbb{Z}_2$  & $2N+1$ &$x_1^2+x_2^2+x^{2N+1}+z^{k+{1\over2}}=0$ & $x^{2N+1}+z^{k+{1\over2}}=0$ & ${4N+2\over 4N+2k+3}$  \\[3pt] \hline
			~& $2N$ & $x_1^2+x_2^2+x^{2N+1}+xz^{k}=0$ & $x^{2N+1}+xz^{k}=0$ & ${2N\over k+2N}$ \\[3pt] \hline
			$A_{2N-1}/\mathbb{Z}_2$&  $2N-1$ & $x_1^2+x_2^2+x^{2N}+xz^{k+{1\over2}}=0$ & $x^{2N}+x z^{k+{1\over2}}=0$ & ${4N-2\over 4N+2k-1}$ \\[3pt] \hline
			~ &   $2N$ &$x_1^2+x_2^2+x^{2N}+z^{k}=0$ & $x^{2N}+z^{k}=0$ & ${2N\over 2N+k}$  \\[3pt] \hline
			$D_{N+1}/\mathbb{Z}_{2}$  & $N+1$ & $x_1^2+x_2^{N}+x_2x_3^2+x_3z^{k+{1\over2}}=0$ & $x^{2N+2}+z^{2k+1}=0$ & ${2N+2\over 2k +2N+3}$ \\[3pt] \hline
			~ &   $2N$ &$x_1^2+x_2^{N}+x_2x_3^2+z^{k}=0$ & $x^{2N+2}+x^{2}z^{k}=0$ & ${2N\over k+2N}$  \\[3pt] \hline
			$D_4/\mathbb{Z}_3$ & $4$ &$x_1^2+x_2^{3}+x_2x_3^2+x_3z^{k\pm {1\over3}}=0$ & $x^{8}+z^{2k\pm \frac{2}{3}}=0$ & ${12\over 12+3k\pm1}$  \\[3pt] \hline
			~&   $6$ &$x_1^2+x_2^{3}+x_2x_3^2+z^{k}=0$ & $x^{8}+x^{2}z^{k}=0$ & ${6\over 6+k}$  \\[3pt] \hline
			$E_6/\mathbb{Z}_2$  & $9$ &$x_1^2+x_2^{3}+x_3^4+x_3z^{k+{1\over2}}=0$ & $x^{12}+x^{3}z^{k+{1\over2}}=0$ & ${18\over 18+2k+1}$  \\[3pt] \hline
			~&   $12$ &$x_1^2+x_2^{3}+x_3^4+z^{k}=0$ & $x^{12}+z^{k}=0$ & ${12\over 12+k}$  \\ [3pt]\hline
			~ &   $8$ &$x_1^2+x_2^{3}+x_3^4+x_2z^{k}=0$ & $x^{12}+x^{4}z^{k}=0$ & ${8\over 8+k}$  \\[3pt] \hline
		\end{tabular}
		\caption{SW geometry of twisted theories at the SCFT point. Here we also list the scaling dimension of coordinate $z$. All $k$'s in this table are integer valued and the power of $z$ coordinate in singularity is equal to  $k_t$ used in equation \eqref{eq:irregularPunctureTwisted}. For example, in the case $(D_4,\bbZ_3, b_t=4)$, $k_t$ is $k\pm\frac{1}{3}$. The definition of $k_t$ and $b_t$ are slightly different from \cite{Li:2022njl} but the ration $k_t/b_t$ remains the same.}
		\label{table:SW:twisted}
	\end{center}
\end{table}

To get non-simply laced flavor groups, we need to specify some outer-automorphism twist of ADE Lie algebra $\fkj$. A systematic study of these AD theories 
was performed in \cite{Wang:2018gvb}. Denoting by $\fkg^\vee$ the invariant algebra of $\mathfrak{j}$ under the twist and  $\fg$ its Langlands dual. Outer-automorphisms and invariant algebras  are summarized in table \ref{table:outm}. The irregular singularity of regular semi-simple type is also classified  in table \ref{table:SW:twisted} with the following form,
\begin{equation}
\Phi(z) =\left({T^t\over z^{2+{k_t\over b_t}}}+\cdots\right)dz.
\label{eq:irregularPunctureTwisted}
\end{equation} 
Here $T^t$ is a simi-simple element of Lie algebra $\fg^{\vee}$, and
the novel thing  is that $k_t$  can take half-integer value or 
in $\frac{1}{3}\bbZ$ ($\fg=G_2$) \cite{Wang:2018gvb}. We could again add 
a twisted regular puncture labeled  by a nilpotent orbit $f$ of $\fg$.  A 4d $\CN=2$ theory is then determined by following data $\boxed{<\mathfrak{j}, o, b_t, k_t, f>}$, with $\fkj$ labelling the type of 6d $(2,0)$ SCFT, $o$ being the outer automorphism twist , $b_t$ and $k_t$ together determining the irregular singularity, and finally $f$ fixing the 
regular singularity.

\textbf{Remark}: The label $f$ here is actually the so-called Nahm (Higgs) label. The actual boundary condition of the Higgs field $\Phi$ around the regular singularity looks like
\begin{equation}
\label{eq:regularBC}
\Phi(z) \sim\left(\frac{f^\vee}{z}+\cdots\right)dz,
\end{equation}
where $f^\vee\in\overline{\CO}_{f^\vee}$. The nilpotent orbit $\CO_{f^\vee}$ in $\fkg^\vee$ is the Spaltanstein dual of $\CO_f$. More carefully, one also needs to specify a conjugacy class $c$ in the component group for the Higgs field \cite{Chacaltana:2012zy}, which will be reviewed in section \ref{sec:CBfixedPoints}.

\subsection{Coulomb branch as Hitchin moduli space}
\label{sec:CBdim}

As discussed above, the Coulomb branch of the theory $\CT_{\fkj,b,k,f}$ (resp. $\CT_{\fkj,o,b_t,k_t,f}$) on a circle is specified by the Hitchin moduli space $ \CM_{Hit}(\mathfrak{j}, \nu, (f^\vee,c))$ with $\nu=\frac{k}{b}+1$ (resp. $\CM_{Hit}((\mathfrak{j},o), \nu, (f^\vee,c))$ with $\nu=\frac{k_t}{b_t}+1$). Given a solution $\Phi(z)\in \CM_{Hit}$, its spectral curve
\begin{equation}
\det(x-\Phi(z))=0
\end{equation}
is identified with the SW curve. In certain cases, the spectral curve is equivalent to the mini-versal deformation of the singularity (listed in table \ref{table:sing:b} and \ref{table:SW:twisted}). One can see that
$\CM_{Hit}$ is fibered over $B$ through the Hitchin map
\begin{equation}
\CM_{Hit}\to B,
\end{equation}
where the base $B$ is the moduli space of the spectral (SW) curve which is just the Coulomb branch of the 4d theory on flat spaces.

Properties of $\CM_{Hit}$ with $f$ trivial were recently studied in \cite{bezrukavnikov2022non}. One interesting information is the complex dimension of the base $B$, which is equal to
the dimension of the fibre due to the property of hyper-K\"{a}hler manifold.  Here we provide a way to compute $\dim B$ from physics. Since coordinates of $B$ 
are parameterized by vacuum expectation values (vev's) of 4d Coulomb branch operators, we can find $\dim B$ by counting the number of 4d Coulomb branch operators. This can be done as following: the spectral curve takes the 
form $f_{ADE}(x,y,z,w)+ \sum a_i \phi_i(z)=0$, and the existence of $\bbC^\ast$ action on $\CM_{Hit}$ ensures that one can define a $\bbC^\ast$ action on the coordinates $x,y,z,w$ by requiring that the spectral curve is homogeneuous under the $\bbC^\ast$ action and $[x]+[z]=1$. From these, one can deduce the 
 $\bbC^\ast$ charge of the coordinate $a_i$. Those $a_i$'s with $\bbC^\ast$ charge greater than $1$ are identified as  Coulomb branch operators, then $\dim B$ is the number of such $a_i$'s.

\begin{example}
Consider a theory whose spectral curve is given by $x^2+z^5+u_1 z^3+u_2 z^2+ u_3 z+u_4=0$. The $\bbC^\ast$ charges are $[x]={5\over 7}, [z]={2\over 7}$ , so the scaling dimensions of base coordinates are 
\begin{equation}
[u_1]={4\over 7} ,~~[u_2]={6\over 7}~~,[u_3]={8\over 7}~~,[u_4]={10\over 7},
\end{equation} 
so there are two coordinates with $\bbC^\ast$ charge greater than $1$, then \begin{equation}
\dim B=2.
\end{equation}
\end{example}

One can  compute $\dim B$ by  the Milnor number of singularity. First, the dimension of the charge lattice of the Coulomb branch is  $2\dim B+\fkf$, where $\fkf$ is the rank of flavor symmetries. This dimension is  the same as the Milnor number $\mu$ of the singularity,  so we have the formula \cite{Xie:2015rpa, Chen:2016bzh, Wang:2016yha, Chen:2017wkw}
\begin{equation}
\label{eq:dimCBmilnor}
\dim B= \frac{1}{2} (\mu-\fkf).
\end{equation}
For a quasi-homogeneous singularity, one can assign a weight $q_i$ for the $i$-th coordinate such  that the weight of the singularity is one, then the Milnor number of the singularity is 
\begin{equation}
\mu=\prod_i (1-{1\over q_i}),
\end{equation}
which is always an integer. We then need to find out the number of mass parameters (those coordinates in the mini-versal deformations with scaling dimension one) which gives $\fkf$.

\begin{example}
Consider the singularity which is given as $x^2+y^5=0$ with weight assignments $(x,y)=({1\over2}, {1\over5})$, then the Milnor number is  $\mu=4$, and there 
is also no mass parameter, so 
\begin{equation}
\dim B=\mu/2=2.
\end{equation}
\end{example}

In general, the dimension of $B$ of the theory $\CT_{\fkj,b,k,f}$ and $\CT_{\fkj,o,b_t,k_t,f}$ are specified by the following formula:
\begin{itemize}
\item For the untwisted theory $\CT_{\fkj,b,k,f}$, 
\begin{equation}
\label{eq:CBdimUT}
\dim B={({h {k\over b}-1})\rank(\mathfrak{g})-f_0\over2}-{\dim \CO_{prin} \over 2}+{\dim {\cal O}_{f^\vee}\over 2}.
\end{equation}
Here $h$ is the Coxeter number for the Lie algebra $\mathfrak{j}$. $f_0$ is the number of mass parameter in irregular singularity \cite{Wang:2018gvb, Xie:2017aqx}, and $\dim \CO_{prin}$ is the complex dimension of
principal nilpotent orbit of $\mathfrak{g}$ which is equal to $\dim(\mathfrak{j})-\rank(\mathfrak{j})$.

\item For the twisted theory $\CT_{\fkj,o,b_t,k_t,f}$ ,
\begin{equation}
\label{eq:CBdimT}
\dim B={({h_\theta {k_t'\over b_t}-1})\rank(\mathfrak{g})-f_0\over2}-{\dim \CO_{prin} \over 2}+{\dim {\cal O}_{f^\vee}\over 2}.
\end{equation}
Here  $k_t'=nk_t+nb_t$ and $n$ is the order of outer-outmorphism $o$.
$h_\theta $ is the twisted Coxeter number listed in the last line of \ref{table:outm}. $f_0$ is the number of mass parameters in irregular singularity \cite{Wang:2018gvb, Xie:2017aqx},
and  $\CO_{prin}$ is the principal nilpotent orbit of Lie algebra $\fkg^\vee$.
\end{itemize}
The above formula is found by explicitly computing the graded Coulomb branch dimensions, see \cite{Li:2022njl} for the derivation.
We also give the explicit expression for $\dim B$ when $f$ is trivial or principal orbit.
\begin{itemize}
\item If $f$ is trivial, the dimension of $B$ is  
\begin{equation}
\dim B={({h_\theta {k_t'\over b_t}-1})\rank(\mathfrak{g})-f_0\over2}.
\end{equation}
This is the same as the result in  \cite{bezrukavnikov2022non}.
\item If $f$ is principal, the dimension of $B$ is given by 
\begin{equation}
\dim B={\left({h_\theta {k_t'\over b_t}-h(\mathfrak{g^\vee})-1}\right)\rank(\mathfrak{g})-f_0\over2}.
\end{equation}
\end{itemize}

In order to derive dimension formulae \eqref{eq:CBdimUT} and \eqref{eq:CBdimT}, we start with a non-twisted theory $\CT(\fkj, b, k, f)$, and if there is irregular singularity only (i.e. $f$ is chosen to be principal), the same theory can also be engineered by putting type IIB theory on a three-fold singularity which are listed in the third column of table \ref{table:sing:b}.  One can then compute $\dim B$ using equation \eqref{eq:dimCBmilnor}. The tables of $\mu$ in each cases can also be found in the last column of table I of \cite{Xie:2016evu} and we also reproduce them in the last column of table \ref{table:sing:b} for reader's convenience. Adding a regular singularity with Nahm label $f$, will change $\dim B$ 
into 
\begin{equation}
\dim B=\frac{1}{2}(\mu-f_0)+\frac{1}{2}\dim \CO_{f^\vee}.
\end{equation}
 Finally one can check case by case that the Milnor number $\mu$ for non-twisted cases can also be written uniformly as
\begin{equation}
\label{eq:muformula}
\mu=(h_{\fkj}\frac{k}{b}+h_{\fkj}-1)\rank(\fkj)-\dim\mathcal{O}_{prin}
\end{equation}
The dimension formula for twisted cases is a direct generalization of the untwisted one.

\begin{example} When $\fkj=A_{N-1}$, the number $b$ can be either $N$ or $N-1$ as table \ref{table:sing:b}. If there is no regular puncture, the corresponding 3-fold singularity is 
\begin{align}
&x^2+y^2+z^N+w^k=0,~~~b=N \nonumber\\
& x^2+y^2+z^N+z w^k=0,~~~b=N-1
\end{align}
For $b=N$,  the Milnor number $\mu=(N-1)(k-1)$. On the other hand, since $h=N$, $\rank(\fkj)=N-1$ and $\dim\CO_{prin}=N^2-N$, we have
\begin{equation}
(h_{\fkj}\frac{k}{b}+h_{\fkj}-1)\rank(\fkj)-\dim\mathcal{O}_{prin}=(k-1)(N-1)=\mu.
\end{equation}
  For $b=N-1$, the Milnor number is $\mu= N(k-1)+1$, which also agrees with equation \eqref{eq:muformula}
  \begin{equation}
  \begin{split}
  &(h_{\fkj}\frac{k}{b}+h_{\fkj}-1)\rank(\fkj)-\dim\mathcal{O}_{prin}\\
  &=Nk+(N-1)^2-N(N-1)
  =N(k-1)+1=\mu.
  \end{split}
  \end{equation}
  \end{example}

  There is a different way of counting $\dim B$ by using the fact that the dimension of the fibre is the same as the dimension of the base $B$. 
The dimension formula of the Hitchin fibre can also be found in math literature \cite{kazhdan1988fixed,bezrukavnikov1996dimension, oblomkov2016geometric} for both untwisted and twisted cases, which is exactly the formula we found using physics arguments. This provids a cross check of \eqref{eq:CBdimUT} and \eqref{eq:CBdimT}.

\subsection{Schur sector and W-algebra}
The Higgs branch of a 4d $\mathcal{N}=2$ theory is given by a Hyper-K\"{a}hler manifold. Unlike the Coulomb branch, there are many $\mathcal{N}=2$ theories which do not
have Higgs branch. However, all  $\mathcal{N}=2$ theories do have a Schur sector, which includes the Higgs branch when exists. For  general $\mathcal{N}=2$ theories especially strongly coupled theories,  direct computations of Higgs (Schur) sector are very difficult. 
Luckily 
one can get a 2d $\text{VOA}(\CT)$ from the Schur sector of a 4d $\mathcal{N}=2$ SCFT $\CT$ with the following properties  \cite{Beem:2013sza}:
\begin{itemize}
	\item There is a subalgebra $V_{k_{2d}}(\fg_F)$ in $\text{VOA}(\CT)$, where $V_{k_{2d}}(\fg_F)$ is the simple quotient of the affine vertex algebra of the affine Kac-Moody (AKM) algebra $\hat{\fg}_F$ at level $k_{2d}$, and $\fg_F$ is the Lie algebra of 4d flavor symmetry $G_F$.
	\item The 2d central charge $c_{2d}$ and the level of the AKM algebra $k_{2d}$ are related to the 4d central charge $c_{4d}$ and the flavor central charge $k_F$ as\footnote{Our normalization of $k_F$ is half of that of \cite{Beem:2013sza,Beem:2014rza}.}
	\begin{equation}
	c_{2d}=-12 c_{4d},~~k_{2d}=-k_F.
	\label{eq:centralchargerelation}
	\end{equation}
	\item The (normalized) vacuum character of $\text{VOA}(\CT)$ is the 4d Schur index $\CI(q)$. The growth function $G$ of the vacuum character is related to 4d central charges by
	\begin{equation}
	-48(a_{4d}-c_{4d})=G
	\end{equation}
	\item The associated variety $X_{\text{VOA}(\CT)}$ is the Higgs branch $\CM_H$ of $\CT$ \cite{Song:2017oew,Beem:2017ooy,arakawa2018chiral}. 
\end{itemize}
If we can find the VOA for a given 4d $\mathcal{N}=2$ SCFT, then the Higgs (Schur) sector can be solved. 

In general there is no systematical way to get $\text{VOA}(\CT)$ from a given $\CT$, However, for our theory $\CT_{\fkj,k,b,f}$ and $\CT_{\fkj,o,k,b,f}$, if the irregular singularity carries no flavor symmetry, the corresponding  VOA are respectively the following W-algebra \cite{Xie:2016evu, Song:2017oew, Wang:2018gvb,Xie:2019yds}
\begin{equation}
\label{eq:WUT}
W_{-h^\vee+{b\over b+k}}(\mathfrak{j}, f),
\end{equation}
and
\begin{equation}
\label{eq:WT}
W_{-h^\vee(\mathfrak{g})+{1\over n}{b_t\over b_t+k_t}}(\mathfrak{g}, f).
\end{equation}
Here $h^\vee$ is the dual Coxeter number of $\fkj$,  $h^\vee(\mathfrak{g})$ is the dual Coxeter number for $\mathfrak{g}$, and  $n$ is the lacety listed in table \ref{table:outm}. The constraints on the irregular singularity $\fkj^b[k]$ which has no mass deformation are summarized in tables \ref{table:constraintADEirregular} and \ref{table:constraintwist} \cite{Xie:2016evu, Xie:2019vzr}.

\begin{table}[h]
\begin{center}
\begin{tabular}{|c|l|c|l|}
\hline
  $\fkj^b[k]$ &no mass &$\fkj^b[k]$&no mass  \\ \hline
     $A_{N-1}^N[k]$ &$(k,N)=1$& $A_{N-1}^{N-1}[k]$ &$\text{No solution}$\\ \hline
          $D_{N}^{2N-2}[k]$ &${2N-2\over \mathrm{gcd}(k,2N-2)}$  even, $\mathrm{gcd}(k,2N-2)$ odd& $D_{N}^{N}[k]$&${N\over \mathrm{gcd}(k,N)}$ even\\ \hline
     $E_{6}^{12}[k]$ &$k\neq 3n$& $E_6^{9}[k]$ &$k\neq 9n$\\ \hline
     $E_{6}^8[k]$ &$\text{No solution}$& $E_{7}^{18}[k]$ &$k\neq 2n$\\ \hline
     $E_7^{14}[k]$ &$k\neq 2n,n>1$& $E_{8}^{30}[k]$ &$k\neq 30n$\\ \hline
     $E_{8}^{24}[k]$ &$k\neq 24n$& $E_{8}^{20}[k]$ &$k\neq 20 n$\\ \hline
\end{tabular}
\end{center}
\caption{Constraints on $k$ so that irregular singularity denoted by $\fkj^b[k]$ has no mass deformation.}
  \label{table:constraintADEirregular}
\end{table}

\begin{table}[h]
	\begin{center}
		\begin{tabular}{|c|c|c|}
			\hline
			$\fkj/o$ &  $b_t$ & no mass \\ \hline
			$A_{2N}/\bbZ_2$  & $2N+1$ & ${4N+2\over \mathrm{gcd}(4N+2,2k+1)}$   even  \\ \hline
			~ & $2N$ & $2N\over \mathrm{gcd}(2N,k) $  even \\ \hline
			$A_{2N-1}/\bbZ_2$ & $2N-1$ & ${4N-2\over \mathrm{gcd}(4N-2,2k+1)}$   even  \\ \hline
			~ &$2N$ & $2N\over \mathrm{gcd}(2N,k) $ even  \\ \hline
			$D_{N}/\bbZ_2$ & $N+1$ & $2N\over \mathrm{gcd}(2k+1,2N)$ even \\ \hline
			~  &$2N$ & ${2N-2\over \mathrm{gcd}(k,2N-2)}$, $\mathrm{gcd}(k,2N-2)$  even   \\ \hline
			$D_4/\bbZ_3$  &$4$ & No constraint  \\ \hline
			~ &$6$ & $k\neq 6n$  \\ \hline
			$E_6/\bbZ_2$ &$9$ &  No constraint\\ \hline
			~ &$12$ & $k\neq 12n$  \\ \hline
			~  &$8$ & $k\neq 8n$, $k$ even  \\ \hline
		\end{tabular}
		\caption{Constraints $k_t$ so that the twisted irregular singularity has no mass deformation.}
		\label{table:constraintwist}
	\end{center}
\end{table}

From tables \ref{table:sing:b} and  \ref{table:SW:twisted}, one can see that given the irregular singularity $\fkj^b[k]$, the allowed values of $b$ is always smaller or equal to the dual Coxeter number $h^\vee$ of $\fkj$. Also recall that a level of the W-algebra $W_{\kappa}(\fkg,f)$ is called {\bf admissible} if it has the form
\begin{equation}
\kappa=-h^\vee+\frac{p}{q},\quad p\geq h^\vee,\ q\in\bbZ_{\geq0},\ \mathrm{gcd}(p,q)=1.
\end{equation}
When $p=h^\vee$ the corresponding level is called {\bf boundary admissible}. Then the W-algebra \eqref{eq:WUT} and  \eqref{eq:WT} are always boundary admissible or {\bf non-admissible}.

\section{Representation theory of admissible W-algebras}
\label{sec:repAdmWalgebra}

As mentioned in the introduction, the core correspondence of the mirror symmetry  here is the bijection between simple modules and fixed points. In this section, we will review key information on the representation theory of W-algebras $W_\kappa(\fkg,f)$ at boundary admissible level, which will provide crucial examples for our duality.

\subsection{Principal admissible modules of $V_\kappa(\fkg)$}
\label{sec:admAKM}

Let $\hat{\fkg}$ be the (non-twisted) affine Lie algebra of $\fkg$.
Let us start from the representation theory of the simple VOA $V_\kappa(\fkg)$ given by the unique simple quotient of the universal vertex algebra associated with $\hat{\fkg}$ at level $\kappa$. The level $\kappa$ is called admissible if it has the following form \cite{Kac:1988tf}
\begin{equation}
\kappa=-h^\vee+\frac{p}{u},\quad p\geq h^\vee,\ u\in \bbZ_{>0},\ \mathrm{gcd}(p,u)=\mathrm{gcd}(n,u)=1.
\end{equation}
Here $h^\vee$ is the dual Coxeter number of $\fkg$. By \cite{Arakawa_2016_rational}, simple modules of $V_\kappa(\fkg)$ at admissible level in the category $\CO$ of $\hat{\fkg}$ are the so-called admissible modules defined in \cite{Kac:1988tf}. Admissible modules have many properties similar to modules at integeral levels, therefore are interesting objects in VOA research. 

From now on, we fix $\kappa$ to be the boundary admissible level, i.e.,
\begin{equation}
\kappa=-h^\vee+{h^\vee\over u},\quad u\in\bbZ_{>0},\ \mathrm{gcd}(h^\vee,u)=\mathrm{gcd}(n,u).
\end{equation}
In this case, the highest weight of admissible modules are given as follows.
One first defines a set of affine coroots $S_u$ depending on $u$ \footnote{We identify $\fkh$ with $\fkh^\vee$  using the natural pairing between roots.}
\begin{equation}
\label{eq:defSu}
S_u\equiv\{-\theta^\vee+u\delta, \alpha^\vee_1,\ldots, \alpha^\vee_r\},
\end{equation}
where $\theta^\vee$ is the coroot corresponding to the highest root $\theta$ of $\fkg$, and $\delta$ is the imaginary root, $\{\alpha^\vee_1,\cdots,\alpha^\vee_r\}$ is the set of simple coroots of $\fkg$.  The set of admissible weights at level $\kappa$ is given by
\begin{equation}
\label{eq:admDef}
\mathrm{Adm}_\kappa=\{w.(\kappa\Lambda_0)\ |\ w\in W_{ext},\ w(S_u)\subset\hat{\Delta}^\vee_+\},
\end{equation}
where $W_{ext}$ is the extended affine Weyl group, $\Lambda_0$ is the $0$-th affine fundamental weight, and $\hat{\Delta}^\vee_+$ is the set of positive real coroots.  The dot action $w.\Lambda$ is defined as
\begin{equation}
w.\Lambda\equiv w(\Lambda+\hat{\rho})-\hat{\rho},
\end{equation}
 with $\hat{\rho}=\sum_{i=0}^{r}\Lambda_i$ being the affine Weyl vector. Here $\Lambda_i$'s are affine fundamental weights of $\hat{\fkg}$ and $r=\mathrm{rank}\fkg$. 
 Moreover, $w.(\kappa\Lambda_0)=w'.(\kappa\Lambda_0)$ if and only if $w^{-1}w'(S_u)=S_u$. Let
 \begin{equation}
 \begin{split}
 W_u=&\{w\in W_{ext}\ |\ w(S_u)\subset\hat{\Delta}^\vee_u\},\\
 \Omega_u=&\{w\in W_{ext}\ |\ w(S_u)=S_u\},\\
 \end{split}
 \end{equation}
 then there is a bijection
 \begin{equation}
 W_u/\Omega_u \xrightarrow{\sim} \mathrm{Adm}_\kappa.
 \end{equation}
 The number of admissible weights at level $\kappa=-h^\vee+h^\vee/u$ is $u^r$.
 An admissible module $L(\Lambda)$ is just the simple highest weight module of $\hat{\fkg}$ with the highest weight $\Lambda\in\mathrm{Adm}_\kappa$. The conformal weight $h_\Lambda$ of the highest weight state of $L(\Lambda)$ is
\begin{equation}
h_\Lambda=\frac{(\Lambda,\Lambda+2\hat{\rho})}{2(\kappa+ h^\vee)}.
\end{equation}

Since $W_{ext}$ is a semi-direct product of the coweight lattice $P^\vee$ and the Weyl group of $\fkg$,  we can also write each $w\in W_{ext}$ uniquely as a composition of a translation in $\beta\in P^\vee$ and a Weyl transformation $y\in W$
\begin{equation}
w=t_\beta y,
\end{equation}
with 
\begin{equation}
t_\beta (\lambda)=\lambda+\lambda(K)\beta-\left((\lambda,\beta)+\frac{1}{2}\lambda(K)(\beta,\beta)\right)\delta.
\end{equation}
Here $K$ is the central element in $\hat{\fkg}$.
 We will also denote $w=t_\beta y$ by $(\beta,y)$. Each $\Lambda\in\mathrm{Adm}_{\kappa}$ can also be written as $(t_\beta w).(\kappa\Lambda_0)$ for some $(\beta,w)$.


Given $\Lambda\in \mathrm{Adm}_\kappa$, let $\ch_\Lambda(z;\tau,t)$ be the character of the admissible module $L(\Lambda)$. The space spanned by characters of admissible modules carries modular transformations generated by
\begin{equation}
\begin{split}
&T:(z,\tau,t)\mapsto (z,\tau+1,t),\\
&S:(z,\tau,t)\mapsto \left(\frac{z}{\tau},-\frac{1}{\tau},t-\frac{(z,z)}{2\tau}\right).
\end{split}
\end{equation}
Explicitly, we have
\begin{equation}
\begin{split}
\ch_\Lambda(z;\tau+1,t)&=\sum_{\Lambda'\in\mathrm{Adm}_\kappa}\bbT_{\Lambda,\Lambda'}\ch_{\Lambda'}(z;\tau,t),\\
\ch_\Lambda\left(\frac{z}{\tau},-\frac{1}{\tau},t-\frac{(z,z)}{2\tau}\right)&=\sum_{\Lambda'\in\mathrm{Adm}_\kappa}\bbS_{\Lambda,\Lambda'}\ch_{\Lambda'}(z;\tau,t).
\end{split}
\end{equation}
Given $\Lambda=(t_\beta y).(\kappa\Lambda_0)$ and $\Lambda'=(t_{\beta'}y').(\kappa\Lambda_0)$, entries of matrices $\bbT$ and $\bbS$ are
\begin{equation}
\label{eq:STAKM}
\begin{split}
\bbT_{\Lambda,\Lambda'}&=e^{2\pi i \left(h_{\Lambda}-\frac{c}{24}\right)}\delta_{\Lambda,\Lambda'},\\
\bbS_{\Lambda,\Lambda'}&=
\left|\frac{P^\vee}{uh^\vee Q^\vee}\right|^{-\frac{1}{2}}\epsilon(yy')
\prod_{\alpha\in\Delta_+}2\sin\frac{\pi i u(\rho,\alpha)}{h^\vee}
e^{-2\pi i\left((\rho,\beta+\beta')+\frac{h^\vee(\beta,\beta')}{u}\right)}.
\end{split}
\end{equation}
Here $c=c(V_\kappa(\fkg))=\frac{\kappa\dim\fkg}{\kappa+h^\vee}$ is the central charge of $V_\kappa(\fkg)$,  $\left|\frac{P^\vee}{uh^\vee Q^\vee}\right|$ in the index of the sublattice $uh^\vee Q^\vee$ in $P^\vee$, $\epsilon(yy')$ is the sign of the Weyl group element $yy'$. 

\begin{example}
\label{ex:su2}
Let  $\mathfrak{g}=\fsl_2$ with boundary admissible level  $\kappa=-2+{2\over u}$. Let $\alpha$ be the unique positive coroot. The set $S_u$ is 
\begin{equation}
S_u=\{-\theta+u\delta, \alpha\}=\{-\alpha+u\delta, \alpha\}.
\end{equation}
The set $\hat{\Delta}^\vee_+$ is 
\begin{equation}
\hat{\Delta}^\vee_+=\{\alpha+n\delta~|~n\in\bbZ_{\geq0}\}\cup\{ -\alpha+n\delta~|~n\in\bbZ_{>0}\}.
\end{equation}
The finite Weyl group is generated by $s_\alpha$, and the co-weight lattice is spanned by ${\alpha\over 2}$,  so $W_{ext}$ is
\begin{equation}
W_{ext}=\{t_{-m\alpha/2},t_{-n\alpha/2}s_\alpha\ |\ m,n\in\bbZ\}.
\end{equation} 
And $\Omega_u\simeq \bbZ_2$ is generated by $t_{u\alpha/2}s_\alpha$. Because $t_{u\alpha/2}s_\alpha$ sends $t_{-{m\alpha/ 2}}$ to $t_{-n\alpha/2}s_\alpha$ for some $n$ and vice versa.
We only need to consider $w=t_{-{m\over2}\alpha}$ satisfying $w(S_u)\subset\hat{\Delta}^\vee_+$. 
 The action of $w=t_{-{m\over2}\alpha}$ on elements in $S_u$ is
\begin{align}
& t_{-{m\over2}\alpha} (-\alpha+u\delta) =-\alpha+(-m+u)\delta, \nonumber\\
&t_{-{m\over2}\alpha}(\alpha)=\alpha+m\delta.
\end{align}
The condition $w(S_u)\subset\hat{\Delta}^\vee_+$ constraints the allowed values of $m$ to be $0\leq m<u$, and  the total number admissible weights is indeed $u$. Using \eqref{eq:admDef}, the set $\mathrm{Adm}_\kappa$ is 
\begin{equation}
\label{eq:admsl2}
\mathrm{Adm}_\kappa=\{\Lambda_m\equiv\left(\kappa+\frac{2m}{u}\right)\Lambda_0-\frac{2m}{u}\Lambda_1,\ 0\leq m <u\}.
\end{equation}
\end{example}

\begin{example}Let $\mathfrak{g}=\mathfrak{sl}_3$ with  boundary admissible level $\kappa=-3+{3\over u}$ such that $\mathrm{gcd}(u,3)=1$.  The representatives of $W_u/\Omega_u$ are
\begin{align}
&\{t_{-(k_1\omega_1+k_2\omega_2)}\mid k_1\geq 0,~k_2\geq 0,\ k_1+k_2\leq u-1 \} \cup
\{t_{(k_1\omega_1+k_2\omega_2)}s_{\theta}\mid k_1\geq 1,~k_2\geq 1,\   k_1+k_2\leq u \} .
\end{align}
Here $\omega_1$ and $\omega_2$ are fundamental weights of $\fsl_3$,  and $s_{\theta}$ is the  reflection with repsect to the highest root $\theta=\alpha_1+\alpha_2$. The total number of admissible weights are $u^2$. For $u=4$, there are a total of 16  admissible weights  listed in table \ref{table:sl3adm}.

\begin{table}[htbp]\small
\begin{center}
\begin{tabular}{|c|c||c|c||c|c|} \hline 
$[t_\beta y]$ & $\Lambda$ & $[t_\beta y]$ & $\Lambda$ & $[t_\beta y]$ & $\Lambda$\\ \hline
$1$ & $-{9\over 4}\Lambda_0$ &$t_{ - \omega_2}$ & $ -{6\over 4}\Lambda_0-{3\over 4}\Lambda_2$ & $t_{ -2 \omega_2}$ & $ -{3\over 4}\Lambda_0-{6\over 4}\Lambda_2$ \\ \hline
$t_{-3 \omega_2}$ & $-{9\over 4}\Lambda_2$ &$t_{- \omega_1 }$ & $-{6\over 4}\Lambda_0-{3\over 4}\Lambda_1$ & $t_{- \omega_1 - \omega_2}$ & $ -{3\over 4}\Lambda_0-{3\over 4}\Lambda_1-{3\over 4}\Lambda_2$ \\ \hline
$t_{-\omega_1 -2 \omega_2}$ & $-{3\over 4}\Lambda_1-{6\over 4}\Lambda_2$ &$t_{- 2\omega_1 }$ & $-{3\over 4}\Lambda_0-{6\over 4}\Lambda_1$ & $t_{-2 \omega_1 - \omega_2}$ & $ -{6\over 4}\Lambda_1-{3\over 4}\Lambda_2$ \\ \hline
$t_{-3 \omega_1 }$ & $-{9\over 4}\Lambda_1$ &$t_{\omega_1 +\omega_2} s_\theta $ & $ {1\over 4}\Lambda_0-{5\over 4}\Lambda_1-{5\over4}\Lambda_2$ & $t_{\omega_1 + 2\omega_2} s_\theta$ & $ -{2\over 4}\Lambda_0-{5\over 4}\Lambda_1-{2\over 4}\Lambda_2$ \\ \hline
$t_{\omega_1 +3 \omega_2} s_\theta $ & $ -{5\over 4}\Lambda_0-{5\over 4}\Lambda_1+{1\over4}\Lambda_2$ &$t_{2\omega_1 + \omega_2} s_\theta $ & $ -{2\over 4}\Lambda_0-{2\over 4}\Lambda_1-{5\over4}\Lambda_2$ & $t_{2\omega_1 +2 \omega_2} s_\theta $ & $ -{5\over 4}\Lambda_0-{2\over 4}\Lambda_1-{2\over 4}\Lambda_2$ \\ \hline
$t_{3\omega_1 + \omega_2} s_\theta $ & $ -{5\over 4}\Lambda_0+{1\over 4}\Lambda_1-{5\over4}\Lambda_2$ & & && \\ \hline
\end{tabular}
\end{center}
\caption{\label{table:sl3adm}The list of admissible weights of $V_{-3+3/4}(\fsl_3)$. The first column summarizes the representatives of classes in $W_u/\Omega_u$. The second column gives the admissible weight corresponding to the elements of the first column.}
\end{table}
\label{ex:su3u4}
\end{example}

\subsection{Representation theory of boundary admissible W-algebras}
\label{sec:repWalg}

Let $f$ be a nilpotent element of $\fkg$, and include $f$ in an $\fsl_2$-triple $(e,f,x)$, so that $[x,e]=e$, $[x,f]=-f$ and $[e,f]=x$. Then $\fkg$ admits an eigenvalue decomposition withe respect to the adjoint action of $x$
\begin{equation}\fkg=\oplus_{j\in\bbZ}\fkg_j.
\end{equation}
By definition $f\in\fkg_{-1}$. One can define an affine W-algebra $W_\kappa(\fkg,f)$ associated with $\fkg$, $f$ at level $\kappa$ by the quantum Drinfeld-Sokolov (qDS) reduction \cite{deBoer:1993iz,kac2003quantum}. 
The central charge of $W_\kappa(\fkg,f)$ is \cite{kac2003quantum}
\begin{equation}
c(W_\kappa(\fkg,f))=\text{dim} \mathfrak{g}_0-{1\over 2} \text{dim} \mathfrak{g}_{\frac{1}{2}}-{12\over \kappa+h^\vee}|\rho-(k+h^\vee) x|^2,
\end{equation}
where $\rho$ is the Weyl vector of $\fkg$. Although the vertex algebra structure of $W_\kappa(\fkg,f)$ does not depend on the choices of $(e,f,x)$, the conformal structure does \footnote{Actually,  the data to get a W-algebra can be relaxed to a nilpotent element $f$ and a good grading on $\fkg$ such that $f\in\fkg_{-1}$. The grading obtained from an $\fsl_2$-triple is called Dynkin which is always good. We will not discuss the construction of W-algebra from more general good gradings in this work.}. To match the central charge of the corresponding 4d theory, $(e,f,x)$ is chosen to be the $(X,Y,H/2)$ with $(X,Y,H)$  being the standard $\fsl_2$-triple defined in \cite{Collingwood:1993rr}. 

Simple modules of $W_\kappa(\fkg,f)$ can be obtained from admissible modules of $V_\kappa(\fkg)$ by qDS-reduction. Firstly conjugate $(e,f,x)$ to a new $\fsl_2$-triple $(e',f',h')$ such that $f'$ a regular nilpotent element in a standard L\'{e}vi subalgebra $\fkl$ of $\fkg$. Here $\fkl$ is the centralizer of
\begin{equation}
\fkh^f=\{ h\in \fkh~|~f(h)=0\}.
\end{equation} $Z_\fkg(\fkh^f)$. 
The root system of $\fkl$ is given by
\begin{equation}
\Delta_\fkl \equiv \{\alpha\in \Delta~|~\alpha|_{\fkh^f}=0\}.
\end{equation}
The simple roots of $\Delta_\fkl$ is required to be a subset of simple roots of $\fkg$ because $\fkl$ is standard.
Kac and Wakimoto \cite{kac2008rationality} (Generalizing \cite{Frenkel:1992ju}) defined a functor
\begin{equation}
H_f(-):V_\kappa(\fkg)-\mathrm{mod}\rightarrow W_\kappa(\fkg,f)-\mathrm{mod},
\end{equation}
and they conjectured that this functor sends admissible module $L(\Lambda)$ of $V_\kappa(\fkg)$ to either $0$ or simple modules of $W_\kappa(\fkg,f)$, and all simple modules of $W_\kappa(\fkg,f)$ are obtained in this way \footnote{When $f$ admits an even grading, $H_f(L(\Lambda))$ is a usual module of $W_\kappa(\fkg,f)$. When $f$ does not admit an even grading, $H_f(L(\Lambda))$ is a Ramond twisted module of $W_\kappa(\fkg,f)$ \cite{kac2008rationality}.}. They further conjectured that $H_f(L(\Lambda))\neq 0$ if and only if
\begin{equation}
\label{eq:LmodtoWmod}
t_\beta y(S_u) \subset \hat{\Delta}^\vee_+\backslash \Delta^\vee_\fkl,\quad \Lambda=(t_\beta y).(\kappa\Lambda_0),
\end{equation}
and $H_f(L(\Lambda))$ is isomorphic to $H_f(L(\Lambda'))$ if and only if 
\begin{equation}
\Lambda'\in W_f.\Lambda,
\end{equation}
where $W_f$ is the Weyl group generated by roots of $\Delta_\fkl$. These conjectures are partially proved in \cite{arakawa2008representation, Arakawa2012Rationality, arakawa2021rationality, Fasquel_2022}.
The conformal weight of $H_f(L(\Lambda))$ is \cite{kac2008rationality,arakawa2021rationality}
\begin{equation}
\label{eq:dimWalgMod}
h_{H_f(L(\Lambda) ) }=\frac{u}{2h^\vee}(|\lambda+\rho|^2-|\rho|^2)-\frac{h^\vee}{2u}|x|^2+(x,\rho),
\end{equation}
with $\lambda$ being the finite part of $\Lambda$. Note that the first term of \eqref{eq:dimWalgMod} is invariant under the actiion of $W_f$, and the choice of $x$ only change the conformal dimension by a constant shift.

The characters $\ch_{H_f(L(\Lambda))}$ of $W_\kappa(\fkg,f)$ also enjoy similar modular properties as characters of $V_\kappa(\fkg)$ modules. If $L(\Lambda)$ and $L(\Lambda')$ are two admissible modules which reduce to different W-algebra modules, the elements of modular matrices are
\begin{equation}
\label{eq:STWalg}
\begin{split}
\bbT_{H_f(L(\Lambda)),H_f(L(\Lambda'))} & = e^{2\pi i\left(h_{H_f(L(\Lambda))} - \frac{c}{24}\right)} \delta_{H_f(L(\Lambda)),H_f(L(\Lambda'))}, \\
\bbS_{H_f(L(\Lambda)),H_f(L(\Lambda'))} & =(-i)^{\frac{1}{2}(\dim\fkg-\dim\fkg^f)}\sum_{y\in W^f} \bbS_{\Lambda,y.\Lambda'},
\end{split}
\end{equation}
where $\bbS_{\Lambda,y.\Lambda'}$ is the modular $S$ matrix of the parent affine vertex algebra, and $\fkg^f=\dim\fkg_0+\dim\fkg_{1/2}$.

%

\begin{example} Let $\fkg=\fsl_2$, $\kappa=-2+2/u$ and $f\in\CO_{[2]}$  an element in the principal nilpotent orbit. Choose $(e,f,h)=(e_{\alpha},f_{\alpha},x)$, then $\Delta_\fkl=\{\alpha\}$, and $W_f$ is just the Weyl group of $\fsl_2$. The condition when the admissible weight $\Lambda=(t_\beta y).(\kappa\Lambda_0)$ does not reduce to zero becomes
\begin{equation}
t_\beta y(S_u)\subset \hat{\Delta}_+\backslash \Delta_{\fkl} = \{ \pm\alpha+m\delta~|~m\in\bbZ_{>0}\}.
\end{equation}
Using admissible modules of $V_{2+\frac{2}{u}}(\fsl_2)$ worked out in example \ref{ex:su2}, one can see that the module $L(\kappa\Lambda_0)$ reduces to $0$, while $L(\Lambda_{m})$ and $L(\Lambda_{u-m})$ reduces to the same $W_{-2+2/u}(\fsl_2,[2])$ module, so the total number of simple modules are $(u-1)/2$. The algebra $W_{-2+2/ u}(\fsl_2,[2])$ is  isomorphic to the $(2,u)$ minimal model (the minimal series representation of the Virasoro algebra with central charge $c=1-\frac{3(u-2)^2}{u}$). The conformal dimension of $H_f(L(\Lambda_m))$ is
\begin{equation}
h_{H_f(L(\Lambda_m))} =-\frac{1}{2u}(m-1)(u-m-1),
\end{equation}
which is symmetric under the exchange $m\leftrightarrow u-m$ and matches with the $(m,1)$ module of the $(2,u)$ minimal model.
\end{example}

\begin{example} Let $\fkg=\fsl_3$, $\kappa=-3+3/u$ and $f\in\CO_{[2,1]}$ an element of the minimal nilpotent orbit. 
To match the 2d central charge with the 4d central charge, one should choose $f$ to be $f_{\theta}$ and $x=\frac{1}{2}(\omega_1+\omega_2)$, with the price that $f$ is not regular in a standard L\'{e}vi. However, we can choose $(f',x')=(f_{\alpha_1},\omega_1)$ which are conjugate to $(f,x)$, such that  $\Delta_\fkl=\{\pm \alpha_1 \}$ defines a standard L\'{e}vi. Now $W_f$ is generated by $s_1 $. The condition for the admissible module $L(\Lambda)$ with $\Lambda=(t_\beta y).(\kappa\Lambda_0)$  not reducing to $0$ becomes
\begin{equation}
t_\beta y(S_u)\subset \hat{\Delta}_+\backslash \Delta_{\fkl}  =\hat{\Delta}_+\backslash\{\alpha_1\}.
\end{equation}
When $u=4$, one can use results in table \ref{table:sl3adm} to work out the simple modules of $W_{-3+3/4}(\fsl_2,[2,1])$ explicitly. There are three modules with conformal weight $0$, one modules with conformal weight $-1/4$, and two modules with conformal weight $-1/2$ (Computed using $x'$). Results are summarized in table \ref{table:sl3min}. One can then map the modules obtained above to modules defined by $x$ using the method in \cite{kac2008rationality}. 
 
\begin{table}[htbp]
\begin{center}
\begin{tabular}{|c|c||c|c|} \hline 
$[t_\beta y], \Lambda$ & $h$ & $[t_\beta y], \Lambda$ & $h$  \\ \hline
$[t_{-\omega_1}]$, $-\frac{6}{4}\Lambda_0-\frac{3}{4}\Lambda_1$  & \multirow{2}{*}{$0$} & $[t_{-3\omega_1}]$, $-\frac{9}{4}\Lambda_1$ & \multirow{2}{*}{$0$} \\

$[t_{\omega_1+3\omega_2}s_\theta]$, $-\frac{5}{4}\Lambda_0-\frac{5}{4}\Lambda_1+\frac{1}{4}\Lambda_2$ &  & $[t_{3\omega_1+\omega_2}s_\theta]$, $-\frac{5}{4}\Lambda_0 +\frac{1}{4} \Lambda_1-\frac{5}{4}\Lambda_2$ & \\ \hline
$[t_{-2\omega_1}]$, $-\frac{3}{4}\Lambda_0-\frac{6}{4}\Lambda_1$ & \multirow{2}{*}{$-1/4$} & $[t_{-\omega_1-\omega_2}]$, $-\frac{3}{4}\Lambda_0-\frac{3}{4}\Lambda_1-\frac{3}{4}\Lambda_2$ & \multirow{2}{*}{$-1/2$} \\ 
$[t_{2\omega_1+2\omega_2}s_\theta]$, $-\frac{5}{4}\Lambda_0-\frac{2}{4}\Lambda_1-\frac{2}{4}\Lambda_2$ &   & $[t_{ \omega_1+ 2\omega_2}s_\theta]$, $\frac{-2}{4}\Lambda_0 -\frac{5}{4}\Lambda_1 -\frac{2}{4}\Lambda_2$ &  \\ \hline
$[t_{-2\omega_1-\omega_2}]$, $-\frac{6}{4}\Lambda_1-\frac{3}{4}\Lambda_2$ & \multirow{2}{*}{$-1/2$} & $[t_{-\omega_1-2\omega_2}]$, $-\frac{3}{4}\Lambda_1 -\frac{6}{4}\Lambda_2$ & $-1/2$ \\
$[t_{2\omega_1+\omega_2}s_\theta]$, $-\frac{2}{4}\Lambda_0 -\frac{2}{4}\Lambda_1 -\frac{5}{4}\Lambda_2$ &   & $[t_{\omega_1+\omega_2}s_\theta]$, $\frac{1}{4}\Lambda_0-\frac{5}{4}\Lambda_1-\frac{5}{4}\Lambda_2$ &   \\ \hline
\end{tabular}
\end{center}
\caption{\label{table:sl3min}The list of simple modules of $W_{-3+3/4}(\fsl_2,[2,1])$. The first column is the weight of the admissible module $L(\Lambda)$ which does not reduce to $0$, and the second column is the conformal weight of $H_f(L(\Lambda))$. Two weights which reduces to the same W-algebra module are related by the dot action of $s_1$. The $L(\Lambda)$'s which reduce to $0$ are not listed. The conformal dimensions are computed using $x'$.}
\end{table}
\end{example}

\section{Coulomb branch and its $\mathbb{C}^*$-fixed points}
\label{sec:CBfixedPoints}
In the last section, we reviewed the representation theory of W-algebras which gives the information on 
the \textbf{Higgs (Schur)} sector of the 4d theory. When classifying  simple modules, the computation reduces to the counting of 
extended affine Weyl group elements satisfying certain conditions.  In this section, we go back to the Hitchin moduli space which describes the Coulomb branch of the 4d theory on a circle. 
 Our Hitchin moduli space has a $\mathbb{C}^*$ action which is the $U(1)_r$ symmetry in the superconformal group. It was found previously in several class of theories that  $\bbC^*$-fixed points are in one to one correspondence to the simple modules of the corresponding W-algebra \cite{Fredrickson:2017jcf, Fredrickson:2017yka}. In those work, fixed points do not have an 
 obvious representation theory meaning, hence it is difficult to generalize them to more complicated cases. We will show that affine Springer fibers provide an alternative description of this fixed varieties which makes the classification and matching (with the modules) more straightforward. 


\subsection{More on  Higgs bundles and  Higgs fields}
\label{sec:Higgsbundle}

As mentioned in section \ref{sec:ADtheory}, the Coulomb branch  is given by the Hitchin system defined on $\mathbb{P}^1$ with one regular and one irregular singularity. We now review some details on the Higgs bundle and the Higgs field in this setting. 
$\CM_{Hit}$ is the space of solutions to Hitchin equation defined on a Riemann surface $\Sigma$ \cite{hitchin1987self}. It has a hyper-K\"{a}hler structure with three complex structures $I,J,K$. In complex structure $I$, each point of $\CM_{Hit}$ describes a  Higgs bundle $(E, \Phi)$, where 
$E$ is a holomorphic $G^\vee$-vector bundle on $\Sigma$, and $\Phi$ is a Higgs field which is a holomorphic section of $\text{End}(E)\otimes K_\Sigma$. Here  $G^\vee$ be a connected and simply connected Lie group whose Lie algebra is $\fkg^\vee$, and we have $\fkg=\fkg^\vee=\fkj$ for the untwisted case labelled by ADE Lie algebra $\fkj$, while  $\fkg$ and $\fkg^\vee$ defined in table \ref{table:outm} in the twisted case labelled by $(\fkj,o)$.
At each singularity, $E$ is equipped with a level structure (which determines the correct gauge transformation at the singularity) and $\Phi$ satisfies certain boundary condition.

First consider the irregular singularity at $z=\infty$. Choose a $\bbZ/b\bbZ$ grading on $\fkj$ \cite{reeder2012gradings}
\begin{equation}
\label{eq:Zbgrading}
\fkj=\oplus_{i\in\bbZ/b\bbZ}\fkj_{i/b}.
\end{equation}
At $\infty$, $E$ is equipped with a level structure determined by the grading \eqref{eq:Zbgrading} \cite{bezrukavnikov2022non}.
The Higgs field behaves as
\begin{equation}
\Phi(z)  \sim (T_kz^{\frac{k}{b}}+\ldots)dz,
\end{equation}
when $z\rightarrow \infty$. The leading coefficient $T_k$ is regular semi-simple in $\fkj_{k/b}$ and invariant under the action of $o$. Details on the choices of subsequent coefficients can be found in \cite{Xie:2017aqx, Wang:2018gvb}.
For the later purpose, we redefine the Higgs field as
\begin{equation}
\Phi(z) =\frac{\Phi'(z)}{ z}
\end{equation}
and the asymptotical behavior for $\Phi'(z)$ at $z=\infty$ is then
\begin{equation}
\label{eq:boundIrr}
\Phi' \sim (T_kz^{\nu}+\ldots)dz,
\end{equation}
where $\nu=\frac{k}{b}+1$ and $\nu>0$ because $k>-b$. So the irregular singularity is specified by a rational number $\nu$.

The regular singularity at $z=0$ is labeled by a nilpotent element $f$ of $\mathfrak{g}$. Recall that we assume that $f$ is a regular nilpotent element in a L\'{e}vi $\fkl$ with $\fkl$ defined in section \ref{sec:repWalg}. Then on the Hitchin side, we should consider the Langlands dual $\fkl^\vee$. Let $\fkp^\vee=\fkl^\vee+\fkn^\vee\subset \fkg^\vee$ be the parabolic subalgebra with L\'{e}vi factor $\fkl^\vee$, and $\fkn^\vee$ be its nilradical part. Let $P^\vee\subset G^\vee$ be the parabolic subgroup whose Lie algebra is $\fkp^\vee$. Then at $z=0$, the Higgs bundle $E$ is equipped with a $P^\vee$-level structure, which means the allowed gauge transformation around $z=0$ is of the form \cite{gukov2006gauge}
\begin{equation}
g=g_0+g_1z+g_2z^2+\cdots,\quad g_0\in P^\vee,~g_{i>0}\in G^\vee.
\end{equation}
The boundary condition of $\Phi'$ at $z=0$ is
\begin{equation}
\Phi'\sim\left((m+\beta)+\cdots\right)dz,
\end{equation}
where the mass deformation $m$ is in the center of $\fkl^\vee$ and $\beta\in \fkn^\vee$. In the massless limit $m\rightarrow 0$, the boundary condition becomes
\begin{equation}
\label{eq:boundreg}
\lim_{z\rightarrow0}\Phi'\in\fkn^\vee.
\end{equation}
This boundary condition is related  to the boundary condition \eqref{eq:regularBC} because
\begin{equation}
\CO_{f^\vee}=d(\CO_{f})=\mathrm{Ind}^{\fkg^\vee}_{\fkl^\vee}d(\CO_{f}^\fkl)=\mathrm{Ind}^{\fkg^\vee}_{\fkl^\vee}\CO^{\fkl^\vee}_0,
\end{equation}
and $\overline{\mathrm{Ind}^{\fkg^\vee}_{\fkl^\vee}\CO^{\fkl^\vee}_0}=G^\vee\cdot \fkn^\vee$ \cite{Collingwood:1993rr}. 
Here $\mathrm{Ind}$ means the induction of orbit and $d(\CO^\fkl_f)$ is the dual orbit of $\CO^\fkl_f$ in $\fkg^\vee$. Since $f$ is in regular in $\fkl$, $d(\CO^\fkl_f) = \CO^{\fkl^\vee}_0$. To further specify the Coulomb branch operators on Hitchin base $B$, one also needs an element $c$ in the so-called 
component group $A(f^\vee)$ of $f^\vee$ introduced in \cite{achar2003order}, then Coulomb branch operators are  gauge invariant functions of $\Phi$ which are also  invariant under the action of $c$ \cite{Chacaltana:2012zy}.
 In summary, the Hitchin moduli space is specified by a Lie algebra $\mathfrak{j}$ (resp. $(\fkj,o)$), a rational number $\nu=k/b+1$ and a pair $(f^\vee,c)$ together with suitable level structure on the Higgs bundle, and the corresponding moduli space might be labeled as 
\begin{equation}
\CM_{Hit}(\mathfrak{j},\nu,(f^\vee,c))~(\mathrm{resp.}~\CM_{Hit}((\mathfrak{j},o),\nu,(f^\vee,c))).
\end{equation}
 


\subsection{Zero fibre of the Hitchin moduli space and the affine Springer fibre}
 
  The Hitchin system considered in this paper has a positive $\mathbb{C}^*$ action on $(x,z)$ coordinates 
 \begin{equation}
 x\to \lambda^\alpha x,~~z\to \lambda^\beta z,
 \end{equation}
 which  makes the spectral curve $\det(x-\Phi(z))=0$ invariant. This implies that the $\bbC^\ast$ weight of $x$ should be 
 the same as $\Phi(z)$, inducing an action on the AKM algebra where $\Phi(z)$ lives in. The invariance of the spectral curve fixes
 the weight of the leading order coefficient $T$ of Higgs field is $0$, and the weights of $x$ and $z$ are related by 
 $\alpha=\beta \frac{k}{b}$. Because of this weight assignment, 
 the $\bbC^\ast$-fixed points of $\CM_{Hit}$ belong to the fibre   over the $\bbC^\ast$-fixed point on the Hitchin base,
which corresponds to the curve at the SCFT point listed in table \ref{table:sing:b} and \ref{table:SW:twisted}. We call this fibre the {\bf zero fibre} \footnote{This is called the central fibre in \cite{Fredrickson:2017jcf}}.

Below, we consider a local situation in which we may assume that the holomorphic bundle $E$ of the Higgs pair $(E,\Phi)$ is trivial. Now the Hitchin moduli space can be described using 
 the language of affine Lie algebra.
 
 {\bf Untwsited cases:} First consider the untwisted theories $\CT_{\fkj,b,k,f}$ with $\fkj=ADE$. Let $\hat{\fkj}=\fkj[z,z^{-1}]\oplus \bbC K\oplus \bbC d$ be the AKM algebra associated with $\fkj$. Here $\fkj[z,z^{-1}]$ is the polynomials in $z$ and $z^{-1}$ with coefficient valued in $\fkj$.  The modified Higgs field $\Phi'(z)$ is now an element in $\hat{\fkj}$ satisfying 
the boundary condition \eqref{eq:boundIrr} and \eqref{eq:boundreg} in last subsection.

{\bf Twisted cases:} Now consider the twisted case $\CT_{\fkj,o,b_t,k_t,f}$. The space $\fkj$ has a decomposition
\begin{equation}
\fkj=\fkj_0\oplus\fkj_{\omega}\oplus\cdots\oplus\fkj_{\omega^{n-1}},
\end{equation}
under the action of $o$. Here the subscripts denote the eigenvalues under the action of $o$ and $n$ the order of $o$. By definition $\fkj_0=\fkg^\vee$  listed in table \ref{table:outm}. The twisted affine Lie algebra $^n\hat{\fkj}$ corresponding to $(\fkj,o)$ is then
\begin{equation}
^n\hat{\fkj}=\oplus_{k\in \bbZ}\left(\fkj_0z^k\oplus \fkj_{\omega}z^{k+\frac{1}{n}}\oplus\cdots\oplus \fkj_{\omega^{n-1}}z^{k+\frac{n-1}{n}}\right)\oplus \bbC K\oplus \bbC d.
\end{equation}
Below we will set formally $^1\hat{\fkj}$ as $\hat{\fkj}$ so we can treat untwisted and twisted case uniformly.

By construction the modified Higgs field $\Phi'(z)$ is an element in ${^n}\hat{\fkj}$ satisfying the following boundary conditions
\begin{equation}
\begin{split}
\Phi'(z)&\sim (Tz^\nu +\cdots)dz,\quad z\rightarrow\infty,\\
\Phi'(z)&\sim (\beta^\vee  +\cdots)dz,\quad z\rightarrow0.
\end{split}
\end{equation}
Here $\nu = k_t/b_t+1$, and $\beta^\vee$ is an element in $ \fkn^\vee\subset\fkg^\vee$.

 For the purpose of counting the fixed varieties, we only need to consider the zero fibre of the Hitchin moduli space and it is easier to describe it using the affine Springer fibre which we will review in the following.
Choose an elliptic element  $\gamma \in {^n\hat{\fkj}}$  whose spectral curve is the $\bbC^\ast$-fixed point in $B$. Let $\bfG^\vee$ be a connected and simply-connected affine Lie group whose Lie algebra is $^n\hat{\fkj}$ \footnote{We will always put a $\vee$ symbol on objects on the fibre side as they are always the Langlands dual of the corresponding objects on the VOA side.}.  Let $\tilde{\fkn}^\vee$ be the Lie algebra with the root system $\Delta_{\tilde{\fkn}^\vee}\equiv \hat{\Delta}^\vee_+\backslash \Delta^\vee_{\fkl}$. Here $\hat{\Delta}^\vee$ is the set of positive real roots of affine Lie algebra $^n\hat{\fkj}$. Let $\bfP^\vee\subset\bfG^\vee$ be the parahori subgroup whose root system is $\hat{\Delta}^\vee_+ \cup \Delta^\vee_{\fkl}$. Then the affine Spaltenstein variety is \cite{varagnolo2009finite,oblomkov2016geometric} 
\begin{equation}
 Sp_{\gamma,\bfP^\vee}= \{ g\in \bfP^\vee\backslash \mathbf{G}^\vee  ~|~ g \gamma g^{-1}\subset \tilde{\fkn}^\vee\}.
 \end{equation} 
 In  \cite{bezrukavnikov2022non} the authors proved that the zero fibre of the Hitchin moduli space $\CM_{Hit}((\fkj,o),\nu,(f^\vee,c))$ is homeomorphic to $Sp_{\gamma,\bfP^\vee}$, with the relation
 \begin{equation}
 \Phi(z)'=g\gamma g^{-1}dz.
 \end{equation}
 The choice of $\gamma$ ensures that $\Phi'$ satisfies the boundary condition from the irregular singularity at $\infty$ while $g \gamma g^{-1}\subset \tilde{\fkn}^\vee$ ensures that $\Phi'$ satisfies the boundary condition from the regular singularity at $0$.

\begin{example} Consider $\nu={u\over h^\vee}$, the  elliptic element of $^n\hat{\fkj}$ is
  \begin{equation}
\gamma= e_{-\theta}z^{u}+\sum_{i=1}^r e_{\alpha_i}.
 \end{equation}
 Here $\theta$ is the longest root, $\alpha_i$'s are  simple roots, and $e_{\alpha}$ is an element in the Chevalley basis corresponding to the root $\alpha$.
In particular when $\fkj=A_{N-1}$ and $\nu={u\over N}$,  the spectral curve of $\gamma$ is 
\begin{equation}
x^N+z^{u-N}=0,
\end{equation}
so $\gamma$ lies in the central fibre.
 It is useful to redefine the coordinate $x'=xz$, and so the spectral curve takes the form
 \begin{equation}
 x'^{N}+z^{u}=0,
 \end{equation} 
and the SW differential in the new variable is $x' {dz\over z}$. 
 \end{example}

\begin{example} Take the Lie algebra $\mathfrak{g}=A_{N-1}$, and let $e_1,\ldots, e_N$  be the standard basis of $\bbR^N$. We give the explicit description of $\fkn^\vee$. The set of positive roots are $\Delta_+=\{e_i-e_j| 1<i<j\leq N\}$, and the set of simple roots are $\Pi=\{e_1-e_2,\ldots e_{N-1}-e_N\}$.
Given a partition $d=[d_1,\cdots, d_s]$ of $N$, one pick the following set of simple roots corresponding to $d$
\begin{equation}
\Pi_d=\Pi_1\cup \Pi_2 \cup\cdots\cup \Pi_s,
\end{equation}
where 
\begin{equation}
\Pi_{i}=\{e_{\sum_{l=1}^{i-1} d_l} - e_{\sum_{l=1}^{i-1} d_l+1},\cdots, e_{\sum_{l=1}^{i} d_l-1} - e_{\sum_{l=1}^{i} d_l} \}.
\end{equation}
Now let $\Delta_d\subset \Delta$ be the sub-root system generated by $\Pi_d$. The standard L\'{e}vi subalgebra corresponding to $d$ is
\begin{equation}
\fkl_d^\vee=\fkh^\vee\oplus_{\alpha\in\Delta_d}\fkg^\vee_\alpha,
\end{equation}
while the standard parabolic algebra $\fkp^\vee_d$ containing $\fkl^\vee_d$ is
\begin{equation}
\fkp^\vee_d=\fkl^\vee_d\oplus_{\alpha\in\Delta_+\backslash \Delta_d}\fkg^\vee_\alpha.
\end{equation}
$\fkp^\vee_d$ has a L\'{e}vi decomposition $\mathfrak{p}^\vee_d=\mathfrak{l}^\vee_d\oplus \mathfrak{n}^\vee_d$ with
\begin{equation}
\mathfrak{n}^\vee_d = \oplus_{\alpha\in\Delta_+\backslash \Delta_d}\fkg^\vee_\alpha.
\end{equation}
The set of roots of $\tilde{\mathfrak{n}}^\vee$  is then
 \begin{equation}
\Delta_{\tilde{\fkn}^\vee}=\{\alpha+n\delta | \alpha \in \Delta, n\in\bbZ_{>0} \} \cup (\Delta_+^\vee \backslash \Delta_{\fkl^\vee}).
 \end{equation}
 In particular, if $d=[N]$ (so-called trivial puncture in physics literature), $\mathfrak{n}^\vee$ is zero, then $\tilde{\fkn}^\vee$ is the Lie algebra generated by the root system $\hat{\Delta}^\vee_+\backslash \Delta^\vee_+$. If $d=[1^N]$ (so-called full puncture in physics literature), $\tilde{\fkn}^\vee$ is generated by the root system $\hat{\Delta}_+^\vee$. 
\end{example}

 \textbf{The requirement of elliptic element:} In this work we focus on  cases when there are no mass parameters in irregular singularity, which puts 
 the constraint on choices of the rational number $\nu$ which is called slope. Recall the constraints on $(b,k)$ (resp. $(b_t,k_t)$) for irregular singularities without mass deformation listed in table \ref{table:constraintADEirregular} (resp. table \ref{table:constraintwist}), and $\nu= \frac{k}{b}+1$ (resp. $\nu =\frac{k_t}{b_t}+1$). The requirement of no mass deformation imposes constraints on the denominator $m$ of $\nu=u/m$ which are listed in \ref{table:ellnumber}. Interestingly, such choices of $m$ coincides with the so-called elliptic numbers \cite{varagnolo2009finite, oblomkov2016geometric}. 
Similarly, the  allowed elliptic numbers for the twisted case is also given in table \ref{table:ellnumber}. An elliptic number is called regular if it is the same as the dual Coxeter number $h^\vee$. 
 The dimension for the elliptic affine Springer fiber is computed by \cite{kazhdan1988fixed,bezrukavnikov1996dimension,oblomkov2016geometric}, which is the same as our result in section \ref{sec:CBdim}. 
 \begin{table}[htp]
\begin{center}
\begin{tabular}{|c|c|} \hline
$\mathfrak{j}$& Elliptic number $m$ \\ \hline
$A_{n}$& $n+1$ \\  \hline
$D_n$&    $m~\text{even} ,~ \frac{2n-2}{m} ~\text{odd}$  \\  \hline
~& $m~\text{even} ,~ \frac{2n}{m} ~\text{even}$ \\ \hline
$E_6$& $12,9,6,3$ \\ \hline
$E_7$& $18,14,6,2$ \\ \hline
$E_8$& $2, 3, 5, 6, 10, 15, 30$ \\ \hline
~&$4, 8, 12, 24$ \\ \hline
~&$20$ \\ \hline
\end{tabular}
\quad
\begin{tabular}{|c|c|} \hline
$\mathfrak{j},o$& Elliptic number $m$ \\ \hline
$A_{2n},\bbZ_2$& $m=2r,~r~\text{odd} ,~ \frac{2n+1}{r} ~\text{odd}$ \\  \hline
~&$m=2r,~r~\text{odd} ,~ \frac{2n}{r} ~\text{even}$ \\ \hline
$A_{2n-1},\bbZ_2$&    $m=2r,~r~\text{odd} ,~ \frac{2n-1}{r} ~\text{odd}$  \\  \hline
~& $m=2r,~r~\text{odd} ,~ \frac{2n}{r} ~\text{even}$ \\ \hline
$D_n,\bbZ_2$& $m~\text{even} ,~ \frac{2n}{m} ~\text{odd}$  \\ \hline
~&$m~\text{even} ,~ \frac{2n-2}{m} ~\text{even}$  \\ \hline
$D_4,\bbZ_3$& $12,6,3$ \\ \hline
$E_6,\bbZ_2$& $18,12,6,4,2$ \\ \hline
\end{tabular}
\end{center}
\caption{List of elliptic number $m$.}
\label{table:ellnumber}
\end{table}

\subsection{Counting fixed varieties}
\label{sec:fpFibre}

In previous section we argued that the affine Springer fibre can be used to replace the Hitchin moduli space when considering the $\bbC^\ast$-fixed points.  For the elliptic case, there is a nice combinatorial counting algorithm \cite{varagnolo2009finite,oblomkov2016geometric} which we will explain here.  Given an elliptic slope $\nu=u/m$ for an (twisted) affine Lie algebra $\hat\fkj$ ($^n\hat{\fkj}$),   define the set $L_\nu$ as 
\begin{equation}
L_\nu=\{\alpha+l\delta \in \hat\Delta^\vee~|~ \nu \alpha( \rho) +l =0\}, \end{equation}
and the set $S_\nu$ as
\begin{equation}
S_\nu=\{\alpha + l\delta\in \hat\Delta^\vee~|~\nu {\alpha}( \rho)) +l =\nu\}.
\end{equation}
Here $\hat\Delta^\vee$ is the set of real {\bf roots} of $\hat\fkj$ ($^n\hat{\fkj}$), $\rho$  the co-Weyl vector of the finite part of $\hat\fkj$ ($^n\hat\fkj$).
Denoted by $W_\nu$ the Weyl group generated by roots in $S_\nu$. 

With $L_\nu$ and $S_\nu$, the set of fixed varieties $Sp_{\gamma,\bfP^\vee}^T$ of the affine Springer fiber $Sp_{\gamma,\bfP^\vee}$ is labelled by the affine Weyl group element up to the action of $W_{\bfP^\vee}$ and $W_\nu$ \footnote{Our $\tilde{w}$ is $\tilde{w}^{-1}$ in \cite{oblomkov2016geometric}.},
\begin{equation}
\label{eq:fpSpringFiber}
Sp_{\gamma,\bfP^\vee}^T = \sqcup H_{\tilde{w}},\quad \{\tilde{w}\in W_{\bfP^\vee}\backslash W_{aff} / W_\nu~|~ \mathrm{Ad}(\tilde{w})\gamma \in \tilde{\fkn}^\vee\}.
\end{equation}
Here $W_{\bfP^\vee}$ is the  Weyl group for the parahori subgroup $\bfP^\vee$. 
The dimension of each fixed variety is \cite{oblomkov2016geometric}
\begin{equation}
\boxed{
\dim H_{\tilde{w}}=|\tilde{w}L_\nu\backslash \Delta_{\tilde{\fkn}^\vee}|-| \tilde{w}S_\nu\backslash \Delta_{\tilde{\fkn}^\vee} |.
}
\label{dimension}
\end{equation}
In the following, we will apply above formula to several interesting cases.

{\bf Regular elliptic case:} Let $\hat{\fkj}$ be a simply laced AKM algebra, then there is no difference between roots and coroots and $\hat{\Delta}_+=\hat{\Delta}^\vee_+$,   $\nu={u\over h^\vee}$, $f^\vee$ is  the principal nilpotent orbit,
so  $W_{\bfP^\vee}$ is trivial. The group $\bfP^\vee$  in this case is  an Iwahori subgroup and is denoted as $\bfI^\vee$, and  $\tilde{\mathfrak{n}}^\vee$ is the same as the set of positive affine roots $\hat\Delta^\vee_+$.
$L_\nu$ is empty because the maximal height of a finite root $\alpha$ is $h^\vee-1$, so the equation
\begin{equation}
{u\over h^\vee}{{\alpha}(\rho)}+l=0,
\end{equation}
has no solution. 
Elements of $S_\nu$ satisfying the following
\begin{equation}
{u\over h^\vee}{\overline{\alpha}(\rho)}+l={u\over h^\vee},
\end{equation}
and the set of solutions is
\begin{equation}
S_\nu=\{-\theta+u \delta,\alpha_1,\ldots, \alpha_r\},
\end{equation}
which is the same as $S_u$ defined previously in equation \eqref{eq:defSu}. The elliptic element $\gamma$ can be chosen as
\begin{equation}
\gamma=e_{-\theta}z^u+\sum_i e_{\alpha_i}.
\end{equation}

The fixed varieties are labelled by the following elements in the affine Weyl group
\begin{equation}
 \{\tilde{w}\in W_{aff}~|~\tilde{w}S_\nu\subset\hat\Delta_+\}.
\end{equation}
Because $L_\nu$ is empty,  we have
\begin{equation}
|\tilde{w}L_\nu\backslash  \hat{\Delta}_+ |=0.
\end{equation} 
Also because $\tilde{w}S_\nu\subset\hat\Delta_+$, 
\begin{equation}
\tilde{w}S_\nu\backslash  \hat{\Delta}_+ =\emptyset.
\end{equation}
The dimension formula \eqref{dimension} then tells us that each fixed variety $H_{\tilde{w}}$ has dimension $0$. The number of fixed points $|Sp_{\frac{u}{h^\vee},\bfI^\vee}^T|$ is then $u^r$ \cite{varagnolo2009finite}.


\textbf{Sub-regular case}: Again consider $\hat\fkj$ simply laced, but now take $\nu={u\over m}$ with $m$ being the next to maximum value in table \ref{table:ellnumber}, and   $f^\vee$ is still the principal nilpotent one. Notice that now  there is only one finite root $\mu$ of $\hat{\fkj}$ with height $m$, so 
 $L_\nu$  consists two roots
 \begin{equation}
L_\nu=\{\pm(\mu-u\delta)\}.
\end{equation}
The set $S_\nu $ is 
\begin{equation}
S_\nu=\{\alpha + l\delta\in \hat\Delta~|~\frac{u}{m} {\alpha}( \rho^\vee)) +l =\frac{u}{m}\}.
\end{equation}
Now $S_\nu$ contains both positive and negative affine roots. One choice of the elliptic element can be
\begin{equation}
\gamma = \sum_{\alpha+l\delta\in S_\nu, l\geq0} e_{\alpha}z^l.
\end{equation}

%

 Since $L_\nu=\{\pm \tilde{\alpha} \}$, and given $\tilde{w}\in W_{aff}$, the cardinality of $\tilde{w}L_\nu \backslash \hat{\Delta}_+$ is always $1$, 
 the first term in the dimension formula \eqref{dimension} is always $1$. The fixed varieties are separated into two groups by their dimensions.
\begin{enumerate}
\item $\dim H_{\tilde{w}}=1$: so $|\tilde{w}S_\nu \backslash \hat{\Delta}_+|=0$, i.e. $\tilde{w} (S_\nu) \subset \hat\Delta_+$.

\item $\dim H_{\tilde{w}}=0$: so $|\tilde{w}S_\nu \backslash \hat{\Delta}_+|=1$, i.e.  $\tilde{w}S_\nu\cap \hat{\Delta}_-$ has exactly $1$ element.
\end{enumerate}
In next section we will provide explicit examples.

\textbf{Twisted case}: consider a twisted affine Lie algebra $^2\hat{A}_3$ with the slope $\nu={1\over2}$.  The set of real roots is \footnote{Our imaginary root $\delta$ is $n$ times the imaginary root of \cite{carter2005lie}.}
\begin{equation}
\hat{\Delta}^\vee = \{\alpha^\vee+\frac{n}{2}\delta~|~ \alpha^\vee \in \Phi^0_s,~n\in\bbZ \} \cup \{\alpha^\vee+n\delta~|~ \alpha^\vee \in \Phi^0_l,~n\in\bbZ \},
\end{equation} 
where $\Phi^0_s$ and $\Phi^0_l$ are respectively the set of short and long roots of $C_2$ Lie algebra which is the finite part of $^2\hat{A}_3$. In orthogonal basis spanned by $\{\beta_i\}$
\begin{equation}
\Phi^0_l=\{\pm 2\beta_i \},~~~\Phi^0_s=\{\pm \beta_i\pm \beta_j,~~i,j=1,2,~~i\neq j\}.
\end{equation} 
The set of simple roots is 
\begin{equation}
\{\alpha^\vee_1=\beta_1-\beta_2,~\alpha^\vee_2=2\beta_2\}.
\end{equation}
The set $L_\nu$ and $S_\nu$ when $\nu = 1/2$ are
\begin{equation}
L_\nu=\{ \pm(\alpha_1+\alpha_2-\delta)\}\cup \{ \pm(\alpha_1-\frac{1}{2}\delta) \}
\end{equation}
and 
\begin{equation}
S_\nu=\{\alpha_1,~\alpha_2, ~-\alpha_1+\delta, ~-\alpha_2+\delta, ~2\alpha_1+\alpha_2-\delta,~-2\alpha_1-\alpha_2+2\delta,~~\alpha_1+\alpha_2-{\delta\over 2},~-\alpha_1-\alpha_2+{3\delta\over 2} \}.
\end{equation}

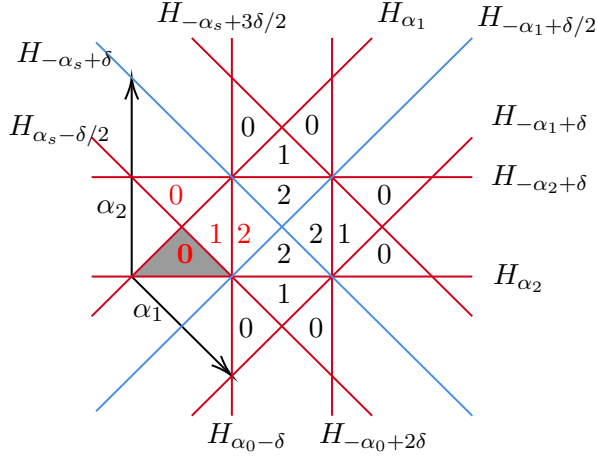
\begin{figure}
\begin{center}

\tikzset{every picture/.style={line width=0.75pt}} 

\begin{tikzpicture}[x=0.75pt,y=0.75pt,yscale=-1,xscale=1]

\draw  [draw opacity=0][fill={rgb, 255:red, 155; green, 155; blue, 155 }  ,fill opacity=1 ] (99,127.55) -- (124,152) -- (74,152) -- cycle ;
\draw    (74,152) -- (74,54) ;
\draw [shift={(74,52)}, rotate = 90] [color={rgb, 255:red, 0; green, 0; blue, 0 }  ][line width=0.75]    (10.93,-3.29) .. controls (6.95,-1.4) and (3.31,-0.3) .. (0,0) .. controls (3.31,0.3) and (6.95,1.4) .. (10.93,3.29)   ;
\draw    (74,152) -- (122.59,200.59) ;
\draw [shift={(124,202)}, rotate = 225] [color={rgb, 255:red, 0; green, 0; blue, 0 }  ][line width=0.75]    (10.93,-3.29) .. controls (6.95,-1.4) and (3.31,-0.3) .. (0,0) .. controls (3.31,0.3) and (6.95,1.4) .. (10.93,3.29)   ;
\draw [color={rgb, 255:red, 208; green, 2; blue, 27 }  ,draw opacity=1 ]   (54,152) -- (244,152) ;
\draw [color={rgb, 255:red, 208; green, 2; blue, 27 }  ,draw opacity=1 ]   (54,102) -- (244,102) ;
\draw [color={rgb, 255:red, 208; green, 2; blue, 27 }  ,draw opacity=1 ]   (124,222) -- (124,32) ;
\draw [color={rgb, 255:red, 208; green, 2; blue, 27 }  ,draw opacity=1 ]   (174,222) -- (174,32) ;
\draw [color={rgb, 255:red, 208; green, 2; blue, 27 }  ,draw opacity=1 ]   (194,32) -- (54,172) ;
\draw [color={rgb, 255:red, 208; green, 2; blue, 27 }  ,draw opacity=1 ]   (244,82) -- (104,222) ;
\draw [color={rgb, 255:red, 208; green, 2; blue, 27 }  ,draw opacity=1 ]   (54,82) -- (194,222) ;
\draw [color={rgb, 255:red, 208; green, 2; blue, 27 }  ,draw opacity=1 ]   (104,32) -- (244,172) ;
\draw [color={rgb, 255:red, 74; green, 144; blue, 226 }  ,draw opacity=1 ]   (244,32) -- (56.22,219.78) ;
\draw [color={rgb, 255:red, 74; green, 144; blue, 226 }  ,draw opacity=1 ]   (56.22,34.22) -- (244,222) ;

\draw (73,164.4) node [anchor=north west][inner sep=0.75pt]    {$\alpha _{1}$};
\draw (55,112.4) node [anchor=north west][inner sep=0.75pt]    {$\alpha _{2}$};
\draw (253,144.4) node [anchor=north west][inner sep=0.75pt]    {$H_{\alpha_2}$};
\draw (252,94.4) node [anchor=north west][inner sep=0.75pt]    {$H_{-\alpha_2+\delta }$};
\draw (252,64.4) node [anchor=north west][inner sep=0.75pt]    {$H_{-\alpha_1+\delta }$};
\draw (195,12.4) node [anchor=north west][inner sep=0.75pt]    {$H_{\alpha_1}$};
\draw (245,14.4) node [anchor=north west][inner sep=0.75pt]    {$H_{-\alpha_1+\delta /2}$};
\draw (166,224.4) node [anchor=north west][inner sep=0.75pt]    {$H_{-\alpha_0+2\delta }$};
\draw (110,224.4) node [anchor=north west][inner sep=0.75pt]    {$H_{\alpha_0-\delta }$};
\draw (85,12.4) node [anchor=north west][inner sep=0.75pt]    {$H_{-\alpha_s+3\delta /2}$};
\draw (15,32.4) node [anchor=north west][inner sep=0.75pt]    {$H_{-\alpha_s+\delta }$};
\draw (11,69.4) node [anchor=north west][inner sep=0.75pt]    {$H_{\alpha_s-\delta /2}$};
\draw (125,124.4) node [anchor=north west][inner sep=0.75pt]    {$\color{red} 2$};
\draw (145,104.4) node [anchor=north west][inner sep=0.75pt]    {$2$};
\draw (161,124.4) node [anchor=north west][inner sep=0.75pt]    {$2$};
\draw (145,134.4) node [anchor=north west][inner sep=0.75pt]    {$2$};
\draw (145,154.4) node [anchor=north west][inner sep=0.75pt]    {$1$};
\draw (175,124.4) node [anchor=north west][inner sep=0.75pt]    {$1$};
\draw (145,84.4) node [anchor=north west][inner sep=0.75pt]    {$1$};
\draw (111,124.4) node [anchor=north west][inner sep=0.75pt]    {$\color{red} 1$};
\draw (126,171.4) node [anchor=north west][inner sep=0.75pt]    {$0$};
\draw (161,171.4) node [anchor=north west][inner sep=0.75pt]    {$0$};
\draw (195,134.4) node [anchor=north west][inner sep=0.75pt]    {$0$};
\draw (195,104.4) node [anchor=north west][inner sep=0.75pt]    {$0$};
\draw (159,69.4) node [anchor=north west][inner sep=0.75pt]    {$0$};
\draw (127,70.4) node [anchor=north west][inner sep=0.75pt]    {$0$};
\draw (91,104.4) node [anchor=north west][inner sep=0.75pt]    {\color{red}$0$};
\draw (95,134.4) node [anchor=north west][inner sep=0.75pt]    {\color{red}$\mathbf{0}$};

\end{tikzpicture}

\end{center}
\caption{Alcoves correspond to fixed varieties of $^2\hat{A}_3$ with $\nu=1/2$ are marked with dimensions,  and the fundamental alcove is the shaded region. Alcoves with dimension marked in red below to different $W_\nu$ orbit. There are one 2d fixed variety, one 1d fixed variety and two fixed points. Redlines are walls from roots in $S_\nu$, blue lines are mirrors from roots in $L_\nu$. Here $\alpha_\theta=2\alpha_1+\alpha_2$, $\alpha_s=\alpha_1+\alpha_2$ is the highest short root, and $\alpha_0=-\alpha_s+\delta/2$.}
\label{fig:2A3nu=2}
\end{figure}

The fixed variety can be found by using the definition \eqref{eq:fpSpringFiber} and the dimension formula \eqref{dimension}. On the other hand, there is also a bijection between fixed varieties and alcoves in the Cartan  $\fkh$, with algorithm listed below \cite{varagnolo2009finite,oblomkov2016geometric}:
\begin{enumerate}
\item For each element $\alpha+l\delta$ in $S_\nu$, one draw a wall which is a hyperplane $H_{\alpha+l\delta}\subset \fkh$ defined by $\{x|(x,\alpha)+l=0\}$ (red lines in figure \ref{fig:2A3nu=2}). For each element $\alpha+l\delta$ in $L_\nu$, one  draws a mirror which is the hyperplane $H_{\alpha+l\delta}$ (blue lines in figure \ref{fig:2A3nu=2}). The fundamental alcove $\Delta_0$ is the region defined by $(x,\alpha)>0$, $i=1,\cdots,r$ and $(x,-\alpha_\theta)+1>0$ (the shaded area of \ref{fig:2A3nu=2}).
\item For an element $\tilde{w}\in W_{aff}$, $H_{\tilde{w}}$ is a fixed variety if $\tilde{w}^{-1}\Delta_0$ lies in a bounded region closed by walls. If any points $x\in \tilde{w}^{-1}\Delta_0$ satisfy $(x,\alpha)>0$ (or $(x,\alpha)<0$), we call $\tilde{w}^{-1}\Delta_0$ is on the positive (or negative) side of the wall $H_\alpha$. $|\tilde{w}S_\nu \backslash \hat{\Delta}_+|$ is the number of walls of which $\tilde{w}^{-1}\Delta_0$ lies on the negative side. Finally, if $\tilde{w_1}^{-1}\Delta_0$ is  the same as  $\tilde{w_2}^{-1}\Delta_0$  reflected by some mirrors, they correspond to the same fixed variety.
\end{enumerate}
Alcoves corresponding to fixed varieties in our example are shown in figure \ref{fig:2A3nu=2} with dimensions labelled. The alcoves marked by red dimension numbers are not reflected by the mirrors so they correspond to different fixed varieties. There are total of $4$ fixed varieties.

\newpage

\section{Mirror symmetry for circle compactified 4d $\mathcal{N}=2$ theory}
\label{sec:4dmirror}

With necessary knowledge reviewed in previous sections, we can finally make precise statements on the mirror symmetry and provide various checks. For a circle compactified 4d $\mathcal{N}=2$ SCFT, we now have two  objects: the first one is the W-algebra $W_{-h^\vee+\frac{1}{n\nu}}(\fkg,f)$ \footnote{Here $n$ is the lacety number which is the same as the rank of the automorphism group when $n>1$ in table \ref{table:outm}.} which describes Schur sector.  The second one is the  Hitchin moduli space $\CM_{hit}((\mathfrak{j},o), \nu, (f^\vee, c))$ for the Coulomb branch. 
Notice that the defining data involves Langlands dual on algebras:
\begin{enumerate}
\item The W-algebra $W_{-h^\vee+\frac{1}{n\nu}}(\fkg,f)$ on the Schur sector is related to the affine Lie algebra  $\hat{\mathfrak{g}}$, while the  Hitchin moduli space involves the twisted affine Lie algebra based $^n\hat{\mathfrak{j}}$ which is  the {\bf Langlands dual} of $\hat{\mathfrak{g}}$. 
\item The $(f^\vee,c) \in \fkg^\vee$ used in the Hitchin moduli space is the dual of the nilpotent element $f\in \fkg$ in $W_{-h^\vee+\frac{1}{n\nu}}(\fkg,f)$, and $\fkg^\vee$ is the Langlands dual of $\fkg$.
\end{enumerate}

\subsection{Simple modules of W-algebra and fixed points}

Our first statement is that there is a natural bijection between simple modules of $W_{-h^\vee+\frac{1}{n\nu}}(\fkg,f)$ and  irreducible components of $\bbC^\ast$-fixed varieties of $\CM_{hit}((\mathfrak{j},o), \nu, (f^\vee,c))$. The case when $\fkg=A_{N-1}$ and $f$ principal was first noticed in \cite{Fredrickson:2017jcf}. Cases when $\fkg=A_1$, $f$ trivial or cases when the W-algebra being $W_N$ and $B_N$ algebras are discussed in \cite{Fredrickson:2017yka}. Our results vastly generalize previous understanding of the bijection between modules and fixed varieties.

Recall that  irreducible components of fixed varieties of $\CM_{Hit}((\mathfrak{j},o), \nu, (f^\vee,c))$ are the same as fixed varieties of corresponding affine Springer fibre $Sp_{\gamma,\bfP^\vee}$ which
are parameterized by 
the affine Weyl elements $\tilde{w}$ satisfying condition in \eqref{eq:fpSpringFiber} up to a double coset. The bijection between fixed varieties of Hitchin moduli space and weights of  simple modules of $W_{-h^\vee+\frac{1}{n\nu}}(\fkg,f)$ is 
\begin{equation}
\label{eq:biFPtoModule}
\begin{split}
&\text{Fixed varieties of}~\CM_{hit}((\mathfrak{j},o), \nu, (f^\vee, C)) \xrightarrow{\simeq} \mathrm{Irrep}(W_{-h^\vee+\frac{1}{n\nu}}(\fkg,f)),\\
&~~H_{\tilde{w}} \mapsto H_f(L(\Lambda_{\tilde{w}})),\quad \mathrm{with}~ \Lambda_{\tilde{w}}=\tilde{w}\cdot(\kappa\Lambda_0),
\end{split}
\end{equation}
Here $H_{\tilde{w}}$ denotes the irreducible component of the fixed varieties, and $\Lambda_{\tilde{w}}$ denotes the affine weight.

\textbf{Remark}: In above proposal, we assume that the module of W-algebra is defined by choosing $f$ regular in a standard L\'{e}vi $\fkl$ which naturally matches the definition of fixed varieties on the Hitchin side. To get
the data for the W-algebra corresponding to the 4d theory (namely the grading has to be given by the standard $\fsl_2$ triple), one need to do a further
transformation resulting a shift in the conformal dimension.  

\subsubsection{W-algebras at boundary admissible level}

Given a Lie algebra $\fkg$, if the level $\kappa$ is at the boundary admissible level $\kappa = -h^\vee+\frac{h^\vee}{u}$ with $\mathrm{gcd}(u,h^\vee)=1$, then the slope $\nu$ of the corresponding fibre is $\frac{u}{h^\vee}$, and the denominator $h^\vee$ is a regular elliptic number. In this situation, $L_\nu$ is always an empty set. By the dimension formula \eqref{dimension} all fixed varieties have dimension $0$ (fixed points).  For boundary admissible case, the bijection can be proved rigorously and is given in our accompanying  paper \cite{Xie:2023pre}.

{\bf AKM cases:} Let $f$ being the trivial nilpotent orbit. One gets the associated vertex algebra $L_{-h^\vee+\frac{h^\vee}{u}}(\fkg)$ on the VOA side.
Following the notation in section \ref{sec:admAKM}, the set of admissible weights are given by
\begin{equation}
\{\tilde{w}.(\kappa\Lambda_0)~|~\tilde{w} \in W_{ext}/\Omega_u,~\tilde{w}(S_u)\subset \hat{\Delta}^\vee_+ \}.
\end{equation}
On the Hitchin side, both $W_{\nu}$ and $W_{\bfP^\vee}$ are trivial, and the set of fixed varieties is labelled by (in this case, $\tilde{\mathfrak{n}}^\vee$ is equal to the set of positive affine roots $\hat{\Delta}^\vee_+$)
\begin{equation}
\{\tilde{w} \in W_{aff}~|~\tilde{w}(S_\nu)\subset \hat{\Delta}^\vee_+\}.
\end{equation}
The dimension of each fixed variety is $0$ (fixed points). Notice that here $S_u=S_\nu$. 
One can show that for each element of the coset $W_{ext}/\Omega_u$, there is one and only one element in $W_{aff}$ \footnote{Note that both $W_{ext}$ and $W_{aff}$ are invariant under Langlands dual.}, hence the bijection. Both the number of admissible modules and fixed points are $u^r$.


\begin{example} $L_{-2+2/u}(\fsl_2)\leftrightarrow \CM_{Hit}(\fsl_2,\frac{u}{2},[2])$. Here $u$ should be an odd integer. We have $\fkg=\fkg^\vee=\fsl_2$,  $S_u=S_\nu=\{\alpha, -\alpha+u\delta\}$. 
An element in the affine Weyl group can always be written as an element of the finite Weyl group followed by a translation in the root lattice $ t_{m\alpha}s$ with $s$ being $1$ or $s_\alpha$. The fixed points are labelled by the following subset of $W_{aff}$
\begin{equation}
\{t_{-m\alpha}~|~0\leq 2m <u\} \cup \{t_{m\alpha} s_\alpha ~|~0<2m\leq u\}.
\end{equation}
The first set has $(u+1)/2$ elements and the second one has $(u-1)/2$ elements, so the total number of fixed points are $u$. Using  the formula \eqref{eq:biFPtoModule}, we found that  weights from the first set are
\begin{equation}
\{ \left(-2+\frac{2}{u}+\frac{4m}{u}\right)\Lambda_0-\frac{4m}{u}\Lambda_1~|~0\leq 2m<u\}
\end{equation}
and weights from the second set are
\begin{equation}
\{   \left(-2+\frac{2}{u}+\frac{2u-4m}{u}\right)\Lambda_0-\frac{2u-4m}{u}\Lambda_1~|~0< 2m<u     \}.
\end{equation}
They give exactly the same weights as \eqref{eq:admsl2} in example \ref{ex:su2}.
\end{example}

\begin{example} $L_{-3+3/u}(\fsl_3)\leftrightarrow \CM_{Hit}(\fsl_3,u/3,[3])$. Here $u$ is coprime with $3$. $S_\nu$ is the same as $S_u$
\begin{equation}
\label{eq:Snusl3}
S_u=S_\nu=\{-\theta+u\delta, \alpha_1, \alpha_2 \}
\end{equation}
 with $\theta=\alpha_1+\alpha_2$.  The condition for  $\tilde{w}=t_{\beta}y$ to give rise to a fixed point is
\begin{equation}
t_\beta y(S_u)\subset \hat{\Delta}_+,\quad \beta\in Q,
\end{equation}
with $Q$ being the root lattice of $\fsl_3$. 
For $u=4$, the list of fixed points and the corresponding affine weights \eqref{eq:biFPtoModule} are listed in table \ref{table:fpsl3}, and we get exactly the same weights  in table \ref{table:sl3adm} in example \ref{ex:su3u4}. We also plot all alcoves corresponding to fixed points in figure \ref{fig:sl3nu43alcove}.
\begin{table}[htp]\small
\begin{center}
\begin{tabular}{|c|c||c|c||c|c|} \hline
$t_\beta y$ & $\Lambda$ & $t_\beta y$ & $\Lambda$ & $t_\beta y$ & $\Lambda$ \\ \hline
$1$ & $ -{9\over 4}\Lambda_0 $& $t_{-\alpha_1-\alpha_2} $ & $ -{3\over4}\Lambda_0-{3\over4}\Lambda_1-{3\over4}\Lambda_2 $& $t_{-\alpha_1-2\alpha_2}$ & $ -{9\over4}\Lambda_2 $  \\ \hline
 $t_{-2\alpha_1-\alpha_2} $ & $ -{9\over 4}\Lambda_1 $& $t_{-\alpha_2}s_1 $ & $ -{2\over4}\Lambda_0-{5\over4}\Lambda_1-{2\over4}\Lambda_2 $ &$t_{\alpha_1-\alpha_2}s_1 $ & $ -{5\over4}\Lambda_0+{1\over4}\Lambda_1-{5\over4}\Lambda_2 $ \\ \hline
 $t_{-\alpha_1}s_2 $ & $ -{2\over4}\Lambda_0-{2\over4}\Lambda_1-{5\over4}\Lambda_2 $& $t_{-\alpha_1+\alpha_2}s_2 $ & $ -{5\over4}\Lambda_0-{5\over4}\Lambda_1+{1\over4}\Lambda_2 $ &$t_{\alpha_2}s_2s_1 $ & $ -{3\over4}\Lambda_1-{6\over4}\Lambda_2 $ \\ \hline
 $t_{2\alpha_2}s_2s_1 $ & $ -{3\over4}\Lambda_0-{6\over4}\Lambda_1 $& $t_{\alpha_1+2\alpha_2}s_2s_1 $ & $ -{6\over4}\Lambda_0-{3\over4}\Lambda_2  $ &$t_{\alpha_1}s_1s_2 $ & $ -{6\over4}\Lambda_1-{3\over4}\Lambda_2 $ \\ \hline
 $t_{2\alpha_1}s_1s_2 $ & $ -{3\over4}\Lambda_0-{6\over4}\Lambda_2  $& $t_{2\alpha_1+\alpha_2}s_1s_2 $ & $ -{6\over4}\Lambda_0-{3\over4}\Lambda_1 $ &$t_{\alpha_1+\alpha_2}s_1s_2s_1 $ & $ {1\over4}\Lambda_0-{5\over4}\Lambda_1-{5\over4}\Lambda_2$ \\ \hline
  $t_{2\alpha_1+2\alpha_2}s_1s_2s_1 $ & $ -{5\over4}\Lambda_0-{2\over4}\Lambda_1-{2\over4}\Lambda_2 $& &&  &\\ \hline
\end{tabular}
\end{center}
\caption{Fixed points of $\CM_{Hit}(\fsl_3,3/u,[3])$ and their images under the bijection \eqref{eq:biFPtoModule}.}
\label{table:fpsl3}
\end{table}

\begin{figure}[h]
\begin{center}
\includegraphics[scale=0.6]{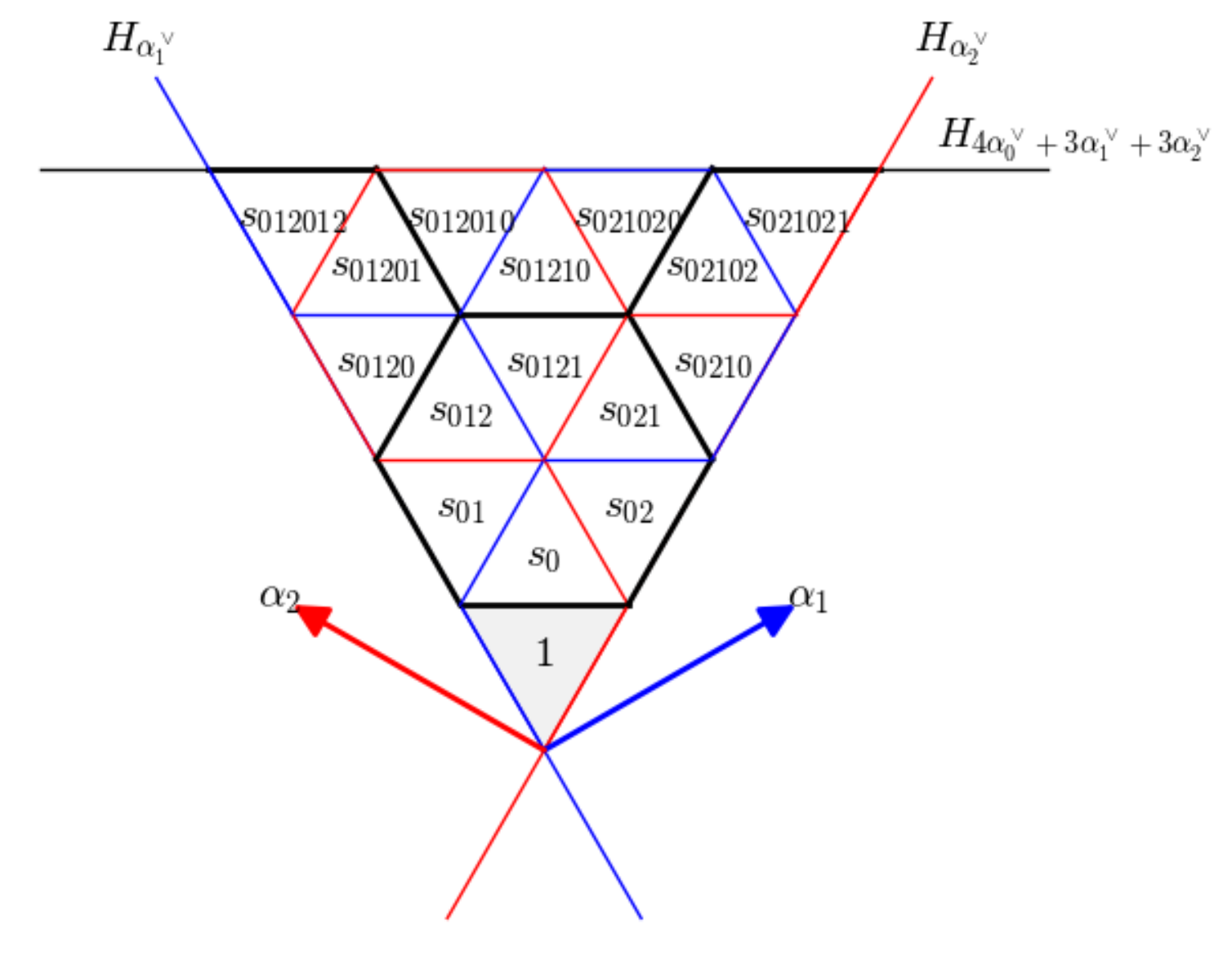}
\end{center}
\caption{\label{fig:sl3nu43alcove}Alcoves corresponding fixed points for $A_2$, $\nu=4/3$. Each alcove gives rise to a affine Weyl group element whose inverse gives rise to a fixed point: here $s_0=s_\theta t_{-\theta}$, and $s_i$ is the Weyl reflection generated by the simple roots of the Lie algebra. For example, the element $s_0$ in the region gives rise to 
an element $s_0^{-1}= t_{\theta}s_\theta=t_{\alpha_1+\alpha_2} s_1 s_2s_1$.}
\end{figure}
\end{example}

{\bf W-algebras case:} Now we consider cases when $f$ is a regular nilpotent element in a L\'{e}vi. As discussed in section \ref{sec:repWalg},  simple modules of $W_{-h^\vee+\frac{h^\vee}{u}}(\fkg,f)$ are reduced from  modules of $L_{-h^\vee+\frac{h^\vee}{u}}(\fkg)$ satisfying the condition \eqref{eq:LmodtoWmod}. In particular, some  modules are projected out, and 
multiple modules of AKM is mapped to the same simple module of the W-algebra. On the Hitchin side, one can easily see the similar pattern: firstly the set $L_\nu$ and $S_\nu$ is not changed, secondly the set of affine roots of $\tilde{\fkn}^\vee$ is now smaller than $\hat{\Delta}^\vee_+$ and so some of 
the previous fixed points will be projected out, thirdly one should quotient by a Weyl group $W_{\bfP^\vee}$ action to get the final results. So the pattern on Hitchin side matches precisely with that on the VOA side.

\begin{example} $W_{-3+3/u}(\fsl_3,[2,1])\leftrightarrow \CM_{Hit}(\fsl_3,u/3,[2,1])$.  $L_\nu$ is again empty and $S_\nu$ is the same as equation \eqref{eq:Snusl3} but 
\begin{equation}
\Delta_{\tilde{\mathfrak{n}}^\vee}=\hat{\Delta}_{+} \backslash \{ \alpha_1\},
\end{equation}
and so $W_{\bfP^\vee}$ is generated by $s_1$.
The condition for fixed points are
\begin{equation}
\tilde{w}(S_u)\subset \Delta_{\tilde{\fkn}^\vee} = \hat{\Delta}^\vee_{+} \backslash \{ \alpha_1\},\quad \tilde{w}\in W_{\bfP^\vee}\backslash W_{aff}.
\end{equation}
The total $6$ fixed points when $u=4$ are listed in table \ref{table:fpsl3u4f21} which are matched to the modules in table \ref{table:sl3min} through the bijection \eqref{eq:biFPtoModule}. Alcoves of fixed points are drawn in figure \ref{fig:alcovesl3u4f21}. In general there are $u(u-1)/2$ fixed points.

\begin{table}[htp]

\begin{center}

\begin{tabular}{|c||c|} \hline 
$t_\beta y,~\Lambda$  & $t_\beta y,~ \Lambda$   \\ \hline
\begin{tabular}{c} 
$t_{2\alpha_1+\alpha_2}s_1s_2$, $-\frac{6}{4}\Lambda_0-\frac{3}{4}\Lambda_1$ \\  $t_{-\alpha_1+\alpha_2}s_2$, $-\frac{5}{4}\Lambda_0-\frac{5}{4}\Lambda_1+\frac{1}{4}\Lambda_2$ \end{tabular} & 
\begin{tabular}{c} $t_{-2\alpha_1-\alpha_2}$, $-\frac{9}{4}\Lambda_1$  \\   $t_{\alpha_1-\alpha_2}s_1$, $-\frac{5}{4}\Lambda_0 +\frac{1}{4} \Lambda_1-\frac{5}{4}\Lambda_2$ \end{tabular}\\
  \hline
\begin{tabular}{c}
$t_{2\alpha_2}s_2s_1$, $-\frac{3}{4}\Lambda_0-\frac{6}{4}\Lambda_1$ \\ $t_{2\alpha_1+2\alpha_2}s_1s_2s_1$, $-\frac{5}{4}\Lambda_0-\frac{2}{4}\Lambda_1-\frac{2}{4}\Lambda_2$ 
\end{tabular}   &
\begin{tabular}{c} $t_{-\alpha_1-\alpha_2}$, $-\frac{3}{4}\Lambda_0-\frac{3}{4}\Lambda_1-\frac{3}{4}\Lambda_2$ \\
 $t_{ -\alpha_2 }s_1$, $\frac{-2}{4}\Lambda_0 -\frac{5}{4}\Lambda_1 -\frac{2}{4}\Lambda_2$ \end{tabular}  \\ 
 \hline
 \begin{tabular}{c}
$t_{\alpha_1}s_1s_2$, $-\frac{6}{4}\Lambda_1-\frac{3}{4}\Lambda_2$ \\ $t_{-\alpha_1}s_2$, $-\frac{2}{4}\Lambda_0 -\frac{2}{4}\Lambda_1 -\frac{5}{4}\Lambda_2$ \end{tabular}
 & \begin{tabular}{c} $t_{\alpha_2}s_2s_1$, $-\frac{3}{4}\Lambda_1 -\frac{6}{4}\Lambda_2$ \\
  $t_{\alpha_1+\alpha_2}s_1s_2s_1$, $\frac{1}{4}\Lambda_0-\frac{5}{4}\Lambda_1-\frac{5}{4}\Lambda_2$ \end{tabular}   \\ \hline
\end{tabular}

%

\end{center}
\caption{Fixed points of $\CM_{Hit}(\fsl_3,3/4,[2,1])$ and their images under the bijection \eqref{eq:biFPtoModule}. The affine Weyl elements in the same box are related by the left action of $s_1$, so they reduces to the same fixed points.}
\label{table:fpsl3u4f21}
\end{table}

\begin{figure}[h]
\begin{center}
\includegraphics[scale=0.6]{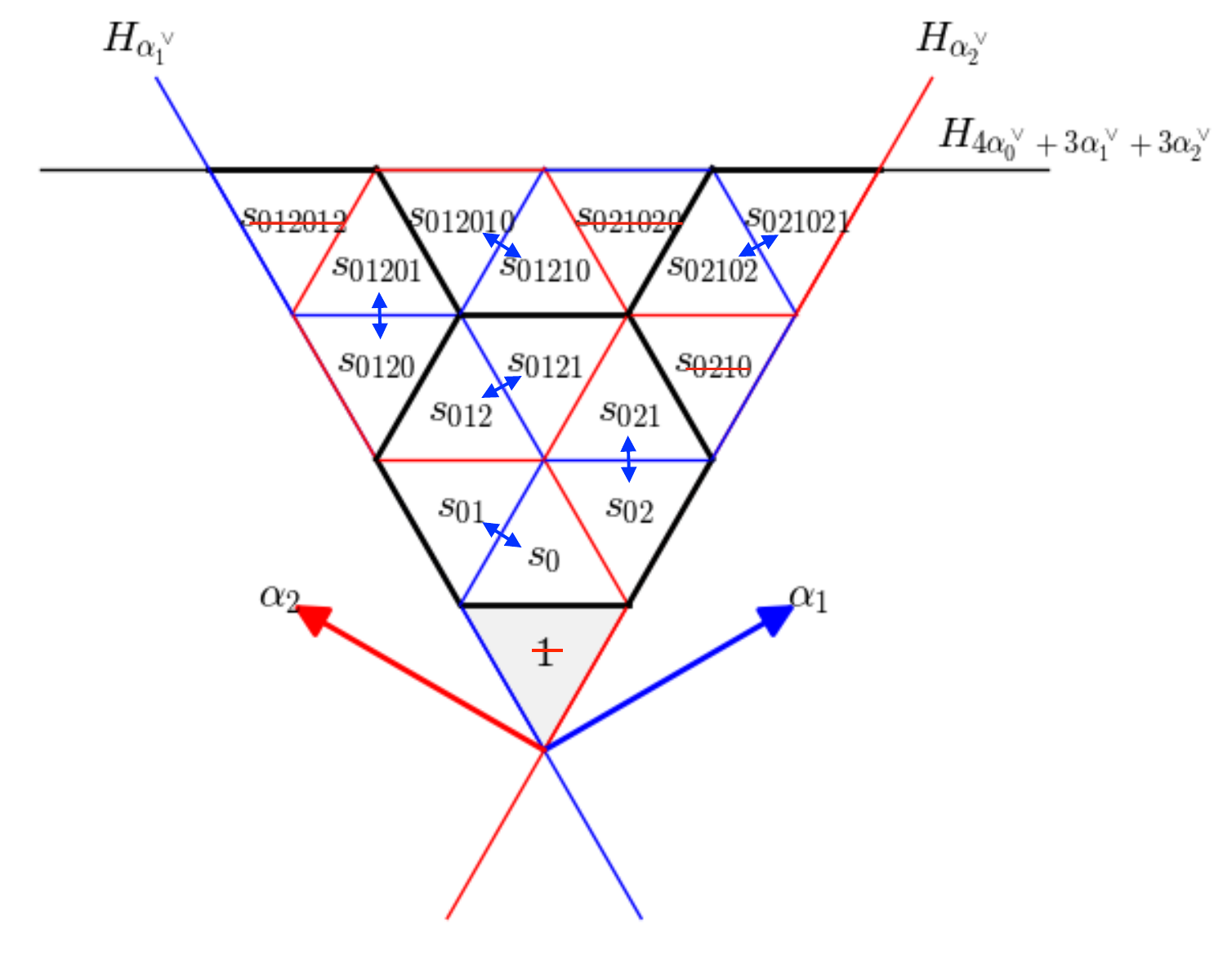}
\end{center}
\caption{\label{fig:alcovesl3u4f21}Alcoves corresponding fixed points for $A_2$, $\nu=4/3$, $f^\vee=[2,1]$. Alcoves with a blue edge (corresponding to reflection $s_1$) on the walls does not reduce to a fixed point (under $s_1$ they are reflected out of the area bounded by walls). Alcoves separated by a blue edge (two alcoves encircled by red edges and black edges) reduce to the same fixed point. Note that the label in each alcove is $w^{-1}$ in terms of simple reflections.}
\end{figure}
\end{example}

\begin{example} $W_{-3+3/u}(\fsl_3,[3])\leftrightarrow \CM_{Hit}(\fsl_3,u/3,[1,1,1])$.  Here $\Delta_{\tilde{\mathfrak{n}}^\vee}=\hat{\Delta}^\vee_+\backslash \{\alpha_1,\alpha_2\}$, and $W_{\bfP^\vee}$ is the full Weyl group of $\fsl_3$, so 
one only has to  consider the affine Weyl group elements of the form $t_{-k_1 \alpha_1-k_2\alpha_2} $. 
Constraints on fixed points are
\begin{equation}
\begin{split}
& t_{-k_1 \alpha_1-k_2\alpha_2} (-\alpha_1-\alpha_2+u\delta) =-\alpha_1-\alpha_2+(-k_1-k_2+u)\delta ,\\
& t_{-k_1\alpha_1-k_2\alpha_2}(\alpha_1)=\alpha_1+(2k_1-k_2)\delta,\\
& t_{-k_1\alpha_1-k_2\alpha_2}(\alpha_2)=\alpha_2+(2k_2-k_1)\delta. 
\end{split}
\end{equation}
The set of allowed $(k_1, k_2)$ is
\begin{equation}
\{(k_1,k_2)\in\bbZ^2~|~u-k_1-k_2> 0,~2k_1-k_2>0,~2k_2-k_1>0\}
\end{equation}
and the number of fixed points are ${(u-2)(u-1)\over 6}$. The corresponding W-algebra $W_{-3+3/u}(\fsl_3,[3])$ is ismorphic to $W_3(3,3+u)$ minimal model and this is the bijection discussed in \cite{Fredrickson:2017jcf}. Alcoves corresponding to fixed points when $u=4$ are shown in figure \ref{fig:alcovesl3u4f111}.

\begin{figure}[h]
\begin{center}
\includegraphics[scale=0.6]{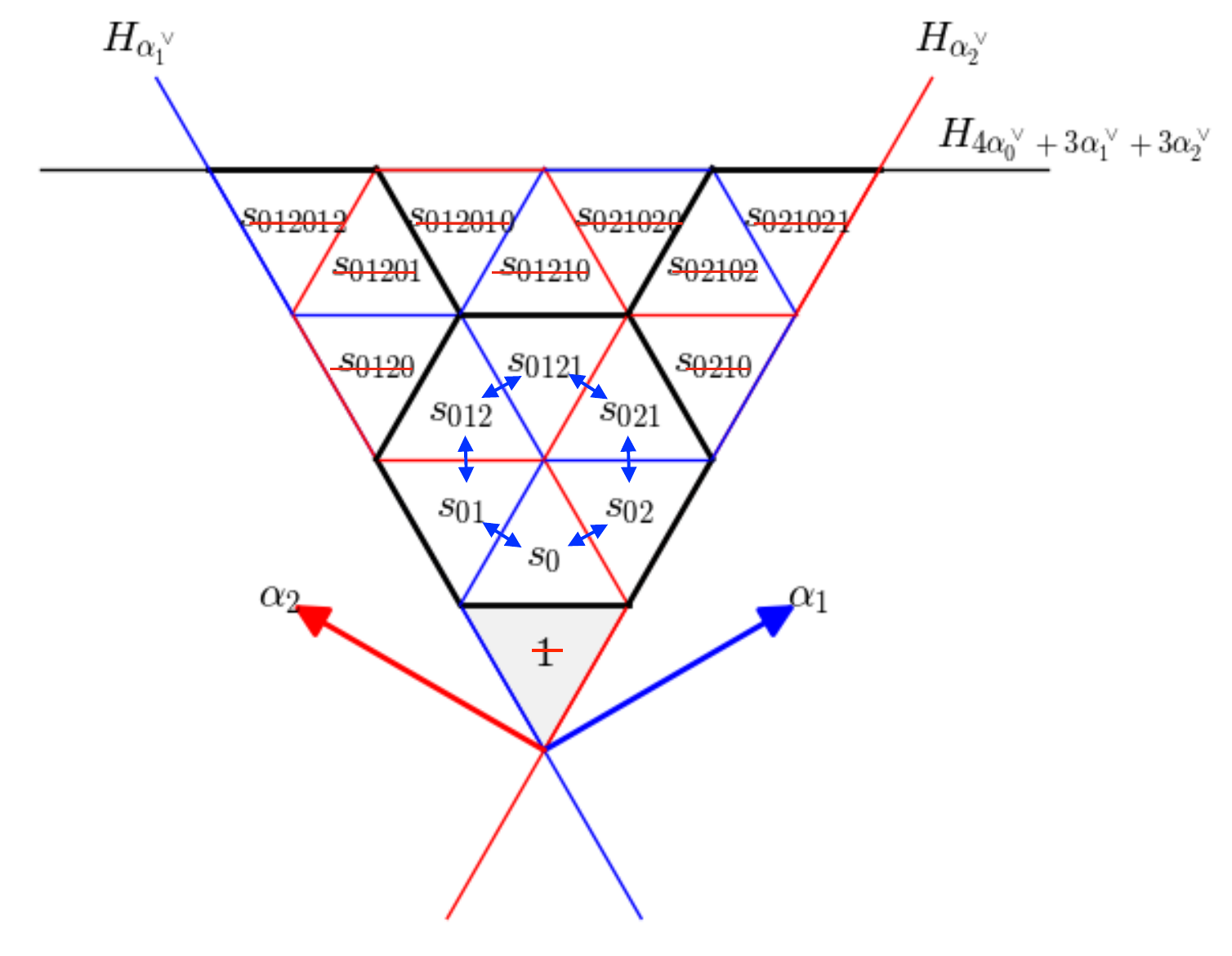}
\end{center}
\caption{\label{fig:alcovesl3u4f111}Alcoves corresponding fixed points for $A_2$, $\nu=4/3$, $f^\vee=[1,1,1]$.  Alcoves encircled by a hexagon of black edges are in the same $W_{\bfP^\vee}$ orbit so reduces to the same fixed point. Since only one such hexagon in the area bounded by walls, so there is only one fixed point. Note that the label in each alcove is $w^{-1}$ in terms of simple reflections.}
\end{figure}
\end{example}

\subsubsection{Non-admissible W-algebras}

In general it is not easy to study the representation theory of non-admissible W-algebras. On the other hand, computing fixed manifolds of corresponding affine Springer fibers is straight forward. Although we will not be able to provide a proof of the bijection like in boundary admissible cases, we can show that the bijection still holds for the few cases when the simple modules of non-admissible W-algebras are known \cite{pervse2013note, Arakawa_2016}, and it is also interesting to use this bijection to predict information on other non-admissible W-algebras.

For example, consider the affine vertex algebra $L_{-2}(D_4)$. Since $h^\vee = 6$, the level $\kappa=-6+4/1$ is non admissible. One the fibre side we have $\mathfrak{g}=D_4, \nu=\frac{1}{4}, \bfP^\vee=\bfI^\vee$. 
%
To compute fixed varieties, we  first find $L_\nu$ and $S_\nu$. The set $L_\nu=\{ \alpha+l\delta ~|~ {1\over 4}(\alpha,\rho^\vee)+l=0 \}$ in this case is non-empty
\begin{equation}
L_\nu=\{\pm(-\mu+\delta) \},
\end{equation}
where $\mu=\alpha_1+\alpha_2+\alpha_3+\alpha_4$, so $W_\nu$ is the Weyl group generated by $s_{-\mu+\delta}$. The set $S_\nu=\{ \alpha+l\delta ~|~  {1\over 4}(\alpha,\rho^\vee)+l={1\over 4} \}$ is also larger than the admissible case: 
\begin{equation}
S_\nu=\{\alpha_1,\alpha_2,\alpha_3,\alpha_4,-\mu+\alpha_1+\delta,-\mu+\alpha_3+\delta, -\mu+\alpha_4+\delta, \theta-\delta \},
\end{equation}
with $\theta=\mu+\alpha_2=\alpha_1+2\alpha_2+\alpha_3+\alpha_4$ being the highest root. We adopt the Bourbaki numbering for simple roots \cite{bourbaki2006groupes,carter2005lie}.

By the discussion in section \ref{sec:fpFibre},  the dimension one fixed variety  is given by the affine Weyl element $\tilde{w}$ such that $\tilde{w}(S_\nu)\subset \hat{\Delta}_+$ up to the right action of $W_\nu$. There are only two such elements in $W_{aff}$, and they are indeed in the same $W_\nu$ orbit
\begin{equation}
\label{eq:fixmanD4}
\begin{split}
&\tilde{w}_0=s_0=t_{\theta} s_2s_3s_1s_2s_4s_2s_3s_1s_2,\\
&\tilde{w}'_0=s_0s_{-\mu+\delta}=t_{\mu} s_3s_1s_2s_4s_2s_3s_1s_2,
\end{split}
\end{equation}
where $s_i$ is the simple reflection corresponding to the simple root $\alpha_i$. Therefore there is only one fixed variety with dimension $1$.

The dimension $0$ fixed points corresponds to affine Weyl group elements $\tilde{w}$ satisfying $|\tilde{w}(S_\nu)\cap \hat{\Delta}_-|=1$ up to the right action by $W_\nu$, and there are four fixed points
\begin{equation}
\label{eq:fpD4}
\begin{split}
\tilde{w}_1&=1\\
\tilde{w}_2&=s_1s_0=t_\theta s_2s_3s_1s_2s_4s_2s_3s_1s_2s_1,\\
\tilde{w}_3&=s_3s_0=t_\theta s_2s_3s_1s_2s_4s_1s_2s_3s_1s_2, \\
\tilde{w}_4&=s_4s_0=t_\theta s_4s_2s_3s_1s_2s_4s_2s_3s_1s_2.
\end{split}
\end{equation}
The weights under the bijection \eqref{eq:biFPtoModule} are given by $t_\beta w.(-2\Lambda_0)$ and results are summarized in  table \ref{table:fpsD4nonadm} ($-2\Lambda_0$ is invariant under the dot action of $W_\nu$ so $\tilde{w}$ and $\tilde{w}s_{-\mu+\delta}$ give the same weight). Indeed they agree with  results in VOA literature \cite{pervse2013note,Arakawa_2016}. If one changes $f$ to be an element in the minimal nilpotent orbit, there will be only one fixed point on the fibre side, and this is also consistent with the fact that $W_{-2}(D_4,\mathrm{min})$ is isomorphic to $\bbC$ \cite{Arakawa_2016}. More examples on the  bijection between fixed varieties and simple modules of non-admissible W-algebras  are discussed in appendix \ref{app:sec:nonAdm}.

\begin{table}
\centering
\begin{tabular}{|c|c|c|} \hline
Dim & $\tilde{w}=t_\beta w$ & $\tilde{w}\cdot (k\Lambda_0)$  \\
\hline
0 & $1$ & $-2\Lambda_0$ \\ \hline
0 & $s_4s_0=t_\theta s_4r_1r_2$ & $-2\Lambda_4$  \\ \hline
0 & $s_1s_0=t_\theta r_1r_2s_1$ & $-2\Lambda_1$  \\ \hline
0 & $s_3s_0=t_\theta r_1s_1r_2$ & $-2\Lambda_3$  \\ \hline
\hline
1 & $s_0=t_\theta r_1r_2$ & $-\Lambda_2$ \\  \hline
\end{tabular}
\caption{\label{table:fpsD4nonadm}Fixed points of $D_4,\nu=\frac{1}{4}$ and their images under the bijection \eqref{eq:biFPtoModule}. Here $r_1=s_2s_3s_1s_2s_4$, $r_2=s_2s_3s_1s_2$ and $\theta=\alpha_1+2\alpha_2+\alpha_3+\alpha_4$.}
\end{table}

\subsubsection{Formula for the number of fixed varieties}
Here we give a formula on the number of fixed varieties of a fibre which will also give the number of simple modules of the corresponding W-algebras under the bijection \eqref{eq:biFPtoModule}.
\begin{enumerate}
\item  For the Hitchin system defined by $^n\hat{\mathfrak{j}}$, $\nu=u/m$ and $\bfI^\vee$, the corresponding VOA is $L_{-h^\vee+{1\over n \nu}}(\fkg)$ where $\hat{\fkg}$ is the Langlands dual of $^o\mathfrak{j}$ whose finite part is $\fkg^\vee$. Let  $a$ be the dimension of the cohomology of fixed varieties when $u=1$, then 
that of the general $u$ is \cite{varagnolo2009finite, oblomkov2016geometric}
\begin{equation}
a u^r
\end{equation} with $r$ being the rank of $\mathfrak{g}$. In particular, when $m$ is a regular elliptic number, $a=1$, and there are only fixed points and so the number is $u^r$. The value of $a$ for the other cases can be found in \cite{oblomkov2016geometric}.
\item  For the Hitchin system defined by non twisted affine Lie algebra $\hat{\mathfrak{j}}$, $\nu=u/ h^\vee$, and general $\bfP^\vee$ which is given by the standard parabolic subalgebra, the corresponding VOA is a W-algebra at boundary admissible number, we show in \cite{Xie:2023pre} the number of fixed points is
\begin{equation}
\label{eq:fpNumReg}
 \frac{(u-m_1)(u-m_2)\cdots(u-m_j)}{(m_1+1)(m_2+1)\cdots(m_i+1)},
\end{equation}
where the set $\{m_1,m_2,\cdots, m_i\}$ is the set of exponents of the Weyl group of $\fkl$ of $\fkp$.
\end{enumerate}



\subsection{Conformal weights and momentum map}

The bijection \eqref{eq:biFPtoModule} also maps geometric data on the Hitchin side to algebra data on the VOA side. On the Hitchin side, one can define a moment map of the $\bbC^\ast$ action, and it was shown in several cases that the value of moment map on each fixed points is equal to the conformal weights of the corresponding module up to shift by a  constant \cite{Fredrickson:2017jcf, Fredrickson:2017yka}. We discuss a generalization of this correspondence in this section.

For simplicity, let us focus on the regular elliptic slope $\nu$,  so the W-algebra is $W_{-h^\vee+\frac{h^\vee}{nu}}(\fkg,f)$. Given a fixed point $\tilde{w}=t_b w$, the corresponding Higgs field is (up to gauge transformation)
\begin{equation}
\Phi_{\tilde{w}}(z)dz=\frac{dz}{z}\sum_{\alpha+l\delta\in S_\nu,l>0}z^{l-(w^{-1}b, \alpha)}e_{w\alpha}.
\end{equation}
 Following \cite{Fredrickson:2017jcf}, one can define the moment map on the Hitchin moduli space \footnote{We add a factor of $\half$ in the definition to better match with the conformal dimension of the simple module. We also set all parabolic weights $\alpha_i$ to zero for simplicity.}
\begin{equation}
\label{eq:momentMapDef}
\mu\equiv \frac{i}{2\pi}\int\mathrm{Tr}\left(\Phi\wedge \Phi^{\dag_h}-\mathrm{Id}|z|^{2(u-h^\vee)/h^\vee}dzd\bar{z}\right),
\end{equation}
where $\Phi^{\dag_h}=h^{-1}\Phi^{\dag}h$ is the Hermitian adjoint of $\Phi$, and $h$ is the Hermitian metric of the Higgs bundle. And we propose the following relation between the moment map of $\Phi_{\tilde{w}}$ and the conformal dimension of $H_f(L(\Lambda_{\tilde{w}}))$
\begin{equation}
\label{eq:momentMaptoDim}
\boxed{h_{H_f(L(\Lambda_{\tilde{w}}))}=\mu(\Phi_{\tilde{w}}(z))-\left[\frac{u}{h^\vee}|\rho|^2-\frac{h^\vee}{u}|x|^2-2(x,\rho)\right].}
\end{equation}
Here $x$ should be chosen to be $H/2$ of the standard triple $(X,Y,H)$ to match with the VOA corresponding to 4d theory. It is straightforward to check that when $\fkg=A_{N-1}$ and $f=[N]$, equation \eqref{eq:momentMaptoDim} reproduces the result in \cite{Fredrickson:2017jcf} and when $\fkg=A_1$ and $f=[1,1]$, equation \eqref{eq:momentMaptoDim} also gives the result in \cite{Fredrickson:2017yka}.

We provide a derivation of \eqref{eq:momentMaptoDim} for $\fkg=A_{N-1}$, which essentially follows from \cite{Fredrickson:2017jcf}. Fixed point then has the following matrix  form
\begin{equation}
\Phi_{\tilde{w}}(z)=M\left(
\begin{array}{cccc}
0 & z^{b_1} &  &   \\
 &  & \ddots &  \\
 &&& z^{b_{N-1}} \\
 z^{b_{N}} &&&
\end{array}
\right)M^{-1}dz,
\end{equation}
where $M$ is a permutation matrix and
\begin{equation}
b_i=-(w^{-1}b,\alpha_i)-1,\ 1\leq i\leq N-1,\quad b_{N}=u-1+(w^{-1}b,\theta).
\label{bvalue}
\end{equation} 
The moment map at this fixed point can be computed explicitly using the definition \eqref{eq:momentMapDef}
\begin{equation}
\mu(\Phi_{(w,b)})=\frac{h^\vee}{2u}|\mathbf{a}|^2,
\end{equation}
where  coordinates of the $N$ dimensional vector $\mathbf{a}$ is related with the coordinates of the vector $\mathbf{b}$ by
\begin{equation}
b_i-\frac{u-h^\vee}{h^\vee}=a_i-a_{i+1},\quad \sum_{i=1}^N a_i=0.
\end{equation}
Here $a_{N+1}$ is identified with $a_1$.
Using the definition \eqref{bvalue} of $b_i$ and $\alpha_i=e_i-e_{i+1}$ in orthogonal basis, one get
\begin{equation}
a_i=(-w^{-1} b-\frac{u}{h^\vee}\rho,e_i).
\end{equation}
Here $\rho$ is the Weyl vector, and $(\rho, \alpha_i)=1$. Therefore the value of moment map at the fixed point is
\begin{align}
&\mu(\Phi_{\tilde{w}}(z))=\frac{h^\vee}{2u}|\mathbf{a}|^2=\frac{h^\vee}{2u}\sum_i (-w^{-1} b-\frac{u}{h^\vee}\rho,e_i)^2 \nonumber\\
&=\frac{h^\vee}{2u}|-w^{-1} b-\frac{u}{h^\vee}\rho|^2=\frac{h^\vee}{2u}|b+\frac{u}{h^\vee}w\rho|^2,
\end{align}
Using the formula of the admissible weight $\Lambda_{\tilde{w}}$
\begin{equation}
\Lambda_{\tilde{w}}=t_bw.\left(-h^\vee+\frac{h^\vee}{u}\right)\Lambda_0,
\end{equation}
the finite part $\lambda_{\tilde{w}}$ of $\Lambda_{\tilde{w}}$ is
\begin{equation}
\lambda_{\tilde{w}}= w\rho+\frac{h^\vee}{u}b-\rho.
\end{equation}
Clearly we have 
\begin{equation}
|b+\frac{u}{h^\vee}w\rho|^2=\frac{u^2}{(h^\vee)^2}|\lambda_{\tilde{w}}+\rho|^2,
\end{equation}
and  the moment map can then be expressed in terms of $\lambda_{\tilde{w}}$
\begin{equation}
\label{eq:momentMapVal}
\mu(\Phi_{\tilde{w}}(z))=\frac{u}{2h^\vee}|\lambda_{\tilde{w}}+\rho|^2.
\end{equation}
Comparing \eqref{eq:momentMapVal} with the formula \eqref{eq:dimWalgMod} of the conformal dimension of $H_f(L(\Lambda_{\tilde{w}}))$
\begin{equation}
h_{H_f(L(\Lambda_{\tilde{w}}) ) }=\frac{u}{2h^\vee}(|\lambda_{\tilde{w}}+\rho|^2-|\rho|^2)-\frac{h^\vee}{2u}|x|^2+(x,\rho),
\end{equation} 
we get the relation \eqref{eq:momentMaptoDim} between moment maps and conformal dimension. 

\subsection{Modular properties}

\textbf{Modular transformation and DAHA}:
One  important aspects of VOA is the modular property on the characters of the modules \cite{kac2008rationality}. It is definitely interesting to 
see whether one can find similar modular transformation on the Hitchin side, which actually indeed exists. 
The cohomology of the Hitchin moduli space (which is related to the data of fixed varieties by using Morse theory) is realized as a finite dimensional representation of double affine Hecke algebra  \cite{varagnolo2009finite, oblomkov2017cohomology}, and the $PSL^c(2,\bbZ)$ action on DAHA \cite{cherednik2005double} will induce a $PSL^c(2,\bbZ)$ action on the cohomology of fixed varieties. 
It is then natural to compare above two sets of modular transformation. This relation will be proved in \cite{Xie:2023pre}. The relation between modular matrices of minimal W-algebras of $A$ type and spherical DAHA of $A$ type was studied in \cite{Gukov:2022gei}.

\textbf{Modular property for non-admissible W-algebras}:
The cohomology group $H^*(\CM_{Hit})$ considered in this paper carries a DAHA action. In good cases there is also a natural $PSL^c(2,\bbZ)$- action on $H^*(\CM_{Hit})$.  
Given the correspondence between the fixed varieties of Hitchin system and the modules of VOA, one would find interesting implication 
for the modular property of non-admissible W-algebra. A crucial fact is that in general the fixed varieties of $\CM_{Hit}$ corresponding to non-admissible W-algebra 
has \textbf{higher} dimensional components. This is in contrast with the admissible case where the fixed varieties are all of dimension zero. 

Now in our correspondence, each irreducible component of fixed varieties gives a simple module (in the category $\CO$) of the corresponding VOA. However, in the Morse theory each  higher dimensional fixed variety would contribute 
more than one basis vector to the cohomology. So the above mismatch suggests that if one want to have the modular property for the VOA, one has to enlarge the set of VOA modules. For instance, 
one might need to add the logarithmic modules to have the modular property which is also observed in some non-admissible VOAs \cite{arakawa2018quasi, Zheng:2022zkm}. In fact, our correspondence suggests the number of added module should be the same as the dimension for the cohomology 
from the fixed varieties. 

\textbf{Modular data and Coulomb branch index}:
One can define a Coulomb branch index $\CI^m_\CT(t)$  (Hitchin character)  of the 4d theory $\CT$ on $L(m,1)\times S^1$ where $L(m,1)$ is the Lens space \cite{Fredrickson:2017yka,Kozcaz:2018usv}.
The Coulomb branch index has an expansion in terms of the fixed varieties, and the geometric data such as momentum map plays a crucial role in computing it. On the other hand, 
the Lens space Coulomb index $\CI^m(t)$ is deeply connected to the modular matrices \eqref{eq:STAKM} and \eqref{eq:STWalg} of the corresponding VOA \cite{Fredrickson:2017yka,Dedushenko:2018bpp,Kozcaz:2018usv}, namely
\begin{equation}
\label{eq:CBItoSTS}
\lim_{t\rightarrow e^{2\pi i}} \CI^m_{\CT}(t) = a (\bbS \bbT^m \bbS)_{vac, vac},
\end{equation} 
where $a$ is a constant determined by $\fkg$, $\bbS$ and $\bbT$ are the modular matrices of characters of the VOA corresponding to theory $\CT$, and $(\bbS \bbT^m \bbS)_{vac, vac}$ means the vacuum-vacuum component of the matrix $\bbS \bbT^m \bbS$. In general, $\CI^m_{\CT}$ is difficult to compute for the theories considered in this paper as most of them lack a Lagrangian description. However, when $n=1$, the Lens space is just the 3-sphere $S^3$, the Coulomb branch index on $S^3\times S^1$ is completely determined by the Coulomb branch spectrum of the 4d theory which can be obtained using the method in \cite{Xie:2012hs,Wang:2015mra,Wang:2018gvb}, allowing us to check the relation \eqref{eq:CBItoSTS} for $m=1$.

{\bf Example:} Consider 4d theory  $\CT_{A_{N-1}, \frac{u}{N},f=\mathrm{trivial}}$ with $\mathrm{gcd}(N,u)=1$ (section \ref{sec:ADandWk}), the corresponding VOA is $L_{-N+\frac{N}{u}}(A_{N-1})$. The Coulomb branch spectrum $\mathrm{CB}_{\CT}$ can be found using the method in \cite{Xie:2012hs} and is the following set of rational numbers
\begin{equation}
\mathrm{CB}_{\CT}=\{ i - j\frac{N}{u} ~|~ i,j\in\bbZ,~2\leq i\leq N, 1\leq j\leq \lfloor (i-1)\frac{u}{n} \rfloor \}.
\end{equation}
Here $\lfloor x\rfloor$ is the maximal integer less or equal to $x$. The Coulomb branch $\CI_{\CT}(t)$ on $S^3\times S^1$ is then
\begin{equation}
\CI_{\CT}(t) = \prod_{d\in\mathrm{CB}_{\CT}} \frac{1}{1-t^{d}}.
\end{equation}
Because all elements in $\mathrm{CB}_{\CT}$ are not integer, the limit $t\rightarrow e^{2\pi i}$ of $\CI_{\CT}$ is not singular, comparing with the modular matrices of $L_{-N+\frac{N}{u}}(A_{N-1})$ \eqref{eq:STAKM} we find that for the theory $\CT_{A_{N-1}, \frac{u}{N},\mathrm{trivial}}$, 
\begin{equation}
\lim_{t\rightarrow e^{2\pi i}}\CI_{\CT} = e^{2i\pi (h_{\mathrm{min}}-\frac{c}{24})} (\bbS\bbT\bbS)_{vac,vac},
\end{equation}
where $h_{\mathrm{min}}=-\frac{\dim \fkg}{24}\left(u-\frac{1}{u}\right)$ is  the smallest conformal weights of all admissible modules of $L_{-N+\frac{N}{u}}(A_{N-1})$. It would be nice to generalize this relation for lens space index $L(m,1)$ in the future.

\subsection{Zhu's $C_2$ algebra and the cohomology ring}
For each VOA $V$ there is a commutative algebra $C_2(V)$ associated to $V$ called  Zhu's $C_2$ algebra. In the following, we will present examples when $C_2(V)$ is isomorphic to the cohomology ring of the corresponding Hitchin system.

Consider $\fkg=A_{N-1}$, $\nu=\frac{u}{N}$ and $f$ principal. The VOA is then the principal W-algebra $W_{-N+N/u}(A_{N-1},\mathrm{prin})$ (i.e. $W_N(N,u)$ minimal model).  Motivated from the character of its vacuum module, $C_2(W_{-N+N/u}(A_{N-1},\mathrm{prin}))$ is conjectured to be the same as the Jacobi algebra
of an isolated hypersurface singularity \cite{Xie:2019zlb}  
\begin{equation}
\label{eq:ringW}
\bbC[T_2, T_3,\cdots, T_N] /\langle{\partial f\over \partial T_2},\cdots,{\partial f\over \partial T_N}\rangle.
\end{equation}
Here $T_2,\cdots T_N$ are generators with degrees $2,\cdots, N$, and $f[T_2,\ldots, T_n]$ is an isolated singularity with degree $u+1$. The generators $\{{\partial f\over \partial T_2},\cdots,{\partial f\over \partial T_N}\}$ of the ideal then have degree $u+1-n,\cdots, u-2$. This construction ensures that the above algebra has the dimension ${(u-1)!\over n!(u-n)!}$, which is just the dimension for the Milnor algebra.

On the other hand, the cohomology ring for the corresponding Hitchin system is given by the following ring \cite{gorsky2013arc, oblomkov2017cohomology}
\begin{equation}
\label{eq:ringHit}
\bbC[e_2, e_3,\cdots, e_N]/\langle g_{u+1-N},\cdots, g_{u-1}\rangle.
\end{equation}
Here generators $e_2,\cdots,e_N$ also have degree $2,\cdots,N$, and the generator $g_{u-n+i}$ of the ideal is the coefficient of $w^{u-n+i}$ in the Taylor expansion of 
\begin{equation}
(1+e_2 w^2+\ldots +e_n w^n)^{u/n}
\end{equation}
at $w=0$. From the descriptions above, one can deduce that the ring \eqref{eq:ringW} and the ring \eqref{eq:ringHit} are isormorphic. This relations has a similar flavor to the Hikita conjecture which relates the \textbf{coordinate ring} of  some scheme coming from a conical symplectic singularity to the \textbf{cohomology ring} of a symplectic resolution of the dual conical symplectic singularity \cite{hikita2017algebro}. In our context, the coordinate 
ring is coming from Zhu's $C_2$ algebra, which would indeed give the coordinate ring of the Higgs branch \cite{arakawa2017representation}. It would be interesting to further study this correspondence in more general setups.


\subsection{Generalization to arbitrary $f$}

So far we assume the nilpootent element $f$ which labels the regular singularity to be a regular (principal) nilpotent element in a L\'{e}vi subalgebra of $\fkg$, however, there are many nilpotent element which is distinguished but not regular in any minimal L\'{e}vi subalgebra containing it (distinguished but not regular for short). For example, when $\fkg$ is of type $CDEFG$, any element $f$ in the subregular nilpotent orbit is distinguished but not regular as the minimal L\'{e}vi subalgebra containing $f$ is $\fkg$ itself.

Given a 4d theory $\CT_{\fkj,b,k,f}$ or $\CT_{\fkj,o,b_t,k_t,f}$ with $f$ distinguished but not regular, we should modify the definition of its corresponding $\CM_{Hit}$ in the following way. Adopting the same notation as in section \ref{sec:Higgsbundle} with the modification that $\fkl$ is the minimal standard L\'{e}vi subalgebra containing $f$, we still consider the Higgs bundle $(E,\Phi)$ with a $P^\vee$-level structure at the regular singularity. However, $\Phi$ should have a new boundary condition around the regular singularity (recall $\Phi'=z \Phi$)
\begin{equation}
\label{eq:regbcgeneralf}
\lim_{z\rightarrow0}\Phi' \in\overline{d(\CO^{\fkl}_f)}\oplus \fkn^\vee,
\end{equation}
which is equivalent to $\lim_{z\rightarrow0}\Phi' \in\overline{\CO}_{f^\vee}$ because $\overline{\CO}_{f^\vee} = G^\vee\cdot(\overline{d(\CO^{\fkl}_f)}\oplus \fkn^\vee)$. 
When $f$ in regular in $\fkl$, $d(\CO^{\fkl}_f)$ is the trivial nilpotent orbit in $\fkl$, therefore \eqref{eq:regbcgeneralf} reduces to \eqref{eq:boundreg}. 

We propose that the corresponding affine Spaltenstein variety should be replaced by the following space
\begin{equation}
\label{eq:fibreMod}
 Sp_{\gamma,\bfP^\vee,f}= \{ g\in \bfP^\vee \backslash \mathbf{G}^\vee  ~|~ g \gamma g^{-1}\in \overline{d(\CO^{\fkl}_f)}\oplus\tilde{\fkn}^\vee\}.
 \end{equation} 
The space $Sp_{\gamma,\bfP^\vee,f}$ is well-defined because $\overline{d(\CO^{\fkl}_f)}\oplus\tilde{\fkn}^\vee$ is stable under the action of $P^\vee$ \cite{losev2021unipotent}.
The fixed varieties of $ Sp_{\gamma,\bfP^\vee,f}$ are
\begin{equation}
Sp_{\gamma,\bfP^\vee,f}^T = \sqcup H_{\tilde{w}},\quad \{\tilde{w}\in W_{\bfP^\vee}\backslash W_{aff} / W_\nu~|~ \mathrm{Ad}(\tilde{w})\gamma \in \overline{d(\CO^{\fkl}_f)}\oplus\tilde{\fkn}^\vee\}.
\end{equation}
We will provide examples to illustrate the matching between fixed varieties of $\CM_{Hit}$ with know results in W-algebras.

\begin{example} $W_{-(2n-2)+\frac{2n-2}{u}}(D_n,[2n-3,3])\leftrightarrow \CM_{Hit}(D_n, \frac{u}{2n-2}, [2^2,2n-4])$. Here $\gcd(u,2n-2)=1$ and the partition $[2n-3,3]$ corresponds to the subregular nilpotent orbit of $D_n$ which is distinguished in $D_n$ itself. The boundary condition \eqref{eq:regbcgeneralf} in this case is
\begin{equation}
\lim_{z\rightarrow0}\Phi'\in\overline{d(\CO_{subreg})}=\overline{\CO_{min}},
\end{equation}
and $P^\vee=G^\vee=SO(8)$.
There are only fixed points in $\CM_{Hit}(D_n, \frac{u}{2n-2}, [2^2,2n-4])$. Extra fixed points comparing to the principal case are labelled by the following elements of the affine Weyl group of $\hat{D}_n$
\begin{equation}
\{\tilde{w}\in W_{G^\vee}\backslash W_{aff}~|~ |\tilde{w}(S_\nu)\cap \Delta^\vee|=1\},
\end{equation}
where $\Delta^\vee$ is the set of all roots of $D_n$. The  number of these extra fixed points is
\begin{equation}
\label{eq:fpDsubreg}
\frac{(u-h_1)(u-h_2)\cdots(u-h_n)}{2^{n-2}(n-2)!},
\end{equation}
where $\{h_1,h_2,\cdots, h_{n}\}$ is the following set of integers
\begin{equation}
\{1,3,\cdots, 2n-5,2n-4,n-1\}.
\end{equation}
This is just the set of exponents of $D_n$ with the maximal exponent $2n-3$ subtracted by $1$. The denominator $2^{n-2}(n-2)!$ is also the order of the Weyl group of the centralizer of an element in $\CO_{min}$. Formula \eqref{eq:fpDsubreg} and \eqref{eq:fpNumReg} together predict the number of  simple modules of subregular W-algebra $W_{-(2n-2)+\frac{2n-2}{u}}(D_n,[2n-3,3])$. When $u=2n-3$, the number of fixed points is $n-2$ which is the same as the number of simple modules of $W_{-(2n-2)+\frac{2n-2}{2n-3}}(D_n,[2n-3,3])$ given in \cite{arakawa2021rationality}.

\end{example}

\begin{example}$W_{-h^\vee+\frac{h^\vee}{u}}(E_n,f_{subreg})\leftrightarrow \CM_{Hit}(E_n, \frac{u}{h^\vee}, f_{min})$. Here $n=6,~7,~8$ and $\gcd(u,h^\vee)=1$. The minimal L\'{e}vi subalgebra containing $f_{subreg}$ is again $E_n$ itself.  Numbers of extra fixed points comparing to the principal case are 
\begin{equation}
\label{eq:fpEsubreg}
\begin{split}
&E_6:~\frac{1}{6!}(u-1)(u-4)(u-5)(u-7)(u-8)(u-10),\\
&E_7:~\frac{1}{2^5 6!}(u-1)(u-5)(u-7)(u-9)(u-11)(u-13)(u-16),\\
&E_8:~\frac{1}{2903040}(u-1)(u-7)(u-11)(u-13)(u-17)(u-19)(u-23)(u-28).\\
\end{split}
\end{equation}
Again $h_i$'s appear in the formula are exponents of $E_n$ with the maximal one subtracted by $1$, and the denominator is the order of the Weyl group of the centralizer of an element in $\CO_{min}$. For example, $2903040$ is  the order of Weyl group of $E_7$ which is the centralizer of an element of the $A_1$ orbit of $E_8$. Formulae \eqref{eq:fpEsubreg} and \eqref{eq:fpNumReg} together predict the number of simple modules of the subregular W-algebra $W_{-h^\vee+\frac{h^\vee}{u}}(E_n,f_{subreg})$. When $u=h^\vee-1$, the number of fixed points matches the number of simple modules of $W_{-h^\vee+\frac{h^\vee}{h^\vee-1}}(E_n,f_{subreg})$ computed in \cite{arakawa2021rationality}.

\end{example}

It was also proved in \cite{arakawa2021rationality} that W-algebras $W_{-h^\vee+\frac{h^\vee}{h^\vee-1}}(\fkg,f_{subreg})$ with $\fkg$ being type $D$ or $E$ are  rational with modular matrices of simple modules worked out explicitly. It would also be nice to match these data from the VOA side with geometric data from the Hitchin side. In general, W-algebras with distinguished $f$ (distinguished W-algebras) play fundamental role among W-algebras. However, the representation theory of distinguished W-algebras that are not of regular type are largely unexplored. Our correspondence provides motivations to study the space $Sp_{\gamma,\bfP^\vee,f}$ and use the geometry to predict representation theories of distinguished W-algebras.

\subsection{Relation with 3d symplectic duality }

When taking the limit that the radius of the circle to be $0$, one can get a 3d $\CN=4$ SCFT $\CT^{3d}$. As mentioned in the introduction, the Higgs branch $X$ of $\CT^{3d}$ is  the same as the Higgs branch of 4d theory, which is identified as the associated variety of the corresponding VOA $V(\CT)$. 
The Coulomb branch $Y$ of $\CT^{3d}$ is also related to the Coulomb branch of the 4d on $S^1$. In the massless limit both $X$ and $Y$ are hyper-K\"{a}hler cones.

In many cases, $Y$ (resp. $X$) can also be realized as the Higgs (resp. Coulomb) branch of another 3d $\CN=4$ quiver gauge  theory $\CT^{3d,mirror}$ which is called the mirror of $\CT^{3d}$ in physics literature \cite{Xie:2021ewm}. Properties of $Y$ can be quite different from its 4d counter part:
\begin{enumerate}
\item Usually there is no flavor symmetries acting on the 4d Coulomb branch. However, there are sometimes emergent global symmetries on $Y$. 
\item $Y$ is not irreducible, i.e. it typically has a component described by free hypermultiplets in the mirror theory.
\end{enumerate}
%

Since $X$ and $Y$ are Higgs or Coulomb branches of the same 3d theory, they form a symplectic pair. Actually, many familiar symplectic pairs arises this way:

\begin{example} Consider the 4d theory $\CT_{A_{N-1},\frac{u}{N}, f=[1^N]}$ with $\gcd(u,N)=1$ and $u>N$. After reducing to 3d, the Higgs branch $X$ is the associated variety of $L_{-N+N/u}(A_{N-1})$ which is the  nilpotent cone ${\CN}$ of $A_{N-1}$ \cite{MR3456698} \footnote{Associated varieties when $u>N$ remain to be the same.}, while the Coulomb brach $Y$ is given by the Higgs branch of  the so-called $T[SU(n)]$ theory \cite{Gaiotto:2008ak} plus $h_{u,N}=\frac{(N-1)(u-N-1)}{2}$ free hypermultiplets, which is ${\CN}$ plus the flat space $\bbC^{h_{u,n}}$. The interacting part 3d theory is self-mirror meaning both its Higgs and Coulomb branch are the same ($\CN$). Notice that when $u>N$, 4d theories $\CT_{A_{N-1},\frac{u}{N}, f=[1^N]}$ with different $u$ give the same symplectic pairs.
\end{example}

\begin{example} Next change $f$ in the above example to be an element in arbitrary nilpotent orbit. Then $X$  becomes $S_f\cap{\CN}$, and $Y$ is the Higgs branch of $T_f[SU(N)]$ theory plus $h_{u,N}$ free hypermultiplets. So $Y$ is $\overline{\CO}_{f^\vee}$ plus $\bbC^{h_{u,n}}$. It is known that $S_f\cap{\CN}$ and $\overline{\CO}_{f^\vee}$ form a symplectic pair.
\end{example}

\begin{example} Now take $N=2l+1$ to be an odd integer, $u=2$ and $f=[1^{2l+1}]$. The 3d mirror for this theory is given in \cite{Xie:2021ewm}, and $X$ is now $\overline{\CO}_{[2^{l},1]}$ and $Y$ is $S_{[l+1,l]}\cap {\CN}$.
\end{example}

In above examples, we see that different 4d theories (VOAs) can have the same Higgs branch (associated variety). Their 4d Coulomb branches are different, however, after reducing to 3d, their 3d Coulomb branch differ only by a $\bbC^h$ factor. It seems that the 4d perspective is a more ``refined" version of 3d symplectic pair. It would also be interesting to see if it can provide new in-sight on symplectic dualities.

Moreoever, one can get a finite W-algebra from the twisted Zhu's algebra of the associated VOA \cite{de2006finite}. The finite W-algebra is precisely those found 
by doing quantization on the Higgs branch of 3d theory,  so from the reduction of 4d theory, one not only get a pair of symplectic singularities, but also an algebra/geometry pair. 

\section{Conclusion and outlook}
\label{sec:concl}

In this paper, we study the mirror symmetry for circle compactified 4d $\mathcal{N}=2$ SCFTs.  This symmetry involves an algebra object which is a VOA capturing the data on the Schur sector, and a geometric object which is the Coulomb branch of the effective 3d theory. We show that the representation theory of the VOA such 
as simple modules, modular transformation, Zhu's algebra can be translated into  geometric properties of the Coulomb branch. Various checks have been 
made in this paper when one can compute things on both sides, and one would get many interesting predictions on each side by using the mirror symmetry map.

Our mirror pair involves W-algebra and the Hitchin's moduli space, which all play important roles in various branches of physics and mathematics, 
and we hope that the mirror proposal in this paper would help understand them further.
While there are many interesting matches in our mirror proposal, a further physical understanding of this mirror symmetry is definitely 
desirable. Hopefully the physical understanding would help us to construct VOA modules and their character.
We mainly focus on regular elliptic slope and special nilpotent orbit in this paper (with a few studies on sub-regular elliptic slope), and  detailed studies of other 
cases will be presented elsewhere.

The mirror symmetry involves an algebra defined using Higgs branch or its generalization and the effective Coulomb branch. Physically, it suggests that
one can study following generalizations:
\begin{enumerate}

\item \textbf{Twisted W-algebras}:  In this paper we encounter non-twisted affine Lie algebra on the VOA side. It is actually possible to find the mirror pair involving twisted affine Lie algebras and non-twisted Hitchin systems. 
Recall that one find the Coulomb branch as Hitchin's moduli space as follows: one get 3d theory by compactify 6d theory on $\Sigma\times S^1$; If we first compactify 6d theory on $\Sigma$, one get a 4d theory on $S^1$, and 
if we first compactify it on $S^1$, one get a 5d theory on $\Sigma$ whose Higgs branch is just the Hitchin's moduli space defined on $\Sigma$. The twisted theory is defined by turning on outer automorphism twist on $\Sigma$.
Now to get 5d theory with non-simply laced gauge group, one need to turn on outer automorphsim twist
around the circle $S^1$, and then one get a non-twisted Hitchin system on $\Sigma$. On the left hand side of figure \ref{intro1}, one first compactify the theory on $\Sigma$ and then on $S^1$ with outer-automorphism twist, and it is naturally to expect that one should get a twisted W-algebra by doing outer automorphism twist. 
\cite{Xie:2023pre} establishes the correspondence for twsited AKM at boundary admissible level.

\item \textbf{Non-elliptic affine Springer fiber}: We mainly focus on the so-called elliptic affine Springer fiber in this paper. The correspondence can 
certainly be generalized to the non-elliptic case. The Hitchin system is well define and  the corresponding VOA can be found using coset construction \cite{Xie:2019yds}. 
On the other, isomorphisms between W-algebras may predict isomorphisms between Hitchin systems. In certain cases,  the non-elliptic Hitchin system is predicted to be isomorphic to elliptic Hitchin system using the isomorphism between W-algebras. 

\begin{example} 

Consider a 4d theory $\CT_{A_1,1, u-1 ,f=[1^2]}$ which is also called the $(A_1, D_{2u})$ AD theory. Hitchin system describing its Coulomb branch is given by the data $(A_1, \nu=u, f^\vee=[2])$, and the corresponding affine Springer
fibre is not elliptic because the denominator of $\nu$ is $1$. However, the same 4d theory can also be realized as $\CT_{A_{u},u+1,-1,[u-1,1^2]}$ by using an irregular singularity which is indeed elliptic and  a special nilpotent orbit \cite{Song:2017oew}. The Hitchin data corresponding to $\CT_{A_{u},u+1,-1,[u-1,1^2]}$ is $(A_{u}, \nu=\frac{u}{u+1}, f^\vee=[3,1^{u-2}])$. The duality of the 4d theory implies the isomorphism between Hitchin moduli space
\begin{equation}
\CM_{Hit}(A_1, u, [2])\simeq \CM_{Hit}(A_{u}, \frac{u}{u+1}, [3,1^{u-2}]).
\end{equation}

\end{example}


\begin{example}
Consider a 4d theory whose spectral curve at SCFT point is \footnote{One should be careful that if the spectral curve at SCFT point gives a non-isolated singularity, one need to specify the mass and relevant deformation for the theory, see \cite{Xie:2019yds} for the discussion on this point.}
\begin{equation}
x^{n+n_1}+x^{n_1}y^k=0,
\end{equation}
where $n_1$, $n$ and $k$ are positive integers such that $\gcd(n,k)=1$. It was proposed that dual descriptions of this theory lead to the following isomorphisms of W-algebras \cite{Xie:2019yds}
\begin{equation}
\begin{split}
&W_{-(n_1(n+k)+n)+\frac{n_1(n+k)+n}{n+k}}(\fsl_{n_1(n+k)+n},[(n+k-1)^{n_1},n+n_1])\\
&\simeq W_{-k+\frac{k}{n+k}}(\fsl_k,[k-n_1,1^{n_1}])
\end{split}
\end{equation}
 and isomorphisms of Hitchin moduli spaces
\begin{equation}
\CM_{Hit}(\fsl_{n_1(n+k)+n},\frac{n+k}{n_1(n+k)+n},[(n+k-1)^{n_1},n+n_1])\simeq \CM_{Hit}(\fsl_k,\frac{n+k}{k},[k-n_1,1^{n_1}]).
\end{equation}
\end{example}

More isomorphisms involving other Lie algebras are also proposed in \cite{Xie:2019yds,Xie:2019vzr}, it would be nice if one can prove these isomorphisms rigorously.

\item \textbf{Class S theory}: VOAs for general class ${\cal S}$ theory has been found in \cite{Lemos:2014lua, Beem:2014rza, arakawa2018chiral}, but little is known about their representation theory. 
On the other hand, the Coulomb branch for circle compactified theory is given by Hitchin system with regular singularities only. There are a lot of studies 
on the cohomology of the moduli space \cite{hausel2008mixed}, and we might learn the representation theory of VOA by using results on Hitchin side.

\item  \textbf{General $\mathcal{N}=2$ theories}:  We hope to push our mirror symmetry to more general $\mathcal{N}=2$ SCFTs and use it to study physical properties of those theories.  It was found recently that one can attach interesting configuration of curves on the central fiber \cite{Xie:2023lko}, and their fixed points and cohomology could be interesting objects to study. One can also consider more general $\mathcal{N}=2$ theories such as pure $SU(N)$ gauge theory compactified on the circle, and study their related mirror symmetry.

\item \textbf{3d $\mathcal {N}=4$ gauge theory  with finite gauge coupling}: The typical feature of 3d $\mathcal{N}=4$ SCFT is that its Coulomb branch can be given by the Higgs branch of 
the mirror quiver gauge theory. However, if one studies the gauge theory with finite gauge coupling,  locally its Coulomb branch has the structure of $\bbR^3\times T$ (ALF space) \cite{Seiberg:1996nz} which can no longer be given 
by the Higgs branch of a quiver gauge theory. Instead, it can be given by the so-called bow diagram \cite{cherkis2011instantons}, which can be viewed as a four dimensional theory on a one dimensional space. 
 To retain the mirror symmetry, the algebra side might also be changed.

\item \textbf{Generalization to five and six dimensional SCFTs}:  One might also consider the $T^2$ compactification of 5d $\mathcal{N}=1$ SCFT or $T^3$ compactification of 6d $(1,0)$ little string theory,
and one would expect to have the similar mirror symmetric between an algebra and  the Coulomb branch $\CM_{C}$ of the effective 3d theory.
For the $T^2$ compactification of rank one 5d theory, locally the effective 3d Coulomb branch would take the form of $T^3 \times \bbR$ (ALH space) \cite{cherkis2012moduli}. 
For the $T^3$ compactifiation of 6d little string theory, the total space of the effective 3d Coulomb branch 
would be compact \cite{intriligator2000compactified}. See table \ref{con} for a summary. We do not know what kind of algebra would be involved for the Higgs branch side yet, and we believe that there are lots of interesting mathematics and physics involved in these mirror symmetries. 

\begin{table}
\begin{center}
		\begin{tabular}{|c|c|c|c|}
			\hline
			Physical theory  & Coulomb branch & Mirror & Algebra  \\ \hline
			3d $\mathcal{N}=4$ SCFT & $ALE$ & Certain quiver variety & Finite W-algebra\\ \hline
			3d gauge theory & $ALF$ & Bow diagram & $?$ \\ \hline
                        4d $\mathcal{N}=2$ on $S^1$ & $ALG$ & Hitchin system & $VOA$\\ \hline
                       5d $\mathcal{N}=2$ on $T^2$     & $ALH$ & Periodic monopole on $T^3$ &$?$ \\ \hline
			 6d $(1,0)$ on $T^3$  & compact & $K_3$ & $?$ \\ \hline

		\end{tabular}
	
		\caption{ Mirror symmetry for theories with eight supercharges.}
		\label{con}
		
	\end{center}
\end{table}

\end{enumerate}

\acknowledgments
PS is supported by NSCF Grant No.~12225108.
WY is supported by Yau Mathematical Science Center at Tsinghua
University. DX and WY are supported by  national key research
and development program of China (NO. 2020YFA0713000), and NNSF of China with
Grant NO: 11847301 and 12047502.

\appendix

\section{Rank one SCFT}
\label{app:sec:nonAdm}

In this section we discuss fixed varieties of the Coulomb branch of rank one SCFTs. The corresponding VOAs are all non-admissible.

{$E_6$ \textbf{theory}}: {$L_{-3}(E_6)\leftrightarrow\CM_{Hit}(E_6,\frac{1}{9},  f^\vee=principal)$}. The two sets are
\begin{equation}
L_\nu=\{\pm(\alpha_1+\alpha_2+2\alpha_3+2\alpha_4+2\alpha_5+\alpha_6-\delta)\},
\end{equation}
and
\begin{equation}
S_\nu=\{\alpha_1,\alpha_2,\alpha_3,\alpha_4,\alpha_5,\alpha_6\}\cup\{\beta_1,\beta_2,\gamma\},
\end{equation}
where $\beta_1=-\alpha_1-\alpha_2-2\alpha_3-2\alpha_4-\alpha_5-\alpha_6+\delta$, $\beta_2=-\alpha_1-\alpha_2-\alpha_3-2\alpha_4-2\alpha_5-\alpha_6+\delta$ and $\gamma=\alpha_1+\alpha_2+2\alpha_3+3\alpha_4+2\alpha_5+\alpha_6-\delta$.
The set of fixed varieties are listed in table \ref{table:fpsE6nonadm} which matches simple modules of $L_{-3}(E_6)$ classified in \cite{Arakawa_2016}.  The fixed variety with dimension $1$ corresponds to the only module with dominant weight $-\Lambda_4$.
\begin{table}[htbp]
\centering
\begin{tabular}{|c|c|c|} \hline
Dim & $\tilde{w}\simeq(w,\beta)$ & $t_\beta w.(k\Lambda_0)$  \\
\hline
0 & $1\simeq(1,0)$ & $-3\Lambda_0$ \\
0 & $s_1s_3s_0\simeq(r_1r_2s_1s_3s_2, \alpha_1+\alpha_2+2\alpha_3+3\alpha_4+2\alpha_5+\alpha_6)$ & $-3\Lambda_1$  \\
0 & $s_6s_5s_0\simeq(s_6r_1s_1r_2s_2, \alpha_1+\alpha_2+2\alpha_3+3\alpha_4+2\alpha_5+\alpha_6)$ & $-3\Lambda_6$  \\
0 & $s_5s_0\simeq(r_1s_1r_2s_2, \alpha_1+\alpha_2+2\alpha_3+3\alpha_4+2\alpha_5+\alpha_6)$ & $-2\Lambda_5+\Lambda_6$  \\
0 & $s_0\simeq(s_2r_1r_2s_2,\alpha_1+2\alpha_2+2\alpha_3+3\alpha_4+2\alpha_5+\alpha_6)$ & $-2\Lambda_2+\Lambda_0$ \\
0 & $s_3s_0\simeq(r_1r_2s_3s_2,\alpha_1+\alpha_2+2\alpha_3+3\alpha_4+2\alpha_5+\alpha_6)$ & $\Lambda_1-2\Lambda_3$ \\
\hline
1 & $s_2s_0\simeq(r_1r_2s_2,\alpha_1+\alpha_2+2\alpha_3+3\alpha_4+2\alpha_5+\alpha_6)$ & $-\Lambda_4$ \\ \hline
\end{tabular}
\caption{Fixed varieties of $\CM_{Hit}(E_6,\frac{1}{9}, principal)$. Here $r_1=s_4s_5s_3s_4s_1s_3s_2s_4s_5s_6s_5s_4$, $r_2=s_3s_2s_4s_5s_1s_3s_4$ and $\lambda_0=\bar{w}_1+\bar{w}_2+\bar{w}_3+\bar{w}_5+\bar{w}_6$.}
\label{table:fpsE6nonadm}
\end{table}

$E_7$ \textbf{theory}: $L_{-4}(E_7)\leftrightarrow\CM_{Hit}(E_7, \frac{1}{14},  f^\vee=principal)$.
Two sets  are: 
\begin{equation}
L_\nu=\{\pm(\alpha_1+2\alpha_2+2\alpha_3+3\alpha_4+3\alpha_5+2\alpha_6+\alpha_7-\delta)\},
\end{equation}
and
\begin{equation}
S_\nu=\{\alpha_i,\ i=1,2,\cdots,7\}\cup\{\beta_1,\beta_2,\gamma\},
\end{equation}
where $\beta_1=-\alpha_1-\alpha_2-2\alpha_3-3\alpha_4-3\alpha_5-2\alpha_6-\alpha_7+\delta$, $\beta_2=-\alpha_1-2\alpha_2-2\alpha_3-3\alpha_4-2\alpha_5-2\alpha_6-\alpha_7+\delta$ and $\gamma=\alpha_1+2\alpha_2+2\alpha_3+4\alpha_4+3\alpha_5+2\alpha_6+\alpha_7-\delta$.

There are one fixed variety of dimension $1$ and seven fixed points of dimension 0 as summarized in \ref{table:fpsE7nonadm}. Again the fixed variety with dimension $1$ corresponds to the simple module with dominant weight of $V_{-4}(E_7)$, while fixed points correspond to other simple modules \cite{Arakawa_2016}.

\begin{table}
\centering
\begin{tabular}{|c|c|c|} \hline
Dim & $\tilde{w}$ & $t_\beta w.(k\Lambda_0)$  \\
\hline
0 & $1$ & $-4\Lambda_0$ \\
0 & $s_0$ & $-3\Lambda_1+2\Lambda_0$ \\
0 & $s_1s_0$ & $\Lambda_1-2\Lambda_3$ \\
0 & $s_2s_3s_1s_0$ & $-2\Lambda_2$  \\
0 & $s_5s_3s_1s_0$ & $-2\Lambda_5+\Lambda_6$  \\
0 & $s_6s_5s_3s_1s_0$ & $-3\Lambda_6+2\Lambda_7$  \\
0 & $s_7s_6s_5s_3s_1s_0$ & $-4\Lambda_7$  \\
\hline
1 & $s_3s_1s_0$ & $-\Lambda_4$ \\\hline
\end{tabular}
\quad
\begin{tabular}{|c|c|c||c} \hline
Dim & $\tilde{w}$ & $t_\beta w.(k\Lambda_0)$  \\
\hline
0 & $1$ & $-6\Lambda_0$ \\
0 & $s_0$ & $-5\Lambda_8+4\Lambda_0$ \\
0 & $s_8s_0$ & $-4\Lambda_7+3\Lambda_8$ \\
0 & $s_7s_8s_0$ & $-3\Lambda_6+2\Lambda_7$  \\
0 & $s_6s_7s_8s_0$ & $-2\Lambda_5+\Lambda_6$  \\
0 & $s_2s_5s_6s_7s_8s_0$ & $-2\Lambda_2$  \\
0 & $s_3s_5s_6s_7s_8s_0$ & $\Lambda_1-2\Lambda_3$ \\
0 & $s_1s_3s_5s_6s_7s_8s_0$ & $-3\Lambda_1$  \\
\hline
1 & $s_5s_6s_7s_8s_0$ & $-\Lambda_4$  \\ \hline
\end{tabular}

\caption{Left: Fixed varieities of $E_7,\frac{1}{14}$. Right: Fixed varieties of $E_8,\frac{1}{24}$.}
\label{table:fpsE7nonadm}
\end{table}

\textbf{$E_8$ theory}: $L_{-6}(E_8)\leftrightarrow \CM_{Hit}(E_8,\frac{1}{24},  f^\vee=principal)$.
Two sets of affine roots are
\begin{equation}
L_\nu=\{\pm(2\alpha_1+3\alpha_2+4\alpha_3+5\alpha_4+4\alpha_5+3\alpha_6+2\alpha_7+\alpha_8-\delta)\},
\end{equation}
and
\begin{equation}
S_\nu=\{\alpha_i,\ i=1,2,\cdots,8\}\cup\{\beta_1,\beta_2,\gamma\},
\end{equation}
where $\beta_1=-2\alpha_1-3\alpha_2-3\alpha_3-5\alpha_4-4\alpha_5-3\alpha_6-2\alpha_7-\alpha_8+\delta$, $\beta_2=-2\alpha_1-2\alpha_2-4\alpha_3-5\alpha_4-4\alpha_5-3\alpha_6-2\alpha_7-\alpha_8+\delta$ and $\gamma=2\alpha_1+3\alpha_2+4\alpha_3+6\alpha_4+4\alpha_5+3\alpha_6+2\alpha_7+\alpha_8-\delta$.

There are one fixed manifold of dimension 1 and eight fixed points of dimension 0 as show in table \ref{table:fpsE7nonadm}. Again the fixed manifold corresponds to the simple module with dominant weight of $V_{-6}(E_8)$, while fixed points correspond to other simple modules  \cite{Arakawa_2016}. 

$G_2$ \textbf{theory}:~$L_{-2}(G_2)\leftrightarrow\CM_{Hit}((\hat{D}_4,\bbZ_3),\frac{1}{6},  f^\vee=principal)$. The set of real affine roots of the twisted affine Lie algebra $^3D_4$ is $\hat{\Delta}^\vee=\Phi_{s}^{re}\cup \Phi_{l}^{re}$ with 
\begin{align}
&\Phi_{s}^{re}=\{\alpha+\frac{r}{3}\delta~|~r\in\bbZ,~\alpha\in \Phi_s^0\}, \nonumber\\
&\Phi_{l}^{re}=\{\alpha+r\delta~|~r\in\bbZ,~\alpha\in \Phi_l^0\}.
\end{align}
Here $\Phi_0$ denotes the root system of $G_2$ which is also the finite part of $^3D_4$, 
\begin{align}
\Phi_s^0&=\{\pm(\beta_1-\beta_2),~\pm(\beta_2-\beta_3),~\pm(\beta_1-\beta_3)\}=\{\pm \alpha_2,~\pm(\alpha_1+2\alpha_2),~\pm(\alpha_1+\alpha_2)\},  \nonumber\\
\Phi_l^0&=\{\pm(-2\beta_1+\beta_2+\beta_3),~\pm(\beta_1-2\beta_2+\beta_3),~\pm(\beta_1+\beta_2-2\beta_3) \}= \nonumber\\
&=\{\pm \alpha_1,~ \pm(\alpha_1+3\alpha_2),~ \pm(2\alpha_1+3\alpha_2)  \}.
\end{align}
Here $\beta_1, \beta_2, \beta_3$ are orthogonal basis, and the {simple} roots are $\alpha_1=-2\beta_1+\beta_2+\beta_3$ and $\alpha_2=\beta_1-\beta_2$.
The highest root is $\theta_l=2\alpha_1+3\alpha_2$, and the highest short root is $\theta_s=\alpha_1+2\alpha_2$. 
The set 
\begin{equation}
L_\nu=\{\alpha+l\delta \in \hat\Delta^\vee~|~ \frac{1}{6} \alpha( \rho^\vee) +l =0\} \to L_\nu=\{\pm(\alpha_1+\alpha_2-\frac{1}{3}\delta) \}
\end{equation}
and the set $S_\nu$ are
\begin{align}
&S_\nu=\{\alpha + l\delta\in \hat\Delta^\vee~|~\frac{1}{6} {\alpha}( \rho^\vee) +l =\frac{1}{6}\} \to 
S_\nu=\{\alpha_1, \alpha_2,  -\theta_s+\frac{2}{3}\delta,~ \theta_s-\frac{1}{3}\delta,-\theta_l+\delta,-\alpha_2+\frac{1}{3}\delta\}
\end{align}
One can then use the graphic method to find the affine Weyl group elements corresponding to fixed varieties.
This fibre has one fixed variety of dimension $1$ labelled by $s_0$, and two fixed varieties labelled by $1$ and $s_1s_0$ \cite{oblomkov2016geometric}, and here $s_0$ is given by the simple reflection of the affine root $-\theta_s+\frac{1}{3}\delta$. Assuming the bijection is still true in the non-admissible case, we conjecture that there are three  simple modules of $L_{-2}(G_2)$ with weights
\begin{equation}
-2\Lambda_0,\quad s_0.(-2\Lambda_0)=-\Lambda_2,\quad (s_1s_0).(-2\Lambda_0) = -2\Lambda_1.
\end{equation} 
It would be nice to check if this statement is true from VOA side.

$SO(7)$ \textbf{theory}: $L_{-2}(SO(7))\leftrightarrow \CM_{Hit}((A_5,\bbZ_2),\frac{1}{6},  f^\vee=principal)$. The set of real affine roots of the twisted Lie algebra $^2\hat{A}_5$ is $\hat{\Delta}^\vee=\Phi_{s}^{re}\cup \Phi_{l}^{re}$ with
\begin{align}
& \Phi_{s}^{re}=\{\alpha+\frac{n}{2}\delta~|~ n\in\bbZ,~\alpha \in \Phi^0_s \}, \nonumber\\
&  \Phi_{l}^{re}=\{\alpha+n\delta~|~n\in\bbZ,~ \alpha \in \Phi^0_l \}.
\end{align}
Here the set of finite roots $\Phi^0_s\cup \Phi^0_l$ is that of $C_3$  Lie algebra:
\begin{align}
\Phi^0_s&=\{\pm \beta_i\pm \beta_j,~~i,j=1,2,3~~i\neq j\}\nonumber\\
&=\{\pm \alpha^\vee_1, \pm \alpha^\vee_2, \pm(\alpha^\vee_2+\alpha^\vee_3),\pm(\alpha^\vee_1+\alpha^\vee_2+\alpha^\vee_3),\pm(\alpha^\vee_1+\alpha^\vee_2), \pm(\alpha^\vee_1+2\alpha^\vee_2+\alpha^\vee_3)\}, \nonumber\\
\Phi^0_l&=\{\pm 2\beta_i,~~i=1,2,3 \} \nonumber\\
&=\{\pm \alpha^\vee_3, \pm(2\alpha^\vee_2+\alpha^\vee_3),~\pm(2\alpha^\vee_1+2\alpha^\vee_2+\alpha^\vee_3)\}.
\end{align} 
Here $\beta_i$ are the orthogonal basis. The set of simple roots are 
\begin{equation}
\alpha^\vee_1=\beta_1-\beta_2,~~~\alpha^\vee_2=\beta_2-\beta_3,~~~~\alpha^\vee_3=2\beta_3,
\end{equation}
which are simple coroots of $B_3$.
The highest root is $\theta^\vee_l=2\alpha^\vee_1+2\alpha^\vee_2+\alpha^\vee_3$, and the highest short root is $\theta^\vee_s=\alpha^\vee_1+2\alpha^\vee_2+\alpha^\vee_3$. 

The set $L_\nu$ is
\begin{equation}
L_\nu=\{\alpha+l\delta \in \hat\Delta~|~ \frac{1}{6} \alpha( \rho) +l =0\} \to L_\nu=\{\pm(\alpha^\vee_1+\alpha^\vee_2+\alpha^\vee_3-\frac{1}{2}\delta) \}
\end{equation}
and the set $S_\nu$ is
\begin{align}
&S_\nu=\{\alpha + l\delta\in \hat\Delta~|~\frac{1}{6} {\alpha}( \rho) +l =\frac{1}{6}\} \to \nonumber\\
&S_\nu=\{\alpha^\vee_1, \alpha^\vee_2, \alpha^\vee_3, \theta^\vee_s-\frac{1}{2}\delta,-\theta^\vee_l+\delta,-(\alpha^\vee_1+\alpha^\vee_2)+\frac{1}{2}\delta,-(\alpha^\vee_2+\alpha^\vee_3)+\frac{1}{2}\delta\}
\end{align}
There is one dimension $1$ fixed variety labelled by $s_0$, and three dimensional $0$ fixed points labelled by $1$, $s_1s_0$ and $s_3s_0$. We predict that the corresponding simple modules have weights
\begin{equation}
-2\Lambda_0, ~s_0.(-2\Lambda_0)=-\Lambda_2, ~(s_1s_0).(-2\Lambda_0)=-2\Lambda_1,~(s_3s_0).(-2\Lambda_0)=-2\Lambda_3.
\end{equation}

$F_4$ \textbf{theory}: $L_{-3}(F_4)\leftrightarrow \CM_{Hit}((E_6,\bbZ_2),\frac{1}{12},  f^\vee=principal)$. The set of real affine roots of the twisted Lie algebra $^2\hat{E}_6$ is $\hat{\Delta}^\vee=\Phi_{s}^{re}\cup \Phi_{l}^{re}$ with
\begin{align}
& \Phi_{s}^{re}=\{\alpha+\frac{n}{2}\delta~|~n\in\bbZ,~ \alpha \in \Phi^0_s \}, \nonumber\\
&  \Phi_{l}^{re}=\{\alpha+n\delta~|~n\in\bbZ,~ \alpha \in \Phi^0_l \}.
\end{align}
Here the set of finite roots $\Phi^0_s\cup \Phi^0_l$ is that of $F_4$  Lie algebra. The simple roots are:
\begin{equation}
\alpha_1=\beta_1-\beta_2,~\alpha_2=\beta_2-\beta_3,~\alpha_3=\beta_3,~\alpha_4=\frac{1}{2}(-\beta_1-\beta_2-\beta_3+\beta_4).
\end{equation}
Here $\beta_i$ are orthogonal basis. The set of roots are
\begin{align}
\Phi_s=&\{ \pm \alpha_3, \pm(\alpha_2+\alpha_3),~\pm(\alpha_1+\alpha_2+\alpha_3),~~\pm(\alpha_1+2\alpha_2+3\alpha_3+2\alpha_4)\} \nonumber\\
&\cup\{\pm(\alpha_2+2\alpha_3+\alpha_4),\pm(\alpha_1+\alpha_2+\alpha_3+\alpha_4),\pm(\alpha_1+\alpha_2+2\alpha_3+\alpha_4),\pm(\alpha_2+\alpha_3+\alpha_4) \} \nonumber\\
&\cup\{\pm(\alpha_3+\alpha_4),\pm(\alpha_1+2\alpha_2+2\alpha_3+\alpha_4),\pm(\alpha_4),\pm(\alpha_1+2\alpha_2+3\alpha_3+\alpha_4) \} ,\nonumber\\
\Phi_l=&\{\pm \alpha_1, \pm(\alpha_1+2\alpha_2+2\alpha_3),\pm(\alpha_1+\alpha_2),\pm(\alpha_1+\alpha_2+2\alpha_3),~\pm(\alpha_1+2\alpha_2+2\alpha_3) \nonumber\\
&\cup \{\pm(2\alpha_1+3\alpha_2+4\alpha_3+2\alpha_4), \pm \alpha_2, \pm(\alpha_2+2\alpha_3),\pm(\alpha_1+\alpha_2+2\alpha_3+2\alpha_4) \} \nonumber\\
&\cup\{\pm(\alpha_1+3\alpha_2+4\alpha_3+2\alpha_4),\pm(\alpha_1+2\alpha_2+2\alpha_3+2\alpha_4),\pm(\alpha_1+2\alpha_2+4\alpha_3+2\alpha_4)\}.
\end{align}
The highest root root is $\theta_l=2\alpha_1+3\alpha_2+4\alpha_3+2\alpha_4$, and the highest short root is $\theta_s=\alpha_1+2\alpha_2+3\alpha_3+2\alpha_4$. The set $L_\nu$ is
\begin{equation}
L_\nu=\{\alpha+l\delta \in \hat\Delta^\vee~|~ \frac{1}{12} \alpha( \rho^\vee) +l =0\} \to L_\nu=\{\pm(\alpha_1+2\alpha_2+2\alpha_3+\alpha_4-\frac{1}{2}\delta) \},
\end{equation}
and the set $S_\nu$ is
\begin{align}
&S_\nu=\{\alpha + l\delta\in \hat\Delta^\vee~|~\frac{1}{12} {\alpha}( \rho^\vee ) +l =\frac{1}{12}\} \to \nonumber\\
&S_\nu=\{\alpha_1, \alpha_2, \alpha_3, \alpha_4, (\alpha_1+2\alpha_2+3\alpha_3+\alpha_4)-\frac{1}{2}\delta,-\theta_l+\delta, \nonumber\\
&-(\alpha_1+\alpha_2+2\alpha_3+\alpha_4)+\frac{1}{2}\delta\}.
\end{align}

The fixed variety of dimension $1$ is labelled by $s_{4^{\vee}}s_{0^\vee}$, while fixed points are labelled by $1$, $s_{0^\vee}$, $s_{2^\vee}s_{4^\vee}s_{0^\vee}$, $s_{1^\vee}s_{2^\vee}s_{4^\vee}s_{0^\vee}$ with $s_{i^\vee}$ being the reflection of the simple root $\alpha^\vee_i$ of  $^2E_6$. Assuming the correspondence between fixed varieties and simple modules still holds,  we predict that the simple modules are
\begin{equation}
\begin{split}
&1.(-3\Lambda_0)=-3\Lambda_0,~s_0.(-3\Lambda_0)=-2\Lambda_1-\Lambda_0,~(s_1s_0).(-3\Lambda_0)=-3\Lambda_2,~\\
&(s_3s_1s_0).(-3\Lambda_0)=-2\Lambda_3+\Lambda_4,~(s_4s_3s_1s_0).(-3\Lambda_0)=-3\Lambda_4.
\end{split}
\end{equation}
There is a flip from $1,2,3,4$ to $4,3,2,1$ because our labelling of roots in $^2\hat{E}_6$ is such that short roots are still $\alpha_3$ and $\alpha_4$ so that the node of $\alpha_0$ is connected to the node of $\alpha_4$ in the Dynkin diagram of $^2\hat{E}_6$.

\section{Twisted theory}
%
%
%
In this section we give an example when the VOA side is the affine vertex algebra of a non-simply laced AKM. On the fibre side we need to consider the twisted affine Lie algebra.

\begin{example} $L_{-(2l-1)+\frac{2l-1}{u}}(B_l)\leftrightarrow \CM_{Hit}((A_{2l-1}, \bbZ_2),\frac{u}{2(2l-1)}, f^\vee=principal)$. On the VOA side, the simple roots of $B_l$ in orthogonal basis are
\begin{equation}
\Delta_+=\{\alpha_1=\beta_1-\beta_2,\cdots,\alpha_{l-1}=\beta_{l-1}-\beta_l,\alpha_l=\beta_l\},
\end{equation}
and the highest long root $\theta=\beta_1+\beta_2$. Therefore, following the definition of $S_u$, we have
\begin{equation}
\begin{split}
S_u=&\{\alpha^\vee_1,\cdots,\alpha_l^\vee,-\theta_l^\vee+u\delta\}\\
=&\{\beta_1-\beta_2,\cdots,\beta_{l-1}-\beta_{l},2\beta_l,-\beta_1-\beta_2+u\delta\}.
\end{split}
\end{equation}
The set of real roots of $\hat{B}_l$ is
\begin{equation}
\hat{\Delta}=\{\alpha+n\delta ~|~ \alpha\in \Delta,~n\in\bbZ\}.
\end{equation}
Here $\Delta$ is the set of roots of $B_l$.

One the fibre side, we need to consider the twisted affine Lie algebra $^2\hat{A}_{2l-1}$ which is the Langlands dual of $\hat{B}_l$. The set of real roots of $^2\hat{A}_{2l-1}$ is $\hat{\Delta}^C=\Phi^{re}_s\cup\Phi^{re}_l$ with
\begin{align}
& \Phi^{re}_{s}=\{\alpha+\frac{n}{2}\delta ~|~ \alpha \in \Phi^0_s ,~n\in\bbZ\}, \nonumber\\
&  \Phi^{re}_{l}=\{\alpha+n\delta~|~ \alpha \in \Phi^0_l,~n\in\bbZ \}.
\end{align}
Here $\Phi^0$ is the set of roots of $C_l$  which is the finite part of $^2\hat{A}_{2l-1}$. In orthogonal basis
\begin{equation}
\Phi^0_l=\{\pm 2\beta_i \},~~~\Phi^0_s=\{\pm \beta_i\pm \beta_j,~~i,j=1,\cdots, l,~~i\neq j\}.
\end{equation} 
The set of simple roots of $C_l$ are 
\begin{equation}
\{\alpha^\vee_1=\beta_1-\beta_2,\ \alpha^\vee_2=\beta_2-\beta_3,\cdots,\alpha^\vee_{l-1}=\beta_{l-1}-\beta_l,\ \alpha^\vee_l=2\beta_l\}.
\end{equation}
By definition there is a natural bijection between $\hat{\Delta}^\vee$ and $\hat{\Delta}^C$ which simply sends $\alpha+n\delta\in\hat{\Delta}^\vee$ into $\alpha+\frac{n}{2}\delta \in\hat{\Delta}^C$\footnote{This is because our $\delta$ is chosen to be $n$ times the $\delta$ in literature like \cite{carter2005lie}.}.

To find the fixed points, we compute the following two sets
\begin{align}
L_\nu=\{\alpha+l\delta \in \hat\Delta^C~|~ \frac{u}{2(2l-1)} \alpha( \rho^\vee) +l =0\} \to L_\nu=\emptyset.
\end{align}
and 
\begin{equation}
S_\nu=\{\alpha + l\delta\in \hat\Delta~|~\frac{u}{2(2l-1)} {\alpha}( \rho^\vee)) +l =\frac{u}{2(2l-1)}\} \to S_\nu=\{ \alpha^\vee_1, \alpha^\vee_2,\cdots,\alpha^\vee_l, -\theta^\vee+\frac{u}{2}\delta\}.
\end{equation}
Here $\theta^\vee=\beta_1+\beta_2$ is the highest short root of $C_l$ which is Langlands dual to the highest long root of $B_l$ and has height $2l-2$. 

We see that the bijection between $\hat{\Delta}$ and $\hat{\Delta}^C$ also sends $S^\vee_u$ into $S_\nu$. Also using the fact that both the affine Weyl group and extended Weyl group of $\hat{B}_l$ and $^2\hat{A}_{2l-1}$ are the same. One can see the natural isomorphisms between admissible modules and fixed points.
\end{example}

\bibliographystyle{JHEP}
\bibliography{ref}

\providecommand{\href}[2]{#2}\begingroup\raggedright\begin{thebibliography}{100}

\bibitem{Greene:1990ud}
B.R.~Greene and M.R.~Plesser, \emph{{Duality in {Calabi-Yau} Moduli Space}},
  \href{https://doi.org/10.1016/0550-3213(90)90622-K}{\emph{Nucl. Phys. B}
  {\bfseries 338} (1990) 15}.

\bibitem{Candelas:1990rm}
P.~Candelas, X.C.~De~La~Ossa, P.S.~Green and L.~Parkes, \emph{{A Pair of
  Calabi-Yau manifolds as an exactly soluble superconformal theory}},
  \href{https://doi.org/10.1016/0550-3213(91)90292-6}{\emph{Nucl. Phys. B}
  {\bfseries 359} (1991) 21}.

\bibitem{Intriligator:1996ex}
K.A.~Intriligator and N.~Seiberg, \emph{{Mirror symmetry in three-dimensional
  gauge theories}},
  \href{https://doi.org/10.1016/0370-2693(96)01088-X}{\emph{Phys. Lett. B}
  {\bfseries 387} (1996) 513}
  [\href{https://arxiv.org/abs/hep-th/9607207}{{\ttfamily hep-th/9607207}}].

\bibitem{braden2010gale}
T.~Braden, A.~Licata, N.~Proudfoot and B.~Webster, \emph{{Gale duality and
  Koszul duality}}, {\emph{Advances in Mathematics} {\bfseries 225} (2010)
  2002}.

\bibitem{braden2012quantizations}
T.~Braden, N.~Proudfoot and B.~Webster, \emph{{Quantizations of conical
  symplectic resolutions I: local and global structure}},
  \href{https://arxiv.org/abs/1208.3863}{{\ttfamily 1208.3863}}.

\bibitem{braden2014quantizations}
T.~Braden, A.~Licata, N.~Proudfoot and B.~Webster, \emph{{Quantizations of
  conical symplectic resolutions II: category $\mathcal O$ and symplectic
  duality}},  \href{https://arxiv.org/abs/1407.0964}{{\ttfamily 1407.0964}}.

\bibitem{Bullimore:2016nji}
M.~Bullimore, T.~Dimofte, D.~Gaiotto and J.~Hilburn, \emph{{Boundaries, Mirror
  Symmetry, and Symplectic Duality in 3d $\mathcal{N}=4$ Gauge Theory}},
  \href{https://doi.org/10.1007/JHEP10(2016)108}{\emph{JHEP} {\bfseries 10}
  (2016) 108} [\href{https://arxiv.org/abs/1603.08382}{{\ttfamily
  1603.08382}}].

\bibitem{Seiberg:1996nz}
N.~Seiberg and E.~Witten, \emph{{Gauge dynamics and compactification to
  three-dimensions}},  in \emph{{Conference on the Mathematical Beauty of
  Physics (In Memory of C. Itzykson)}}, pp.~333--366, 6, 1996
  [\href{https://arxiv.org/abs/hep-th/9607163}{{\ttfamily hep-th/9607163}}].

\bibitem{Fredrickson:2017jcf}
L.~Fredrickson and A.~Neitzke, \emph{{From $S^1$-fixed points to
  $\mathcal{W}$-algebra representations}},
  \href{https://arxiv.org/abs/1709.06142}{{\ttfamily 1709.06142}}.

\bibitem{Fredrickson:2017yka}
L.~Fredrickson, D.~Pei, W.~Yan and K.~Ye, \emph{{Argyres-Douglas Theories,
  Chiral Algebras and Wild Hitchin Characters}},
  \href{https://doi.org/10.1007/JHEP01(2018)150}{\emph{JHEP} {\bfseries 01}
  (2018) 150} [\href{https://arxiv.org/abs/1701.08782}{{\ttfamily
  1701.08782}}].

\bibitem{Dedushenko:2018bpp}
M.~Dedushenko, S.~Gukov, H.~Nakajima, D.~Pei and K.~Ye, \emph{{3d TQFTs from
  Argyres-Douglas theories}},
  \href{https://doi.org/10.1088/1751-8121/abb481}{\emph{J. Phys. A} {\bfseries
  53} (2020) 43LT01} [\href{https://arxiv.org/abs/1809.04638}{{\ttfamily
  1809.04638}}].

\bibitem{Beem:2013sza}
C.~Beem, M.~Lemos, P.~Liendo, W.~Peelaers, L.~Rastelli and B.C.~van Rees,
  \emph{{Infinite Chiral Symmetry in Four Dimensions}},
  \href{https://doi.org/10.1007/s00220-014-2272-x}{\emph{Commun. Math. Phys.}
  {\bfseries 336} (2015) 1359}
  [\href{https://arxiv.org/abs/1312.5344}{{\ttfamily 1312.5344}}].

\bibitem{Song:2017oew}
J.~Song, D.~Xie and W.~Yan, \emph{{Vertex operator algebras of Argyres-Douglas
  theories from M5-branes}},
  \href{https://doi.org/10.1007/JHEP12(2017)123}{\emph{JHEP} {\bfseries 12}
  (2017) 123} [\href{https://arxiv.org/abs/1706.01607}{{\ttfamily
  1706.01607}}].

\bibitem{Beem:2017ooy}
C.~Beem and L.~Rastelli, \emph{{Vertex operator algebras, Higgs branches, and
  modular differential equations}},
  \href{https://doi.org/10.1007/JHEP08(2018)114}{\emph{JHEP} {\bfseries 08}
  (2018) 114} [\href{https://arxiv.org/abs/1707.07679}{{\ttfamily
  1707.07679}}].

\bibitem{arakawa2018chiral}
T.~Arakawa, \emph{{Chiral algebras of class $\mathcal{S}$ and Moore-Tachikawa
  symplectic varieties}},  \href{https://arxiv.org/abs/1811.01577}{{\ttfamily
  1811.01577}}.

\bibitem{Xie:2016evu}
D.~Xie, W.~Yan and S.-T.~Yau, \emph{{Chiral algebra of the Argyres-Douglas
  theory from M5 branes}},
  \href{https://doi.org/10.1103/PhysRevD.103.065003}{\emph{Phys. Rev. D}
  {\bfseries 103} (2021) 065003}
  [\href{https://arxiv.org/abs/1604.02155}{{\ttfamily 1604.02155}}].

\bibitem{Wang:2018gvb}
Y.~Wang and D.~Xie, \emph{{Codimension-two defects and Argyres-Douglas theories
  from outer-automorphism twist in 6d $(2,0)$ theories}},
  \href{https://doi.org/10.1103/PhysRevD.100.025001}{\emph{Phys. Rev. D}
  {\bfseries 100} (2019) 025001}
  [\href{https://arxiv.org/abs/1805.08839}{{\ttfamily 1805.08839}}].

\bibitem{Xie:2019yds}
D.~Xie and W.~Yan, \emph{{W algebras, cosets and VOAs for 4d $ \mathcal{N} $ =
  2 SCFTs from M5 branes}},
  \href{https://doi.org/10.1007/JHEP04(2021)076}{\emph{JHEP} {\bfseries 04}
  (2021) 076} [\href{https://arxiv.org/abs/1902.02838}{{\ttfamily
  1902.02838}}].

\bibitem{Xie:2012hs}
D.~Xie, \emph{{General Argyres-Douglas Theory}},
  \href{https://doi.org/10.1007/JHEP01(2013)100}{\emph{JHEP} {\bfseries 01}
  (2013) 100} [\href{https://arxiv.org/abs/1204.2270}{{\ttfamily 1204.2270}}].

\bibitem{Wang:2015mra}
Y.~Wang and D.~Xie, \emph{{Classification of Argyres-Douglas theories from M5
  branes}}, \href{https://doi.org/10.1103/PhysRevD.94.065012}{\emph{Phys. Rev.
  D} {\bfseries 94} (2016) 065012}
  [\href{https://arxiv.org/abs/1509.00847}{{\ttfamily 1509.00847}}].

\bibitem{Buican:2015ina}
M.~Buican and T.~Nishinaka, \emph{{On the superconformal index of
  Argyres--Douglas theories}},
  \href{https://doi.org/10.1088/1751-8113/49/1/015401}{\emph{J. Phys.}
  {\bfseries A49} (2016) 015401}
  [\href{https://arxiv.org/abs/1505.05884}{{\ttfamily 1505.05884}}].

\bibitem{Buican:2015hsa}
M.~Buican and T.~Nishinaka, \emph{{Argyres-Douglas theories, S$^1$ reductions,
  and topological symmetries}},
  \href{https://doi.org/10.1088/1751-8113/49/4/045401}{\emph{J. Phys.}
  {\bfseries A49} (2016) 045401}
  [\href{https://arxiv.org/abs/1505.06205}{{\ttfamily 1505.06205}}].

\bibitem{Cordova:2015nma}
C.~Cordova and S.-H.~Shao, \emph{{Schur Indices, BPS Particles, and
  Argyres-Douglas Theories}},
  \href{https://doi.org/10.1007/JHEP01(2016)040}{\emph{JHEP} {\bfseries 01}
  (2016) 040} [\href{https://arxiv.org/abs/1506.00265}{{\ttfamily
  1506.00265}}].

\bibitem{Buican:2015tda}
M.~Buican and T.~Nishinaka, \emph{{Argyres-Douglas Theories, the Macdonald
  Index, and an RG Inequality}},
  \href{https://doi.org/10.1007/JHEP02(2016)159}{\emph{JHEP} {\bfseries 02}
  (2016) 159} [\href{https://arxiv.org/abs/1509.05402}{{\ttfamily
  1509.05402}}].

\bibitem{Song:2015wta}
J.~Song, \emph{{Superconformal indices of generalized Argyres-Douglas theories
  from 2d TQFT}}, \href{https://doi.org/10.1007/JHEP02(2016)045}{\emph{JHEP}
  {\bfseries 02} (2016) 045}
  [\href{https://arxiv.org/abs/1509.06730}{{\ttfamily 1509.06730}}].

\bibitem{Cecotti:2015lab}
S.~Cecotti, J.~Song, C.~Vafa and W.~Yan, \emph{{Superconformal Index, BPS
  Monodromy and Chiral Algebras}},
  \href{https://doi.org/10.1007/JHEP11(2017)013}{\emph{JHEP} {\bfseries 11}
  (2017) 013} [\href{https://arxiv.org/abs/1511.01516}{{\ttfamily
  1511.01516}}].

\bibitem{Nishinaka:2016hbw}
T.~Nishinaka and Y.~Tachikawa, \emph{{On 4d rank-one $ \mathcal{N}=3 $
  superconformal field theories}},
  \href{https://doi.org/10.1007/JHEP09(2016)116}{\emph{JHEP} {\bfseries 09}
  (2016) 116} [\href{https://arxiv.org/abs/1602.01503}{{\ttfamily
  1602.01503}}].

\bibitem{Buican:2016arp}
M.~Buican and T.~Nishinaka, \emph{{Conformal Manifolds in Four Dimensions and
  Chiral Algebras}},
  \href{https://doi.org/10.1088/1751-8113/49/46/465401}{\emph{J. Phys.}
  {\bfseries A49} (2016) 465401}
  [\href{https://arxiv.org/abs/1603.00887}{{\ttfamily 1603.00887}}].

\bibitem{Cordova:2016uwk}
C.~Cordova, D.~Gaiotto and S.-H.~Shao, \emph{{Infrared Computations of Defect
  Schur Indices}}, \href{https://doi.org/10.1007/JHEP11(2016)106}{\emph{JHEP}
  {\bfseries 11} (2016) 106}
  [\href{https://arxiv.org/abs/1606.08429}{{\ttfamily 1606.08429}}].

\bibitem{Song:2016yfd}
J.~Song, \emph{{Macdonald Index and Chiral Algebra}},
  \href{https://doi.org/10.1007/JHEP08(2017)044}{\emph{JHEP} {\bfseries 08}
  (2017) 044} [\href{https://arxiv.org/abs/1612.08956}{{\ttfamily
  1612.08956}}].

\bibitem{Creutzig:2017qyf}
T.~Creutzig, \emph{{W-algebras for Argyres-Douglas theories}}, {\emph{European
  Journal of Mathematics} {\bfseries 3} (2017) 659}
  [\href{https://arxiv.org/abs/1701.05926}{{\ttfamily 1701.05926}}].

\bibitem{Cordova:2017ohl}
C.~Cordova, D.~Gaiotto and S.-H.~Shao, \emph{{Surface Defect Indices and 2d-4d
  BPS States}}, \href{https://doi.org/10.1007/JHEP12(2017)078}{\emph{JHEP}
  {\bfseries 12} (2017) 078}
  [\href{https://arxiv.org/abs/1703.02525}{{\ttfamily 1703.02525}}].

\bibitem{Cordova:2017mhb}
C.~Cordova, D.~Gaiotto and S.-H.~Shao, \emph{{Surface Defects and Chiral
  Algebras}}, \href{https://doi.org/10.1007/JHEP05(2017)140}{\emph{JHEP}
  {\bfseries 05} (2017) 140}
  [\href{https://arxiv.org/abs/1704.01955}{{\ttfamily 1704.01955}}].

\bibitem{Buican:2017uka}
M.~Buican and T.~Nishinaka, \emph{{On Irregular Singularity Wave Functions and
  Superconformal Indices}},
  \href{https://doi.org/10.1007/JHEP09(2017)066}{\emph{JHEP} {\bfseries 09}
  (2017) 066} [\href{https://arxiv.org/abs/1705.07173}{{\ttfamily
  1705.07173}}].

\bibitem{Buican:2017fiq}
M.~Buican, Z.~Laczko and T.~Nishinaka, \emph{{$ \mathcal{N} $ = 2 S-duality
  revisited}}, \href{https://doi.org/10.1007/JHEP09(2017)087}{\emph{JHEP}
  {\bfseries 09} (2017) 087}
  [\href{https://arxiv.org/abs/1706.03797}{{\ttfamily 1706.03797}}].

\bibitem{Buican:2017rya}
M.~Buican and Z.~Laczko, \emph{{Nonunitary Lagrangians and unitary
  non-Lagrangian conformal field theories}},
  \href{https://doi.org/10.1103/PhysRevLett.120.081601}{\emph{Phys. Rev. Lett.}
  {\bfseries 120} (2018) 081601}
  [\href{https://arxiv.org/abs/1711.09949}{{\ttfamily 1711.09949}}].

\bibitem{Choi:2017nur}
J.~Choi and T.~Nishinaka, \emph{{On the chiral algebra of Argyres-Douglas
  theories and S-duality}},
  \href{https://doi.org/10.1007/JHEP04(2018)004}{\emph{JHEP} {\bfseries 04}
  (2018) 004} [\href{https://arxiv.org/abs/1711.07941}{{\ttfamily
  1711.07941}}].

\bibitem{Creutzig:2018lbc}
T.~Creutzig, \emph{{Logarithmic W-algebras and Argyres-Douglas theories at
  higher rank}}, \href{https://doi.org/10.1007/JHEP11(2018)188}{\emph{JHEP}
  {\bfseries 11} (2018) 188}
  [\href{https://arxiv.org/abs/1809.01725}{{\ttfamily 1809.01725}}].

\bibitem{Nishinaka:2018zwq}
T.~Nishinaka, S.~Sasa and R.-D.~Zhu, \emph{{On the Correspondence between
  Surface Operators in Argyres-Douglas Theories and Modules of Chiral
  Algebra}}, \href{https://doi.org/10.1007/JHEP03(2019)091}{\emph{JHEP}
  {\bfseries 03} (2019) 091}
  [\href{https://arxiv.org/abs/1811.11772}{{\ttfamily 1811.11772}}].

\bibitem{Beem:2019snk}
C.~Beem, C.~Meneghelli, W.~Peelaers and L.~Rastelli, \emph{{VOAs and rank-two
  instanton SCFTs}},
  \href{https://doi.org/10.1007/s00220-020-03746-9}{\emph{Commun. Math. Phys.}
  {\bfseries 377} (2020) 2553}
  [\href{https://arxiv.org/abs/1907.08629}{{\ttfamily 1907.08629}}].

\bibitem{achar2003order}
P.N.~Achar, \emph{{An order-reversing duality map for conjugacy classes in
  Lusztig's canonical quotient}}, {\emph{Transformation groups} {\bfseries 8}
  (2003) 107}.

\bibitem{Xie:2023pre}
P.~Shan, D.~Xie and W.~Yan, \emph{{Verlinde algebras for W-algebras and DAHA}},
  {\emph{{in preparation}} }.

\bibitem{kac2008rationality}
V.G.~Kac and M.~Wakimoto, \emph{{On rationality of W-algebras}},
  {\emph{Transformation Groups} {\bfseries 13} (2008) 671}.

\bibitem{kac1988modular}
V.G.~Kac and M.~Wakimoto, \emph{{Modular invariant representations of
  infinite-dimensional Lie algebras and superalgebras}}, {\emph{Proceedings of
  the National Academy of Sciences} {\bfseries 85} (1988) 4956}.

\bibitem{varagnolo2009finite}
M.~Varagnolo and E.~Vasserot, \emph{{Finite-dimensional representations of DAHA
  and affine Springer fibers: The spherical case}}, {\emph{Duke Math. J.}
  {\bfseries 146} (2009) 439}.

\bibitem{oblomkov2016geometric}
A.~Oblomkov and Z.~Yun, \emph{{Geometric representations of graded and rational
  Cherednik algebras}}, {\emph{Advances in Mathematics} {\bfseries 292} (2016)
  601}.

\bibitem{cherednik2005double}
I.~Cherednik, \emph{Double Affine Hecke Algebras}, Lecture note series,
  Cambridge University Press (2005).

\bibitem{Gukov:2022gei}
S.~Gukov, P.~Koroteev, S.~Nawata, D.~Pei and I.~Saberi, \emph{{Branes and DAHA
  Representations}},  \href{https://arxiv.org/abs/2206.03565}{{\ttfamily
  2206.03565}}.

\bibitem{Kozcaz:2018usv}
C.~Koz\c{c}az, S.~Shakirov and W.~Yan, \emph{{Argyres\textendash{}Douglas
  theories, modularity of minimal models and refined
  Chern\textendash{}Simons}},
  \href{https://doi.org/10.4310/ATMP.2022.v26.n3.a3}{\emph{Adv. Theor. Math.
  Phys.} {\bfseries 26} (2022) 643}
  [\href{https://arxiv.org/abs/1801.08316}{{\ttfamily 1801.08316}}].

\bibitem{zhu1996modular}
Y.~Zhu, \emph{Modular invariance of characters of vertex operator algebras},
  {\emph{Journal of the American Mathematical Society} {\bfseries 9} (1996)
  237}.

\bibitem{boalch2008irregular}
P.~Boalch, \emph{{Irregular connections and Kac-Moody root systems}},
  \href{https://arxiv.org/abs/0806.1050}{{\ttfamily 0806.1050}}.

\bibitem{Benini:2010uu}
F.~Benini, Y.~Tachikawa and D.~Xie, \emph{{Mirrors of 3d Sicilian theories}},
  \href{https://doi.org/10.1007/JHEP09(2010)063}{\emph{JHEP} {\bfseries 09}
  (2010) 063} [\href{https://arxiv.org/abs/1007.0992}{{\ttfamily 1007.0992}}].

\bibitem{Xie:2021ewm}
D.~Xie, \emph{{3d mirror for Argyres-Douglas theories}},
  \href{https://arxiv.org/abs/2107.05258}{{\ttfamily 2107.05258}}.

\bibitem{de2006finite}
A.~De~Sole and V.G.~Kac, \emph{{Finite vs affine W-algebras}}, {\emph{Japanese
  Journal of Mathematics} {\bfseries 1} (2006) 137}.

\bibitem{Cecotti:2010fi}
S.~Cecotti, A.~Neitzke and C.~Vafa, \emph{{R-twisting and 4d/2d
  correspondences}},  \href{https://arxiv.org/abs/1006.3435}{{\ttfamily
  1006.3435}}.

\bibitem{Seiberg:1994rs}
N.~Seiberg and E.~Witten, \emph{{Electric-magnetic duality, monopole
  condensation, and confinement in N=2 supersymmetric Yang-Mills theory}},
  \href{https://doi.org/10.1016/0550-3213(94)90124-4,
  10.1016/0550-3213(94)00449-8}{\emph{Nucl. Phys.} {\bfseries B426} (1994) 19}
  [\href{https://arxiv.org/abs/hep-th/9407087}{{\ttfamily hep-th/9407087}}].

\bibitem{Seiberg:1994aj}
N.~Seiberg and E.~Witten, \emph{{Monopoles, duality and chiral symmetry
  breaking in N=2 supersymmetric QCD}},
  \href{https://doi.org/10.1016/0550-3213(94)90214-3}{\emph{Nucl. Phys.}
  {\bfseries B431} (1994) 484}
  [\href{https://arxiv.org/abs/hep-th/9408099}{{\ttfamily hep-th/9408099}}].

\bibitem{Gaiotto:2009we}
D.~Gaiotto, \emph{{N=2 dualities}},
  \href{https://doi.org/10.1007/JHEP08(2012)034}{\emph{JHEP} {\bfseries 08}
  (2012) 034} [\href{https://arxiv.org/abs/0904.2715}{{\ttfamily 0904.2715}}].

\bibitem{Gaiotto:2009hg}
D.~Gaiotto, G.W.~Moore and A.~Neitzke, \emph{{Wall-crossing, Hitchin systems,
  and the WKB approximation}},
  \href{https://doi.org/10.1016/j.aim.2012.09.027}{\emph{Adv. Math.} {\bfseries
  234} (2013) 239} [\href{https://arxiv.org/abs/0907.3987}{{\ttfamily
  0907.3987}}].

\bibitem{reeder2012gradings}
M.~Reeder, P.~Levy, J.-K.~Yu and B.H.~Gross, \emph{{Gradings of positive rank
  on simple Lie algebras}}, {\emph{Transformation Groups} {\bfseries 17} (2012)
  1123}.

\bibitem{Xie:2015rpa}
D.~Xie and S.-T.~Yau, \emph{{4d N=2 SCFT and singularity theory Part I:
  Classification}},  \href{https://arxiv.org/abs/1510.01324}{{\ttfamily
  1510.01324}}.

\bibitem{Chacaltana:2012zy}
O.~Chacaltana, J.~Distler and Y.~Tachikawa, \emph{{Nilpotent orbits and
  codimension-two defects of 6d N=(2,0) theories}},
  \href{https://doi.org/10.1142/S0217751X1340006X}{\emph{Int. J. Mod. Phys. A}
  {\bfseries 28} (2013) 1340006}
  [\href{https://arxiv.org/abs/1203.2930}{{\ttfamily 1203.2930}}].

\bibitem{Li:2022njl}
B.~Li, D.~Xie and W.~Yan, \emph{{On low rank 4d $ \mathcal{N} $ = 2 SCFTs}},
  \href{https://doi.org/10.1007/JHEP05(2023)132}{\emph{JHEP} {\bfseries 05}
  (2023) 132} [\href{https://arxiv.org/abs/2212.03089}{{\ttfamily
  2212.03089}}].

\bibitem{bezrukavnikov2022non}
R.~Bezrukavnikov, P.B.~Alvarez, M.~McBreen and Z.~Yun, \emph{{Non-abelian Hodge
  moduli spaces and homogeneous affine Springer fibers}},
  \href{https://arxiv.org/abs/2209.14695}{{\ttfamily 2209.14695}}.

\bibitem{Chen:2016bzh}
B.~Chen, D.~Xie, S.-T.~Yau, S.S.T.~Yau and H.~Zuo, \emph{{4D $\mathcal{N} = 2$
  SCFT and singularity theory. Part II: complete intersection}},
  \href{https://doi.org/10.4310/ATMP.2017.v21.n1.a2}{\emph{Adv. Theor. Math.
  Phys.} {\bfseries 21} (2017) 121}
  [\href{https://arxiv.org/abs/1604.07843}{{\ttfamily 1604.07843}}].

\bibitem{Wang:2016yha}
Y.~Wang, D.~Xie, S.S.T.~Yau and S.-T.~Yau, \emph{{$4d$ $\mathcal{N} = 2$ SCFT
  from complete intersection singularity}},
  \href{https://doi.org/10.4310/ATMP.2017.v21.n3.a6}{\emph{Adv. Theor. Math.
  Phys.} {\bfseries 21} (2017) 801}
  [\href{https://arxiv.org/abs/1606.06306}{{\ttfamily 1606.06306}}].

\bibitem{Chen:2017wkw}
B.~Chen, D.~Xie, S.S.T.~Yau, S.-T.~Yau and H.~Zuo, \emph{{4d $\mathcal{N}=2$
  SCFT and singularity theory Part III: Rigid singularity}},
  \href{https://doi.org/10.4310/ATMP.2018.v22.n8.a2}{\emph{Adv. Theor. Math.
  Phys.} {\bfseries 22} (2018) 1885}
  [\href{https://arxiv.org/abs/1712.00464}{{\ttfamily 1712.00464}}].

\bibitem{Xie:2017aqx}
D.~Xie and K.~Ye, \emph{{Argyres-Douglas matter and S-duality: Part II}},
  \href{https://doi.org/10.1007/JHEP03(2018)186}{\emph{JHEP} {\bfseries 03}
  (2018) 186} [\href{https://arxiv.org/abs/1711.06684}{{\ttfamily
  1711.06684}}].

\bibitem{kazhdan1988fixed}
D.~Kazhdan and G.~Lusztig, \emph{{Fixed point varieties on affine flag
  manifolds}}, {\emph{Israel Journal of Mathematics} {\bfseries 62} (1988)
  129}.

\bibitem{bezrukavnikov1996dimension}
R.~Bezrukavnikov, \emph{{The dimension of the fixed point set on affine flag
  manifolds}}, {\emph{Mathematical Research Letters} {\bfseries 3} (1996) 185}.

\bibitem{Beem:2014rza}
C.~Beem, W.~Peelaers, L.~Rastelli and B.C.~van Rees, \emph{{Chiral algebras of
  class S}}, \href{https://doi.org/10.1007/JHEP05(2015)020}{\emph{JHEP}
  {\bfseries 1505} (2015) 020}
  [\href{https://arxiv.org/abs/1408.6522}{{\ttfamily 1408.6522}}].

\bibitem{Xie:2019vzr}
D.~Xie and W.~Yan, \emph{{4d $\mathcal{N}=2$ SCFTs and lisse W-algebras}},
  \href{https://doi.org/10.1007/JHEP04(2021)271}{\emph{JHEP} {\bfseries 04}
  (2021) 271} [\href{https://arxiv.org/abs/1910.02281}{{\ttfamily
  1910.02281}}].

\bibitem{Kac:1988tf}
V.G.~Kac and M.~Wakimoto, \emph{{{Modular and conformal invariance constraints
  in representation theory of affine algebras}}},
  \href{https://doi.org/10.1016/0001-8708(88)90055-2}{\emph{Adv. Math.}
  {\bfseries 70} (1988) 156}.

\bibitem{Arakawa_2016_rational}
T.~Arakawa, \emph{{Rationality of admissible affine vertex algebras in the
  category O}}, \href{https://doi.org/10.1215/00127094-3165113}{\emph{Duke
  Mathematical Journal} {\bfseries 165} (2016) }.

\bibitem{deBoer:1993iz}
J.~de~Boer and T.~Tjin, \emph{{{The Relation between quantum W algebras and Lie
  algebras}}}, \href{https://doi.org/10.1007/BF02103279}{\emph{Commun. Math.
  Phys.} {\bfseries 160} (1994) 317}
  [\href{https://arxiv.org/abs/hep-th/9302006}{{\ttfamily hep-th/9302006}}].

\bibitem{kac2003quantum}
V.~Kac, S.-S.~Roan and M.~Wakimoto, \emph{Quantum reduction for affine
  superalgebras}, {\emph{Commun. Math. Phys.} {\bfseries 241} (2003) 307}.

\bibitem{Collingwood:1993rr}
D.H.~Collingwood and W.M.~McGovern, \emph{Nilpotent orbits in semisimple Lie
  algebra: an introduction}, CRC Press (1993).

\bibitem{Frenkel:1992ju}
E.~Frenkel, V.~Kac and M.~Wakimoto, \emph{{{Characters and fusion rules for W
  algebras via quantized Drinfeld-Sokolov reductions}}},
  \href{https://doi.org/10.1007/BF02096589}{\emph{Commun. Math. Phys.}
  {\bfseries 147} (1992) 295}.

\bibitem{arakawa2008representation}
T.~Arakawa, \emph{{Representation theory of W-algebras, II: Ramond twisted
  representations}},  \href{https://arxiv.org/abs/0802.1564}{{\ttfamily
  0802.1564}}.

\bibitem{Arakawa2012Rationality}
T.~Arakawa, \emph{{Rationality of W-algebras: principal nilpotent cases}},
  {\emph{Annals of Mathematics} {\bfseries 182} (2012) }.

\bibitem{arakawa2021rationality}
T.~Arakawa and J.~van Ekeren, \emph{{Rationality and Fusion Rules of
  Exceptional W-Algebras}},  \href{https://arxiv.org/abs/1905.11473}{{\ttfamily
  1905.11473}}.

\bibitem{Fasquel_2022}
J.~Fasquel, \emph{{Rationality of the exceptional W-algebras $W_k
  (\mathfrak{sp}_4, f_{subreg})$ associated with subregular nilpotent elements
  of $\mathfrak{sp}_4$}},
  \href{https://doi.org/10.1007/s00220-021-04294-6}{\emph{Commun. Math. Phys.}
  {\bfseries 390} (2022) 33}.

\bibitem{hitchin1987self}
N.J.~Hitchin, \emph{{The self-duality equations on a Riemann surface}},
  {\emph{Proceedings of the London Mathematical Society} {\bfseries 3} (1987)
  59}.

\bibitem{gukov2006gauge}
S.~Gukov and E.~Witten, \emph{{Gauge Theory, Ramification, And The Geometric
  Langlands Program}},  \href{https://arxiv.org/abs/hep-th/0612073}{{\ttfamily
  hep-th/0612073}}.

\bibitem{carter2005lie}
R.~Carter and R.W.~Carter, \emph{Lie algebras of finite and affine type},
  no.~96, Cambridge University Press (2005).

\bibitem{pervse2013note}
O.~Perse, \emph{{A note on representations of some affine vertex algebras of
  type D}}, {\emph{Glasnik matematicki} {\bfseries 48} (2013) 81}
  [\href{https://arxiv.org/abs/1205.3003}{{\ttfamily 1205.3003}}].

\bibitem{Arakawa_2016}
T.~Arakawa and A.~Moreau, \emph{{Joseph Ideals and Lisse Minimal
  $W$-algebras}}, \href{https://doi.org/10.1017/S1474748016000025}{\emph{J.
  Inst. Math. Jussieu} {\bfseries 17} (2018) 397}
  [\href{https://arxiv.org/abs/1506.00710}{{\ttfamily 1506.00710}}].

\bibitem{bourbaki2006groupes}
N.~Bourbaki, \emph{Groupes de Lie}, Springer (2006).

\bibitem{oblomkov2017cohomology}
A.~Oblomkov and Z.~Yun, \emph{{The cohomology ring of certain compactified
  Jacobians}},  \href{https://arxiv.org/abs/1710.05391}{{\ttfamily
  1710.05391}}.

\bibitem{arakawa2018quasi}
T.~Arakawa and K.~Kawasetsu, \emph{Quasi-lisse vertex algebras and modular
  linear differential equations},  in \emph{Lie Groups, Geometry, and
  Representation Theory}, pp.~41--57, Springer (2018).

\bibitem{Zheng:2022zkm}
H.~Zheng, Y.~Pan and Y.~Wang, \emph{{Surface defects, flavored modular
  differential equations, and modularity}},
  \href{https://doi.org/10.1103/PhysRevD.106.105020}{\emph{Phys. Rev. D}
  {\bfseries 106} (2022) 105020}
  [\href{https://arxiv.org/abs/2207.10463}{{\ttfamily 2207.10463}}].

\bibitem{Xie:2019zlb}
D.~Xie and W.~Yan, \emph{{Schur sector of Argyres-Douglas theory and
  $W$-algebra}},
  \href{https://doi.org/10.21468/SciPostPhys.10.3.080}{\emph{SciPost Phys.}
  {\bfseries 10} (2021) 080}
  [\href{https://arxiv.org/abs/1904.09094}{{\ttfamily 1904.09094}}].

\bibitem{gorsky2013arc}
E.~Gorsky, \emph{{Arc spaces and DAHA representations}}, {\emph{Selecta
  Mathematica} {\bfseries 19} (2013) 125}.

\bibitem{hikita2017algebro}
T.~Hikita, \emph{{An algebro-geometric realization of the cohomology ring of
  Hilbert scheme of points in the affine plane}}, {\emph{International
  Mathematics Research Notices} {\bfseries 2017} (2017) 2538}.

\bibitem{arakawa2017representation}
T.~Arakawa, \emph{{Representation theory of W-algebras and Higgs branch
  conjecture}},  in \emph{{International Congress of Mathematicians}},
  pp.~1261--1278, 2018 [\href{https://arxiv.org/abs/1712.07331}{{\ttfamily
  1712.07331}}].

\bibitem{losev2021unipotent}
I.~Losev, L.~Mason-Brown and D.~Matvieievskyi, \emph{{Unipotent Ideals and
  Harish-Chandra Bimodules}},
  \href{https://arxiv.org/abs/2108.03453}{{\ttfamily 2108.03453}}.

\bibitem{MR3456698}
T.~Arakawa, \emph{Associated varieties of modules over {K}ac-{M}oody algebras
  and {$C_2$}-cofiniteness of {$W$}-algebras}, {\emph{Int. Math. Res. Not.
  IMRN} (2015) 11605}.

\bibitem{Gaiotto:2008ak}
D.~Gaiotto and E.~Witten, \emph{{S-Duality of Boundary Conditions In N=4 Super
  Yang-Mills Theory}},
  \href{https://doi.org/10.4310/ATMP.2009.v13.n3.a5}{\emph{Adv. Theor. Math.
  Phys.} {\bfseries 13} (2009) 721}
  [\href{https://arxiv.org/abs/0807.3720}{{\ttfamily 0807.3720}}].

\bibitem{Lemos:2014lua}
M.~Lemos and W.~Peelaers, \emph{{Chiral Algebras for Trinion Theories}},
  \href{https://doi.org/10.1007/JHEP02(2015)113}{\emph{JHEP} {\bfseries 02}
  (2015) 113} [\href{https://arxiv.org/abs/1411.3252}{{\ttfamily 1411.3252}}].

\bibitem{hausel2008mixed}
T.~Hausel and F.~Rodriguez-Villegas, \emph{{Mixed Hodge polynomials of
  character varieties: with an appendix by Nicholas M. Katz}},
  {\emph{Inventiones mathematicae} {\bfseries 174} (2008) 555}.

\bibitem{Xie:2023lko}
D.~Xie, \emph{{Pseudo-periodic map and classification of theories with eight
  supercharges}},  \href{https://arxiv.org/abs/2304.13663}{{\ttfamily
  2304.13663}}.

\bibitem{cherkis2011instantons}
S.A.~Cherkis, \emph{{Instantons on Gravitons}},
  \href{https://doi.org/10.1007/s00220-011-1293-y}{\emph{Commun. Math. Phys.}
  {\bfseries 306} (2011) 449}
  [\href{https://arxiv.org/abs/1007.0044}{{\ttfamily 1007.0044}}].

\bibitem{cherkis2012moduli}
S.A.~Cherkis and R.S.~Ward, \emph{{Moduli of Monopole Walls and Amoebas}},
  \href{https://doi.org/10.1007/JHEP05(2012)090}{\emph{JHEP} {\bfseries 05}
  (2012) 090} [\href{https://arxiv.org/abs/1202.1294}{{\ttfamily 1202.1294}}].

\bibitem{intriligator2000compactified}
K.A.~Intriligator, \emph{{Compactified little string theories and compact
  moduli spaces of vacua}},
  \href{https://doi.org/10.1103/PhysRevD.61.106005}{\emph{Phys. Rev. D}
  {\bfseries 61} (2000) 106005}
  [\href{https://arxiv.org/abs/hep-th/9909219}{{\ttfamily hep-th/9909219}}].

\end{thebibliography}\endgroup

\end{document}